\documentclass[aps,prx,twocolumn,superscriptaddress,showpacs,floatfix]{revtex4-1}
\usepackage{color}
\usepackage{mathtools}
\usepackage[caption=false]{subfig}
\usepackage{bm}        % for math
\usepackage{amssymb}   % for math
\usepackage{amsmath}
\usepackage[colorlinks=false,citecolor=blue,linkcolor=magenta]{hyperref}
\usepackage{graphicx}
    \usepackage[caption=false]{subfig}
\usepackage{braket}
\usepackage{natbib}
\usepackage [latin1]{inputenc}
\usepackage{siunitx}
\usepackage{blkarray, bigstrut}

\newcommand{\pdg}{{\vphantom\dagger}}
\newcommand{\bea}{\begin{eqnarray}}
\newcommand{\I}{\text{i}}
\newcommand{\eea}{\end{eqnarray}}

\newcommand{\bsigma}{\boldsymbol{\sigma}}
\newcommand{\bdelta}{\boldsymbol{\delta}}
\newcommand{\bS}{\mathbf{S}}
\newcommand{\br}{\mathbf{r}}
\newcommand{\bd}{\mathbf{d}}

\newcommand{\bk}{\mathbf{k}}

\newcommand{\bfm}{\mathbf{m}}

\newcommand{\bj}{\mathbf{j}}
\newcommand{\bs}{\mathbf{s}}
\newcommand{\bp}{\mathbf{p}}

\newcommand{\bD}{\mathbf{D}}

\newcommand{\td}{\text{d}}

\begin{document}

\title{Resonant optical topological Hall conductivity from skyrmions}
\author{Sopheak Sorn}
\email{ssorn@physics.utoronto.ca}
\affiliation{Department of Physics, University of Toronto, Toronto, Ontario M5S1A7, Canada.}
\author{Luyi Yang}
\affiliation{State Key Laboratory of Low Dimensional Quantum Physics, Department of Physics, Tsinghua University, Beijing 100084, China}
\author{Arun Paramekanti}
\email{arunp@physics.utoronto.ca}
\affiliation{Department of Physics, University of Toronto, Toronto, Ontario M5S1A7, Canada.}

\date{\today}

\begin{abstract}
We study the high frequency Hall conductivity in a two-dimensional (2D) model of conduction electrons coupled to a background magnetic skyrmion texture via an effective
Hund's coupling term. For an ordered skyrmion crystal, a Kubo formula calculation using the basis of 
skyrmion crystal Chern bands reveals a resonant Hall response at a frequency set by the Hund's coupling: $\hbar\omega_{\rm res} \approx J_H$.
A complementary real-space Kubo formula calculation for an isolated skyrmion in a box reveals a similar resonant Hall response. A linear relation between the area under the Hall resonant curve and the skyrmion density is discovered numerically and is further elucidated using a gradient expansion which is valid for smooth textures
and a local approximation based on a spin-trimer calculation.
We point out the 
issue of distinguishing this skyrmion contribution from a similar feature arising from spin-orbit interactions, as demonstrated in a model for Rashba spin-orbit coupled electrons in a collinear ferromagnet,
which is analogous to the difficulty of unambiguously separating the d.c. topological 
Hall effect from the anomalous Hall effect.
The resonant feature in the high frequency topological Hall effect is proposed to provide a potentially
useful local optical signature of skyrmions via probes such as scanning magneto-optical Kerr microscopy. 
\end{abstract}

\maketitle

\section{Introduction}
Originally proposed in particle physics \cite{skyrme}, skyrmions are realized in many condensed matter systems as non-coplanar magnetic swirls with a non-trivial integer topological invariant. The topological nature of skyrmions provides them with a sense of robustness against small perturbations, which, together with their small sizes, have made them promising candidates for realizing future dense data-storage technology \cite{Nagaosa2013, Fert2013, Jiang2017, Back2020}. The stabilization, manipulation, and detection of skyrmions have thus emerged as a central theme which has attracted immense research interest in the fundamental and applied aspects of skyrmion physics \cite{Back2020}. Skyrmions have been found in both isolated and crystalline forms in a wide range of magnetic solids \cite{Tokura2021} including quantum Hall systems \cite{Barrett1995, Shkolnikov2005}, non-centrosymmetric magnetic systems \cite{Neubauer2009, Lee2009, Muhlbauer2009, Tokura2012, Yu2010, Seki2012, Moreau2016, Wang2018}, antiferromagnets \cite{Gao2020}, and frustrated magnets \cite{Okubo2012, Kurumaji2019, Hirschberger2019, Khanh2020}. Recently, skyrmions have even been proposed in two-channel Kondo lattice systems where two channels of conduction electron are Kondo coupled symmetrically to a lattice of Kondo spins \cite{Coleman2020}. At low temperature, the channel $SU(2)$ symmetry is spontaneously broken by the Kondo effect, leading to a ``ferromagnetic" ordering of a spinor-like order parameter. Skyrmions then appear as topological defects in the ``ferromagnetic" phase \cite{Coleman2020}, and they have been shown to support movable Majorana zero modes, providing a new, potential platform for realizing topological quantum computation \cite{Flint2021}. Skyrmions have also been proposed to drive superconductivity in twisted bilayer graphene \cite{Vishwanath2021} and there has been interest in studying skyrmions at topological insulator (TI) surfaces and graphene
\cite{nomura2010,Hurst2015,Lado2015,Nomura2017,Nogueira2018,Tiwari2019,wang2020scattering,Paul2021,divic2021magnetic,li2021twisted}.

In experiments, magnetic skyrmions can be detected directly by visualizing the real-space profile of their out-of-plane and in-plane spin components using magnetic force microscopy \cite{Wang2018, Soumyanarayanan2017, Maccariello2018} and Lorentz transmission electron microscopy \cite{Yu2010, Yu2011, Khanh2020} respectively. Other techniques suitable for detecting and studying crystals of skyrmions include X-ray \cite{Kurumaji2019, Hirschberger2019, Khanh2020} and neutron diffraction \cite{Muhlbauer2009, Tokura2012}. In metallic magnets, the noncoplanarity of the skyrmion spin texture imprints a real-space Berry phase on the conduction electrons \cite{He1999, Schulz2012}. The extra Berry phase can be regarded as an emergent magnetic field seen by the electrons, thereby inducing an additional Hall effect known as the topological Hall effect (THE), first observed in MnSi \cite{Neubauer2009, Lee2009}, which provides an indirect transport probe of skyrmions. In addition to charge transport, the real-space Berry phase also affects heat flow through topological Nernst \cite{Tokura2013, Tokura2020} and thermal Hall effects \cite{Kim2019}. However, there is growing evidence of examples in thin films where the ostensible appearance of THE may in fact be an additional Hall effect contribution 
stemming from inhomogeneous magnetic domains and domain wall scattering \cite{Kan2018, Gerber2018, Wang2020, Keimer2020, Wu2020, Bartram2020, Sorn2021}. Hence, one generally needs complementary imaging probes to confirm the existence of skyrmions. 

While the impact of skyrmions on electronic charge and heat transport properties has been actively studied, their impact on electronic optical properties remains 
less explored. To fill in this gap, we examine in this paper the nonzero-frequency Hall conductivity in a model of conduction electrons coupled to skyrmions via an 
effective Hund's coupling $J_H$. 

The band structure of electrons coupled to local moments is conveniently divided into two sectors containing electronic states whose spins are locally parallel or anti-parallel to the underlying skyrmion spins. For a skyrmion crystal (SkX), 
each sector supports bands with nontrivial Chern numbers as a result of the real-space Berry phase; this Berry phase
can be regarded as a fictitious Aharonov-Bohm flux which leads to an effective electronic Hofstadter model. We use the Kubo formula to compute the resulting a.c. Hall conductivity, uncovering nontrivial frequency dependence in two frequency windows. The structure at low frequency is set by the energy gap between neighboring Chern bands, while the high frequency response near $\hbar\omega\! \approx\! J_H$ exhibits a resonance originating from a large number of intersector spin-flip transition channels between pairs of Chern bands which disperse similarly and differ in energy by a similar amount of $J_H$. 
We show that the high frequency resonance 
is robust, occurring also for isolated skyrmions, as demonstrated by a real-space Kubo formula computation for a skyrmion in a box. We provide an analytical understanding of this resonance through  a smooth texture approximation and a local spin-trimer calculation. These show that the area under resonance of 
Im\,$\sigma_{xy}$ scales linearly with the skyrmion density $\rho_{sk}$.

Introducing a Rashba-type spin-orbit coupling (SOC) term into the electronic tight-binding model and considering a ferromagnetic spin texture, we show that SOC also produces a similar Hall effect resonance. The analogy between the Rashba spin-orbit coupling and the coupling to skyrmions may be understood as two analogous ways to mix the orbital and the spin degree of freedom of the electron \cite{Rashba1964, Rashba2020, Egorov2021}. While the mixing takes place in momentum space for the Rashba spin-orbit coupling, it occurs in real space for the coupling to skyrmions. We will discuss how the momentum-space SOC affects the manifestation of the resonant topological Hall effect. Such resonant features in the Hall and Kerr effects have been previously discussed at charged impurities on 
magnetized TI surfaces \cite{Wilson2014} and for time-reversal breaking multipolar magnetic orders in solids \cite{Motome2021}.

The nontrivial high-frequency topological Hall effect is expected to be manifest in experiments such as scanning Kerr microscopy, magneto-optical Kerr effect, and 
Faraday effect. 
Materials hosting a large density of skyrmions, such as Gd$_2$PdSi$_3$\cite{Kurumaji2019}, Gd$_3$Ru$_4$Al$_{12}$ \cite{Hirschberger2019}, GdRu$_2$Si$_2$ \cite{Khanh2020}, and MnGe \cite{Tokura2012}, are likely to be viable candidates for observing the non-zero-frequency topological Hall effect and its resonant feature.

The paper is organized as the following. Section \ref{sec:model} introduces a triangular-lattice model of conduction electrons coupled to a spin texture, and the spin textures studied in this work. Section \ref{sec:skx} is devoted to properties of the SkX spin texture, encompassing its electronic band structures, conductivity 
spectra featuring the Hall resonance, and the numerical linear scaling of the area under the resonance with the skyrmion density.
In Section \ref{sec:resonance}, this scaling relation is explained by an analytical expression obtained from a smooth texture approximation. In Section \ref{sec:flake},
we present the study of a ``skyrmion in a box'' and show that the resonance is robust and occurs even when the spin texture contains only a single skyrmion. 
Section \ref{sec:trimer} discusses an analytical understanding of how this resonance occurs even in the minimal setting with three sites, i.e. a spin trimer,
hosting a noncoplanar spin configuration. Based on this analytical spin trimer result, we try to understand the Hall resonance in the SkX case and the 
skyrmion-in-a-box case as a local spatial average over triangular plaquettes.
In Section \ref{sec:soc}, we discuss how spin orbit coupling in a uniform ferromagnet can also produce a resonance feature similar to skyrmions,
and how it may complicate the issue of extracting the skyrmion contribution to the resonant optical Hall response. 
We conclude in Section \ref{sec:conclusion} with a discussion of how the spatial variation of the imaginary part of the Hall conductivity at the resonant frequency, 
using tools such as scanning Kerr microscopy, might serve as a useful optical probe of skyrmions and SkX.

\section{Model}
\label{sec:model}

\begin{figure}[t]
\centering
\includegraphics[width=0.5\textwidth]{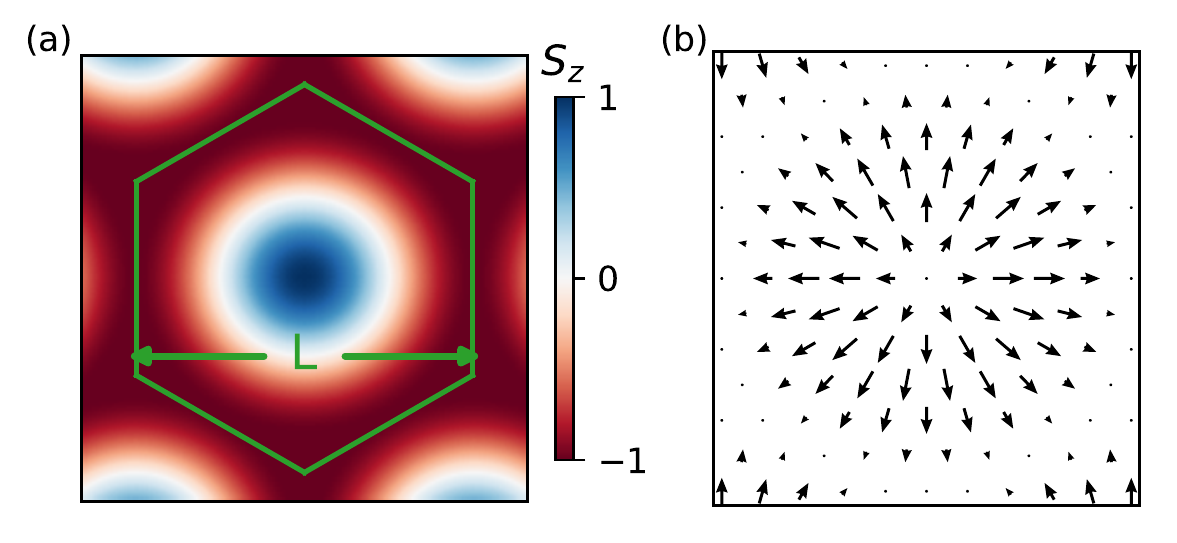}
\caption{Triangular SkX spin texture. (a) Color plot illustrating the profile of the z-component of the spin texture with a hexagonal unit cell whose width is given by $L$. Each unit cell contains a N\'eel skyrmion with a radius $R = L/2$. (b) Profile of the in-plane spin components.}
\label{fig:neel_sk}
\end{figure}

We consider a triangular-lattice model of electrons hopping on nearest-neighbor bonds, sensing a spin-dependent potential of a background 
spin texture $\{\bS_i\}$ via an effective ferromagnetic Hund's coupling $J_H > 0$. The Hamiltonian is given by \cite{Anderson1955, Nagaosa2000, Nagaosa2015}
\bea
H_0 &=& - t \sum_{\langle ij \rangle} \left( c^{\dagger}_{i \sigma}c_{j\sigma} + c^{\dagger}_{j\sigma}c_{i\sigma} \right) - J_H \sum_i \hat{\bs}_i \cdot \bS_i,
\label{eq:model}
\eea
where $\hat{\bs}_i = \frac{1}{2}c^{\dagger}_{i\sigma}\boldsymbol{\sigma}_{\sigma\sigma'}c_{i\sigma'}$ is the electron spin operator, $\bsigma = (\sigma_x, \sigma_y, \sigma_z)$ are the Pauli matrices, and $\bS_i$ is a unit-norm classical spin vector at site $i$ of the spin texture. We set the lattice constant of the triangular
lattice $a\!=\!1$.

We consider two configurations to numerically solve for the electronic properties: 
(i) a triangular crystal of N\'eel skyrmions commensurate with the underlying triangular lattice, and (ii)
an open-boundary hexagonal box containing one skyrmion. 

Figure \ref{fig:neel_sk} illustrates the SkX spin texture with a hexagonal unit cell of width $L$ containing one N\'eel skyrmion and enclosing
$L^2$ sites. The hexagonal box is obtained
by simply cutting out a single unit cell from the lattice. Within the hexagonal box (unit cell), the skyrmion texture is defined by a circular 
ansatz \cite{Banerjee2014}
\bea
\bS_i &=& \left(\sin \theta(\mathfrak{r}_i) \cos \phi(\mathfrak{r}_i), \sin \theta(\mathfrak{r}_i) \sin \phi(\mathfrak{r}_i), \cos \theta(\mathfrak{r}_i) \right),
\eea
where $\mathfrak{r}_i = \br_i - \br_{\rm hex}$ is the position of the $i$-th site relative to the center $\br_{rm hex}$ of the hexagonal box (unit cell).
\bea
\theta(\mathfrak{r}_i) &=& \begin{cases} \frac{\pi|\mathfrak{r}_i|}{R} &\text{ for } |\mathfrak{r}_i| < R,\\
0 &\text{ otherwise},
\end{cases},\\
\cos \phi(\mathfrak{r}_i) &=& \frac{\mathfrak{r}_{i, x}}{|\mathfrak{r}_i|};~~\sin \phi(\mathfrak{r}_i) = \frac{\mathfrak{r}_{i, y}}{|\mathfrak{r}_i|},
\eea
where $\mathfrak{r}_{i, x(y)}$ is the x(y)-component of $\mathfrak{r}_i$. 

In the remainder of the paper, we fix the skyrmion radius $R\!=\!L/2$, with
$R\! \gg\! 1 $ for large skyrmions. We will quote the value of $R$ when discussing variation of the optical response with skyrmion size.

\section{Skyrmion crystal}
\label{sec:skx}

In this section, we focus on the properties of a SkX, including its electronic bands, high frequency resonance in the Hall response, and the scaling of the
resonant feature with the skyrmion density.

\subsection{Electronic bands}
 
\begin{figure}
	\centering
	\includegraphics[width=0.47\textwidth]{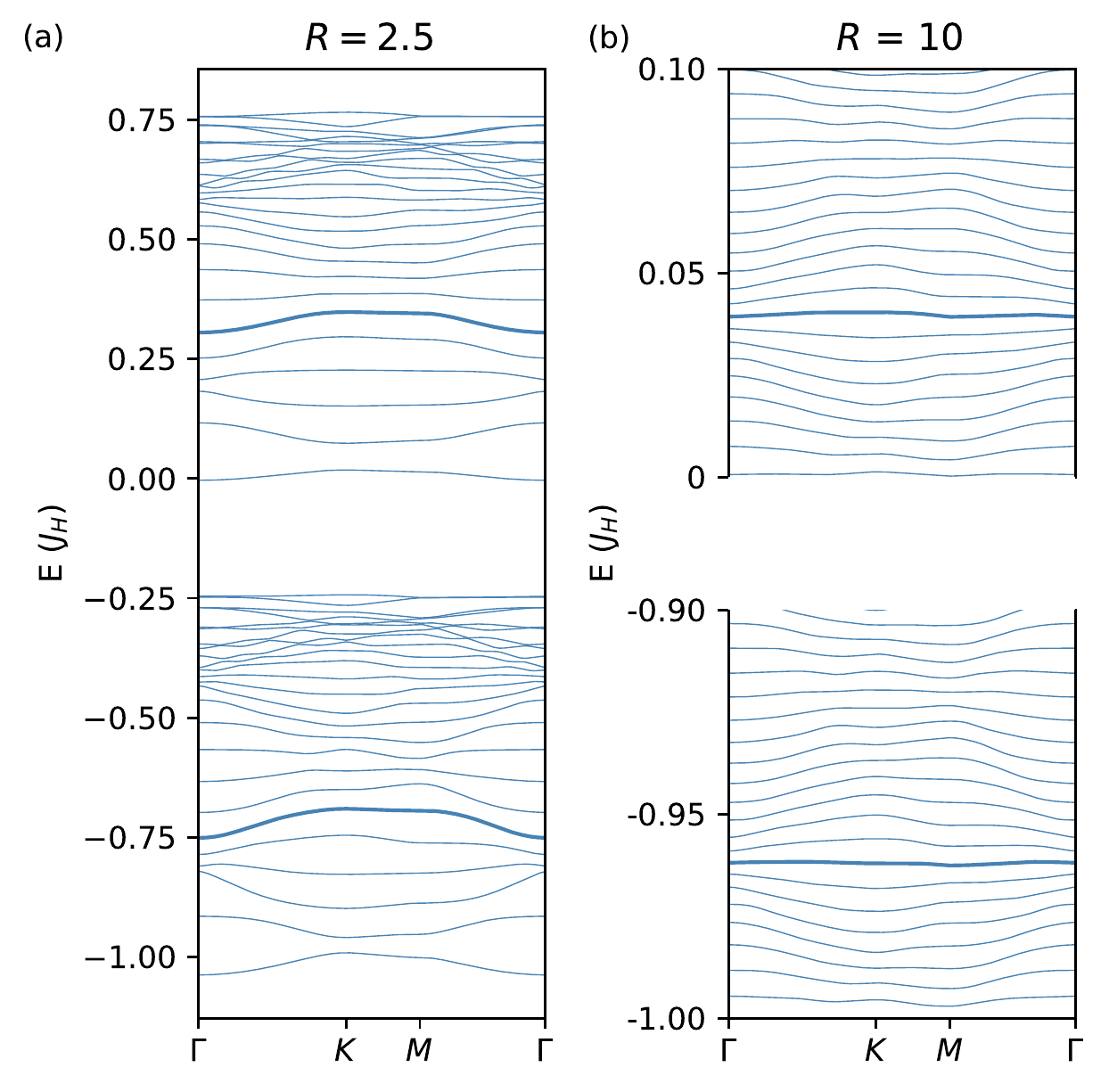}
	\caption{Band structures corresponding to the SkX with (a) $R\!=\! 2.5$ and (b) $R\!=\! 10$, with $J_H\!=\!10 t$. The band structures consist of Chern bands grouped into a low energy parallel sector and a high energy anti-parallel sector. Bold bands illustrate an example of pairs of similarly dispersing low and high energy Chern bands differing in energy by an amount  $\sim\! J_H$, which leads to a resonant feature in the a.c. conductivity at $\hbar\omega \approx J_H$ as shown later.}
	\label{fig:bandstructure}
\end{figure}

Figure \ref{fig:bandstructure} shows the band structures corresponding to the SkX with $R\!=\! 2.5$ in panel (a) and $R\!=\! 10$ in panel (b) when $J_H \!=\! 10t$. These bands are shown along high-symmetry lines of the mini-Brillouin zone (BZ) obtained from folding the original BZ of the underlying triangular lattice. 
Given the large $J_H/t \!\gg\! 1$, the system is in the adiabatic regime, where the bands are effectively Zeeman-split into a low-energy sector and a high-energy sector, differing in energy by $J_H$, corresponding to states
where the electron spins are locally parallel and anti-parallel, respectively, to the underlying skyrmion spins. To highlight this, the
energy in Figure \ref{fig:bandstructure} is measured in units of $J_H$.
Each sector has $L^2=4 R^2$ bands, with a bandwidth $W \! \approx\! 9 t$, so the average energy spacing between a pair of neighboring bands is
$W/4 R^2$; for our choice of $J_H/t\!=\!10$ this corresponds to a level separation $\sim \! J_H/4R^2$.

These bands have non-trivial Chern numbers due to the real-space Berry phase picked up by electrons as they traverse the skyrmion texture,
which admits an effective description of the two sectors using generalized Hofstadter models carrying opposite fluxes 
\cite{Hofstadter1976, Nagaosa2000, Gobel2017, Gobel2018, Sorn2019}.
We find pairs of similarly dispersing Chern bands, one from the parallel sector and the other from the anti-parallel sector, as illustrated by the bold bands in Fig.\ref{fig:bandstructure}. They differ in energy by a similar amount of roughly $J_H$, resulting in a number of optical transition channels at this energy, 
which will be shown later to cause a resonance feature in the a.c. conductivity at $\hbar\omega_{\rm res} \approx J_H$.

The following linear response Kubo formula is used to study the a.c. conductivity tensor $\sigma_{\alpha\beta}(\omega)$ \cite{Mahan2000, Coleman2015}:
\bea
\!\! \!\!  \!\! \sigma_{\alpha\beta} \!\!&=&\!\! \frac{\I \hbar e^2}{{\cal A}} \!\! \sum_{\bk m n} \!\! \frac{f(E_{\bk n}) \!-\! f(E_{\bk m})}{E_{\bk m} \! -\! E_{\bk n}}\frac{(v_{\bk\alpha})_{nm}(v_{\bk\beta})_{mn}}{\hbar \omega\! +\! i\gamma\! +\! E_{\bk n}\! -\! E_{\bk m}},
\eea
where $\alpha, \beta = x, y$, and $\hbar (v_{\bk\alpha})_{nm} \equiv \bra{\bk n} \frac{\partial \mathcal{H}(\bk)}{\partial k_{\alpha}}\ket{\bk m}$ is a matrix element of the velocity operator between the Bloch states $\ket{\bk n}$ and $\ket{\bk m}$ corresponding to the energy eigenvalues $E_{\bk n}$ and $E_{\bk m}$ respectively. $\mathcal{H}(\bk)$ is the Hamiltonian matrix, $f$ is the Fermi distribution, ${\cal A}$ is the area of the 2D system, and $\gamma$ is a small broadening. 
To study the open-boundary skyrmion in a box, the Kubo formula is modified into a real-space version where the composite label $(\bk, m)$ is replaced with the energy level label. Similar to the SkX case, the current operator $\bj = -e \mathbf{v}$ can be obtained from $\delta H_0[\mathbf{A}]/\delta \mathbf{A}$ (see e.g. Appendix A of Ref.\cite{Sorn2021} for more details), where $\mathbf{A}$ is the vector potential of the externally applied field, and $H_0[\mathbf{A}]$ is the Hamiltonian after the Peierls substitution.
In the rest of the paper, we fix parameters $\gamma \!=\! 0.05t$, $J_H \!=\! 10t$, and set the electron filling to $1/6$. The main results can be straightforwardly generalized to other electron fillings, which is discussed in Sec. \ref{sec:resonance}.

\subsection{Frequency dependent conductivity in a skyrmion crystal}
\label{sec:spectrum}

\begin{figure}[t]
	\centering
	\includegraphics[width=0.495\textwidth]{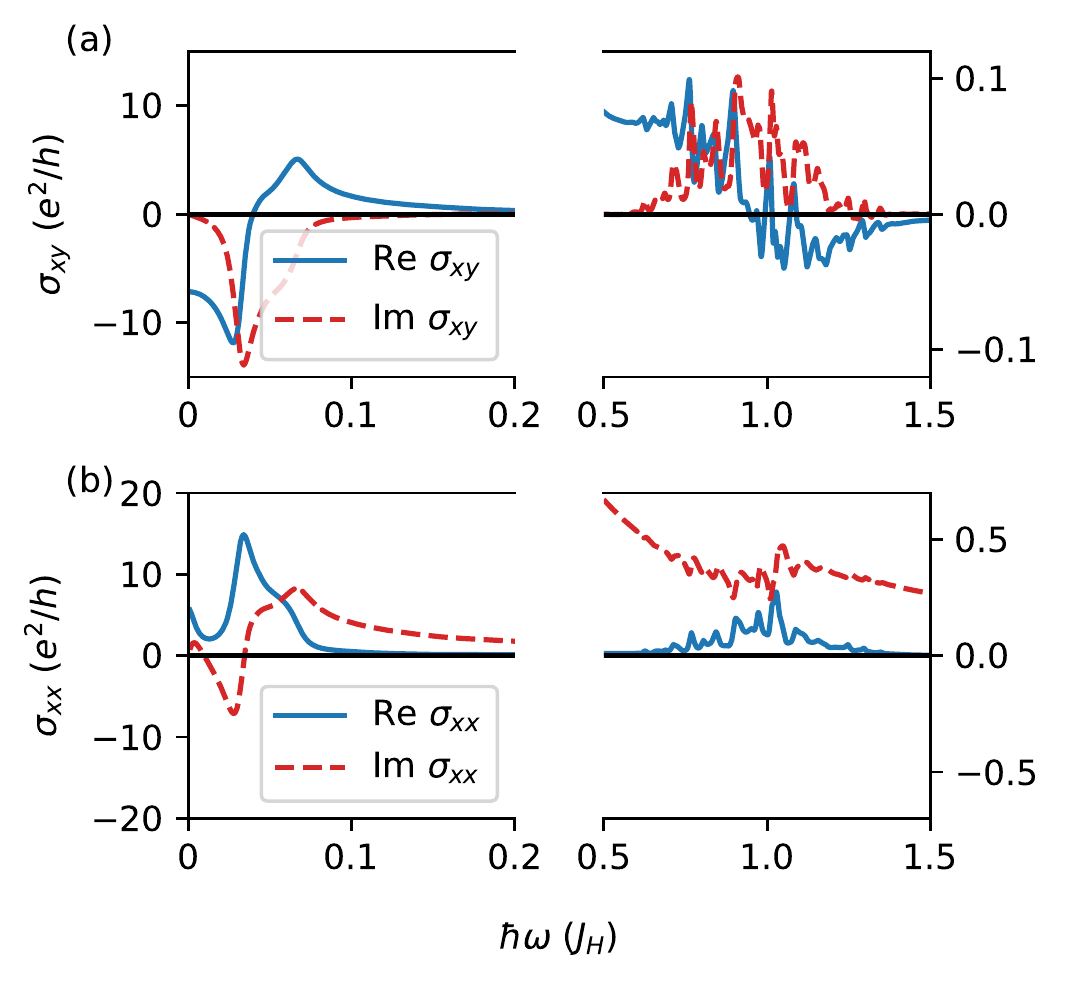}
	\caption{Spectrum of (a) $\sigma_{xy}$ and (b) $\sigma_{xx}$ for a small SkX unit cell $L\!=\!5$ with a skyrmion radius $R\!=\!2.5$ which corresponds to the band structure in Fig.\ref{fig:bandstructure}(a). Two windows of nontrivial frequency dependence are seen: (1) a low frequency regime set by the average energy gap between 
	neighboring Chern bands and (2) a high frequency regime around $\hbar\omega \! \approx\! J_H$. The former originates from transitions among Chern 
	bands within the same, parallel-spin sector, whereas the later arises from spin-flip transitions across pairs of Chern bands which disperse similarly 
	and differ in energy by $\approx\! J_H$.}
	\label{fig:spec_small}
\end{figure}

Figure \ref{fig:spec_small} shows $\sigma_{xy}(\omega)$ and $\sigma_{xx}(\omega)$ obtained from the Kubo formula for the SkX with $R\!=\!2.5$, 
which corresponds to the band structure in Fig.\ref{fig:bandstructure}(a). $\sigma_{xy}$ and $\sigma_{xx}$ exhibit nontrivial frequency dependence in two windows, one at small frequency set by the average energy gap between neighboring Chern bands, and the other occurring around $J_H$. The former arises from intra-sector transitions among Chern bands within the parallel sector, while the latter originates from inter-band spin-flip transitions between Chern bands in the parallel sector and those in the anti-parallel sector. The dissipative parts of the conductivity tensor, Re $\sigma_{xx}$ and Im $\sigma_{xy}$, can be shown to track the joint density of state (JDOS) of the optical transitions $\propto \sum_{\bk}\sum_{E_{\bk n} < \mu}\sum_{E_{\bk m} > \mu}\delta(\hbar \omega - E_{\bk m} + E_{\bk n})$, where $\mu$ is the chemical potential. 
The non-dissipative parts, Im $\sigma_{xx}$ and Re $\sigma_{xy}$ correspondingly exhibit a frequency dependence consistent with Kramers-Kr\"onig relations
given the above dissipative response.
In the high-frequency window, the JDOS peaks at $\hbar\omega \!\approx\! J_H$ as a result of having a number of transition channels due to the presence of many similarly dispersing pairs of Chern bands differing in energy by a similar amount of roughly $J_H$, as mentioned in the previous section.
For a SkX with a small $R$, the average energy gap $W/4R^2$ is significant, so that the Chern bands are well-separated, and the conductivity spectrum displays discrete peaks.

\begin{figure}[t]
	\centering
	\includegraphics[width=0.495\textwidth]{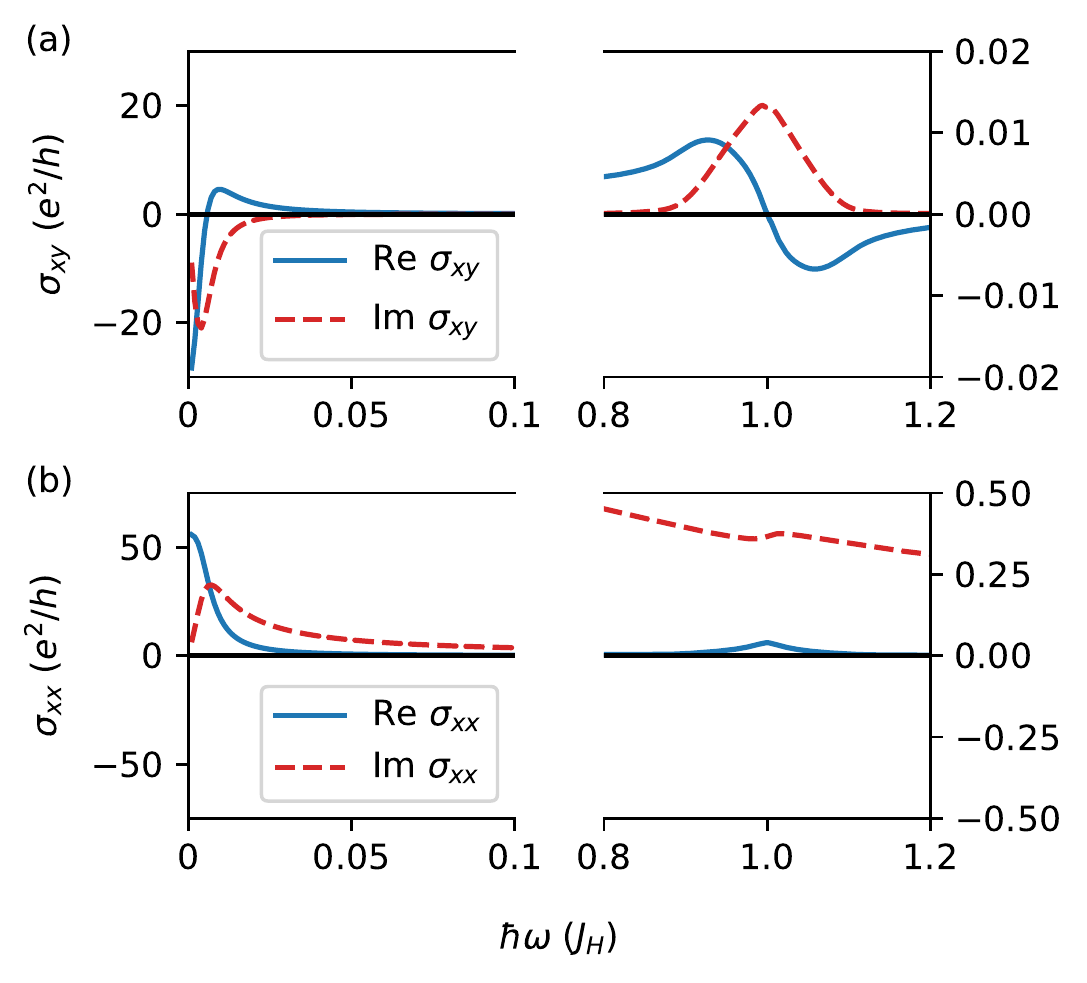}
	\caption{Spectrum of (a) $\sigma_{xy}$ and (b) $\sigma_{xx}$ for a larger SkX unit cell $L\!=\!20$ and larger-radius skyrmion $R\!=\!10$, 
	corresponding to the band structure in Fig.\ref{fig:bandstructure}(b). While the low energy response moves to lower frequency due to the smaller energy separation between neighboring Chern bands, the large-energy response remains around $\hbar\omega\! \approx \! J_H$ . Due to the smaller separation between neighboring Chern bands, the response shows a smooth Lorentzian behavior. The resonant peaks in Im $\sigma_{xy}$ and Re $\sigma_{xx}$ near $J_H$ originate from spin-flip transitions 
	involving many pairs of `energy-nested' Chern bands.}
	\label{fig:spec_large}
\end{figure}

Upon increasing the skyrmion size and the corresponding periodicity of the SkX, the small-frequency window shifts towards zero as the energy gap between 
neighboring Chern bands shrinks.
However, the high-frequency transition remains at around $J_H$, being set by the local spin-flip energy gap. This is illustrated in Fig.\ref{fig:spec_large} which 
shows $\sigma_{xy}(\omega)$ and $\sigma_{xx}(\omega)$ for a SkX with $R\!=\!10$ whose band structure is given by Fig.\ref{fig:bandstructure}(b). 
Due to the shrunken energy gap between adjacent levels, the discrete peaks seen earlier have merged with one another to form a Lorentzian-like resonant 
curve of JDOS peaking around $J_H$, which can again be traced back to the presence of
many similarly dispersing pairs of Chern bands.

\subsection{Dependence on $J_H/t$}
\label{sec:nonadiabatic}

\begin{figure}[t]
	\includegraphics[width = 0.4\textwidth]{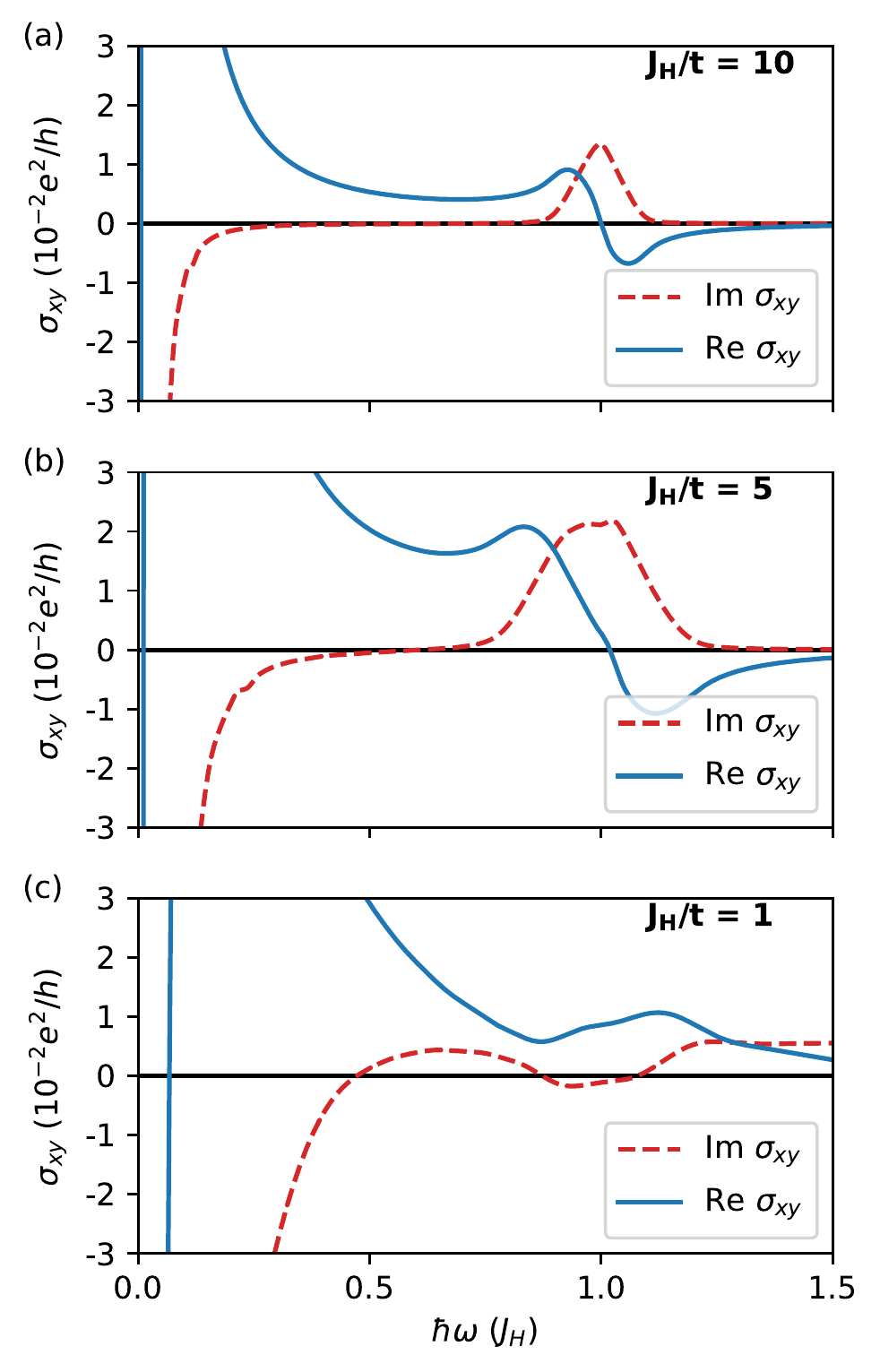}
	\caption{$\sigma_{xy}$ spectra for (a)$J_H/t = 10$, (b)$J_H/t = 5$, and (c) $J_H/t = 1$, illustrating the shift of the Lorentzian-like resonance feature at $\hbar \omega_{\rm res} \approx J_H$ to lower frequencies upon decreasing $J_H$. For small $J_H/t$, the resonance feature becomes inseparable from the low-frequency spectrum. Here, R = 10.}
	\label{fig:nonadia}
\end{figure}

Upon decreasing $J_H$, the bands at the top of parallel sector begin to overlap and hybridize with the bottom bands of the anti-parallel sector. Meanwhile, the feature of having pairs of similarly dispersing Chern bands becomes weaker for those bands that are not involved in the hybridization. The resonance feature shifts to a smaller frequency $\omega_{\rm res}$ following the decreasing $J_H$ and becomes eventually inseparable from the low-frequency spectrum. Figure \ref{fig:nonadia} illustrates the evolution of the resonance feature.
For a reasonably large $J_H/t$, e.g. panel (b), the resonance feature is fairly isolated from the low-frequency spectrum, at least for the imaginary part.
For $J_H/t = 1$, the resonance contribution, identifiable with the Lorentzian-like dip of Im $\sigma_{xy}$ at $J_H$, is still visible but differs by a sign from that of the adiabatic limit. The sign change may be caused by the changes in the band occupations, namely some ``antiparallel-sector" bands are now occupied, and the changes in the wave functions as a result of a significant hybridization between the parallel and the anti-parallel sector. For correlated transition metal oxides, we expect $t \sim 200$-$300$\,meV while the Hund's coupling
is $J_H \sim 0.7$-$1$\,eV, so we expect $J_H/t \sim 2$-$5$. For materials such as MnGe, $J_H/t \sim 1$ \cite{Zhu2019}.
The rest of the paper focuses on the adiabatic limit, where the resonance feature is well-separated from the low-frequency spectrum, which enables a tractable analysis in Section \ref{sec:resonance}.

\subsection{Scaling of resonance with skyrmion density}
\label{sec:scaling}
\begin{figure}[t]
\centering
\includegraphics[width = 0.35\textwidth]{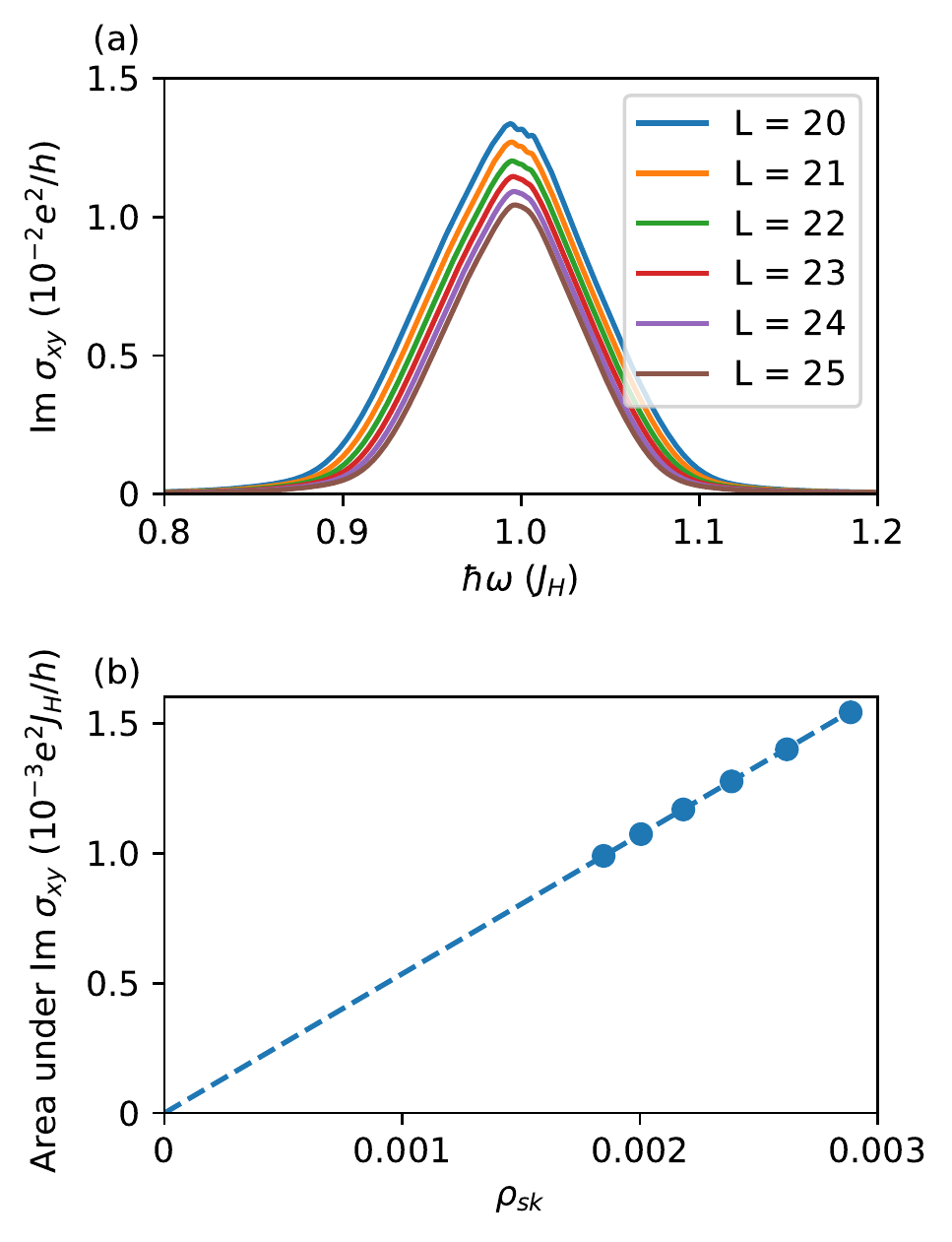}
\caption{(a) Resonant feature in Im $\sigma_{xy}$ near $J_H$ from numerical Kubo calculation for SkXs with varying unit cell size $L$. 
(b) Linear relation between $\rho_{sk} = 2/\sqrt{3}L^2$ and area under Im $\sigma_{xy}$ resonance obtained from the curves shown in panel (a). The
value of $\rho_{sk}=0.002$ corresponds to $L=24$.}
\label{fig:kubo_skx}
\end{figure}

We next examine the scaling of the Hall resonance with the areal skyrmion density $\rho_{sk}=2/\sqrt{3}L^2$ by varying 
$L=2 R$. Figure \ref{fig:kubo_skx}(a) shows the Kubo results for Im $\sigma_{xy}$ near $J_H$, for $L = 20, ..., 25$. 
As shown in Fig.\ref{fig:kubo_skx}(b), the areas under the resonant curves in panel (a) vary linearly with 
$\rho_{sk}$. In the next section, we derive this result using a smooth-texture approximation on the Kubo formula.

Lastly, it is worth mentioning that it is the imaginary part of $\sigma_{xy}(\omega)$, instead of the real part, that has a direct connection with the skyrmion density, in contrast with the dc topological Hall contribution, which is real-valued. Moreover, the linear relation discovered in this paper occurs only after $\sigma_{xy}(\omega)$ is integrated over the frequency rather than at any generic high frequency. For instance, it is found that Im $\sigma_{xy}(\omega)$ at the resonant frequency $J_H$, though monotonically increasing with the skyrmion density, deviates from a linear relation even in the adiabatic regime.

\section{Smooth-texture approximation for the Hall resonance}
\label{sec:resonance}

In this section, we elucidate the Lorentzian-like shape of the Hall resonance and the linear relation between $\rho_{sk}$ and the area under resonance of 
Im $\sigma_{xy}$ which can be estimated by applying a smooth texture approximation on the Kubo formula.  Our main result in this section is that the 
area under the resonant imaginary Hall conductivity is given by
\bea
{\cal S} \equiv \int \!\! d\omega ~{\mathrm{Im}} \sigma^{\rm res}_{xy}(\omega)  \approx \frac{e^2}{\hbar^2} \frac{\pi t^2}{J_H} \frac{N_{sk}}{\mathcal{A}} {\cal F}
\label{eq:ressum}
\eea
where the numerical factor ${\cal F}$ involves a momentum integral which depends on the electron filling; we give its explicit form in Appendix \ref{app:matrix_element}.
This result reproduces the linear scaling with skyrmion density $N_{sk}/{\mathcal{A}}$ which we have observed numerically in the previous section; we will 
see later that a similar numerical scaling is also obtained 
for a single ``skyrmion in a box'' problem.

With somewhat less rigor, we can show that the
Hall
conductivity near the resonance behaves as
\bea
\sigma^{\rm res}_{xy}(\omega) \approx - \frac{e^2}{\hbar} \frac{N_{sk}}{\mathcal{A}} \frac{t^2}{J_H} \frac{{\cal F}}{\hbar\omega-J_H + \I \tilde{\gamma}}
\label{eq:res}
\eea
where $\tilde{\gamma} \propto t$ is an effective broadening (which decreases roughly as $1/R$ with increasing skyrmion size).

The above results are obtained in two steps.
We begin by expressing the Kubo formula in a local basis, where the spin quantization axis at each site is taken to be the direction of the skyrmion spin. In a second step, we replace certain matrix elements by their simplified values for a corresponding uniform ferromagnet.
This second step becomes increasingly accurate in the limit where the SkX spin texture varies slowly in space, namely large $L$, when the 
neighboring spins are almost parallel.
The local basis we use is defined via eigenfunctions of the Hund's coupling term, so that
\bea
\begin{pmatrix}
c_{i\uparrow}\\
c_{i\downarrow}
\end{pmatrix} &=& \bfm_i\cdot \boldsymbol{\sigma} \begin{pmatrix}
p_i\\
a_i
\end{pmatrix},\\
\bfm_i &=& \left(\sin\frac{\theta_i}{2}\cos\phi_i, \sin\frac{\theta_i}{2}\sin\phi_i, \cos\frac{\theta_i}{2}\right)^T,
\eea
where $p_i$ and $a_i$ denote the electron annihilation operators for electron at site $i$ whose spin is parallel and antiparallel to $\bS_i$ 
respectively, while $\theta_i$ and $\phi_i$ are defined in Section \ref{sec:model}. In this basis, the Hamiltonian is
\bea
H_0 \! &=& \!-\! J_H \!\! \sum_{i} (p^{\dagger}_i p^\pdg_i \!-\! a^{\dagger}_i a^\pdg_i) \!+\! H_{pp} \!+\! H_{aa} \!+\! H_{ap} \!+\! H_{pa} \\
%&& + H_{ap} + H_{pa},\\
H_{pp} &=& \sum_{\langle ij \rangle} (T_{pp, ij} p^{\dagger}_i p_j + h.c.),\\
H_{aa} &=& \sum_{\langle ij \rangle} (T_{aa, ij} a^{\dagger}_i a_j + h.c.),\\
H_{pa} &=& H_{ap}^\dagger = \sum_{\langle ij \rangle} \left( T_{pa, ij} p_i^{\dagger} a_j + T_{pa, ji} p^{\dagger}_j a_i \right)
\eea
The local unitary rotation leads to the spin texture being absorbed into effective complex hopping amplitudes $T_{ij}$. $H_{pp}$ and $H_{aa}$ then appear as
generalized Hofstadter models of spinless fermions, $p^{\dagger}$ and $a^{\dagger}$, which see opposite Berry fluxes as determined by
these hopping integrals \cite{Nagaosa2000}. These fluxes can be viewed as an emergent magnetic field, which is responsible for the topological Hall 
effect. The hopping integrals are given by
\begin{widetext}
\bea
T_{ij} &=& \begin{pmatrix}
T_{pp,ij} & T_{pa,ij}\\
T_{ap,ij} & T_{aa,ij}
\end{pmatrix} \nonumber\\
&\!\!\!\!\!\!=\!\!\!\!\!\!\!& -t\left(
\begin{array}{cc}
 \cos \left(\frac{\text{$\theta_i $}}{2}\right) \cos \left(\frac{\text{$\theta_j $}}{2}\right)+e^{-i (\text{$\phi_i $}-\text{$\phi_j $})} \sin \left(\frac{\text{$\theta_i $}}{2}\right) \sin \left(\frac{\text{$\theta_j $}}{2}\right) & e^{-i \text{$\phi_j $}} \cos \left(\frac{\text{$\theta_i $}}{2}\right) \sin \left(\frac{\text{$\theta_j $}}{2}\right)-e^{-i \text{$\phi_i $}} \cos \left(\frac{\text{$\theta_j $}}{2}\right) \sin \left(\frac{\text{$\theta_i $}}{2}\right) \\
 e^{i \text{$\phi_i $}} \cos \left(\frac{\text{$\theta_j $}}{2}\right) \sin \left(\frac{\text{$\theta_i $}}{2}\right)-e^{i \text{$\phi_j $}} \cos \left(\frac{\text{$\theta_i $}}{2}\right) \sin \left(\frac{\text{$\theta_j $}}{2}\right) & \cos \left(\frac{\text{$\theta_i $}}{2}\right) \cos \left(\frac{\text{$\theta_j $}}{2}\right)+e^{i (\text{$\phi_i $}-\text{$\phi_j $})} \sin \left(\frac{\text{$\theta_i $}}{2}\right) \sin \left(\frac{\text{$\theta_j $}}{2}\right) \\
\end{array}
\right)
\eea
The following form of the Kubo formula for the Hall conductivity is useful for the discussion in this section \cite{Tong2016}.
\bea
\label{eq:kubo0}
\sigma_{xy}(\omega) &=& \frac{\I}{{\cal A}\omega} \sum_{\bk}\sum_{E_{\bk n}<\mu}\sum_{E_{\bk m}>\mu} \left[\frac{\bra{\bk n}j_{x}\ket{\bk m}\bra{\bk m}j_{y}\ket{\bk n}}{\hbar\omega + \I \gamma + E_{\bk n} - E_{\bk m}} - \frac{\bra{\bk n}j_{y}\ket{\bk m}\bra{\bk m}j_{x}\ket{\bk n}}{\hbar\omega + \I \gamma - E_{\bk n} + E_{\bk m}}\right],
\eea
where $\bj$ is the current operator. For large $\omega > 0$ relevant to the resonance, we can drop the second term in the square bracket. In the local basis, the current operator is $\bj = \bj_{pp} + \bj_{aa} + \bj_{ap} + \bj_{pa}$. 
Near resonance, the optical transitions between the parallel and anti-parallel sectors are dominant, so that $\bj_{pp} + \bj_{aa}$ can be neglected. The
Kubo formula may thus be approximated by
\bea
\label{eq:kubo1}
\sigma^{\rm res}_{xy}(\omega) &\approx& \frac{\I\hbar}{{\cal A} J_H} \sum_{\bk,m,n} {}^{\!\!\!'} \left[\frac{\bra{P \bk n} j^x_{pa}\ket{A \bk m}\bra{A \bk m} j^y_{ap}\ket{P \bk n}}{\hbar\omega + \I \gamma + E^P_{\bk n} - E^A_{\bk m}}\right],
\eea
\end{widetext}
where we have replaced the frequency in the prefactor by its value $J_H$ on resonance, and the prime on the sum indicates a restriction to 
$E_{\bk n}<\mu$ and $E_{\bk m}>\mu$. The current operators are explicitly given by
\bea
\bj_{pa} &=& - \frac{\I e}{\hbar} \sum_{\langle ij \rangle} T_{pa, ij}\left(p^{\dagger}_ia_j + p^{\dagger}_ja_i\right)\br_{ij},\\
\bj_{ap} &=& - \frac{\I e}{\hbar}\sum_{\langle ij \rangle} T_{ap, ij}\left(a^{\dagger}_ip_j + a^{\dagger}_jp_i\right)\br_{ij},
\eea
with $\br_{ij} = \br_j - \br_i$. We have replaced $(\ket{\bk n}, \ket{\bk m})$ by $(\ket{P\bk n}, \ket{A\bk m})$ to denote that these are Bloch states of the effective Hofstadter models for the parallel sector and anti-parallel sector respectively. Therefore, $m,n$ now run over $1, 2, \cdots, L^2$ rather than upto $2 L^2$. We have restricted the initial states to the parallel sector which is appropriate for our electron filling, and restricted the summation over intermediate states to just the anti-parallel bands since they are 
the dominant terms for the resonance.  In the large $J_H/t$ limit, $\ket{P\bk n}$ and $\ket{A\bk m}$ are annihilated by $a$ and $p$ operators respectively, 
which will be used below.

At this stage, we can integrate the imaginary part of the response from Eq.~\ref{eq:kubo1} to get
\bea
\!\!\!\! {\cal S}
& \approx & \frac{\pi}{{\cal A} J_H} \!\! \sum_{\bk m n} {}^{\!\!\!'} {\mathrm{Im}}\,{\cal N}_{\bk nm}\\
{\cal N}_{\bk n m} &=& \bra{P \bk n} j^x_{pa}\ket{A \bk m}\bra{A \bk m} j^y_{ap}\ket{P \bk n}
\label{eq:sumrule}
\eea
As shown in Appendix \ref{app:matrix_element}, the sum over $m$ can be carried out exactly, 
which leads to the expectation value of a ``kinetic energy''-type operators $p^\dagger_i p^\pdg_j$
in the state $\ket{P\bk n}$. For a smooth, slowly varying skyrmion texture, the slow gradients of the spin direction are mainly captured by the prefactors
$T_{pa, ij}, T_{ap, ij}$. The operator expectation values can be replaced, to leading order, by the corresponding expectation values in a uniform 
ferromagnet, which leads to the final result in Eq.~\ref{eq:ressum}.

To make progress on understanding the frequency dependence of the resonant Hall effect, 
we have examined the current matrix elements in the numerator of the Hall conductivity. For any fixed $\bk$, our numerical calculation shows that 
$|\bra{P \bk n} j^x_{pa}\ket{A \bk m}\bra{A \bk m} j^y_{ap}\ket{P \bk n}|$ has its largest magnitude around
$m \!\approx\! n$, and it decays rapidly as $|m-n|$ increases (see Appendix \ref{app:matrix_element} for an illustrative plot of the matrix elements). The decay occurs 
over an energy window $|(E^A_{\bk m}-E^P_{\bk n})-J_H| < \alpha t$, where $\alpha\! \approx\! 1$ for $R\!=\! 10$.
Since the matrix elements are peaked, while the denominator is a smooth function
of the energy difference, it is a reasonable approximation to set 
\bea
\label{eq:kuboapprox}
\!\!\!\!\!\! \sigma^{\rm res}_{xy}(\omega) \!&\approx&\! - \frac{\hbar}{{\cal A} J_H} ~ \frac{1}{{\hbar\omega + \I \tilde{\gamma} -J_H}}
\sum_{\bk m n} {}^{\!\!\!'} {\mathrm{Im}}\,{\cal N}_{\bk n m}
\eea
where the effective broadening $\tilde{\gamma}\!=\! \alpha t \ll J_H$ is determined by the energy window discussed above. This leads to the result
in Eq.\ref{eq:res}, in agreement with the resonant Lorentzian response we find from our numerics.
From our numerical results, we also find that $\tilde{\gamma}$ scales inversely with the unit cell width, roughly like $1/L$, so the resonant 
peak height of Im $\sigma^{\rm res}_{xy}$ scales roughly like the square root of the skyrmion density.

\begin{figure}
\centering
\includegraphics[width=0.9\columnwidth]{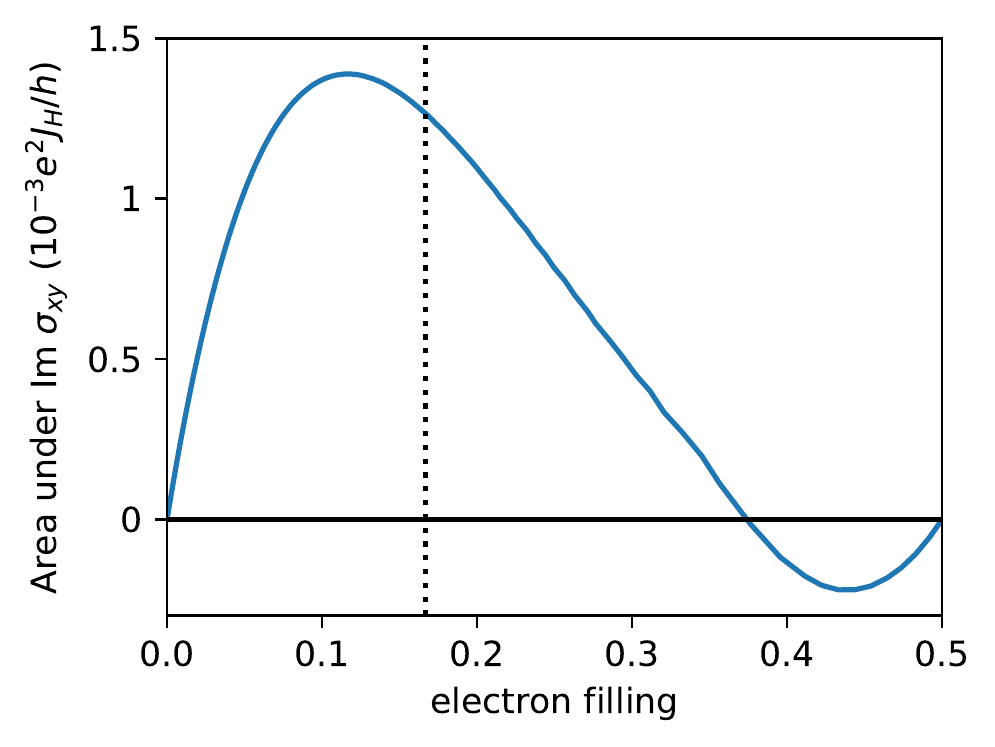}
\caption{Electron filling dependence of the area under Im $\sigma_{xy}$ computed from the smooth texture approximation Eq.\ref{eq:ressum} at skyrmion density $\rho_{sk} = 0.002$.}
\label{fig:electron_filling}
\end{figure}

Finally, Fig.\ref{fig:electron_filling} shows the electron filling dependence of the area under Im $\sigma_{xy}$ at a fixed $L=24$ 
(equivalently $\rho_{sk} = 0.002$) computed from the smooth texture approximation Eq.\ref{eq:ressum}.
The dotted line marks the 1/6 filling used in the Kubo formula calculation, showing a quantitative agreement with the Kubo formula result in Fig.\ref{fig:kubo_skx}(b).
The sign switching of the area implies a sign switching of the resonant peak, according to Eq.\ref{eq:res}. This occurs at the electron filling where the chemical potential falls into the region of the band structure with many band crossings associated with a van Hove singularity, analogous to the sign switching behavior in the d.c. topological Hall conductivity \cite{Gobel2017}.

The analysis in this section is possible in the limit of large $J_H/t$, which supports a well-separated resonance feature from the low-frequency spectrum. The inseparability at small $J_H/t$ leads to an impediment to an analysis, in contrast with the dc topological Hall effect, which can shown to be proportional to the topological charge density regardless of the magnitude of $J_H/t$ \cite{Kawamura2002, Kohno2014, Denisov2016}.

\section{Hall resonance of isolated skyrmion}
\label{sec:flake}

\begin{figure}[t]
\centering
\includegraphics[width = 0.49\textwidth]{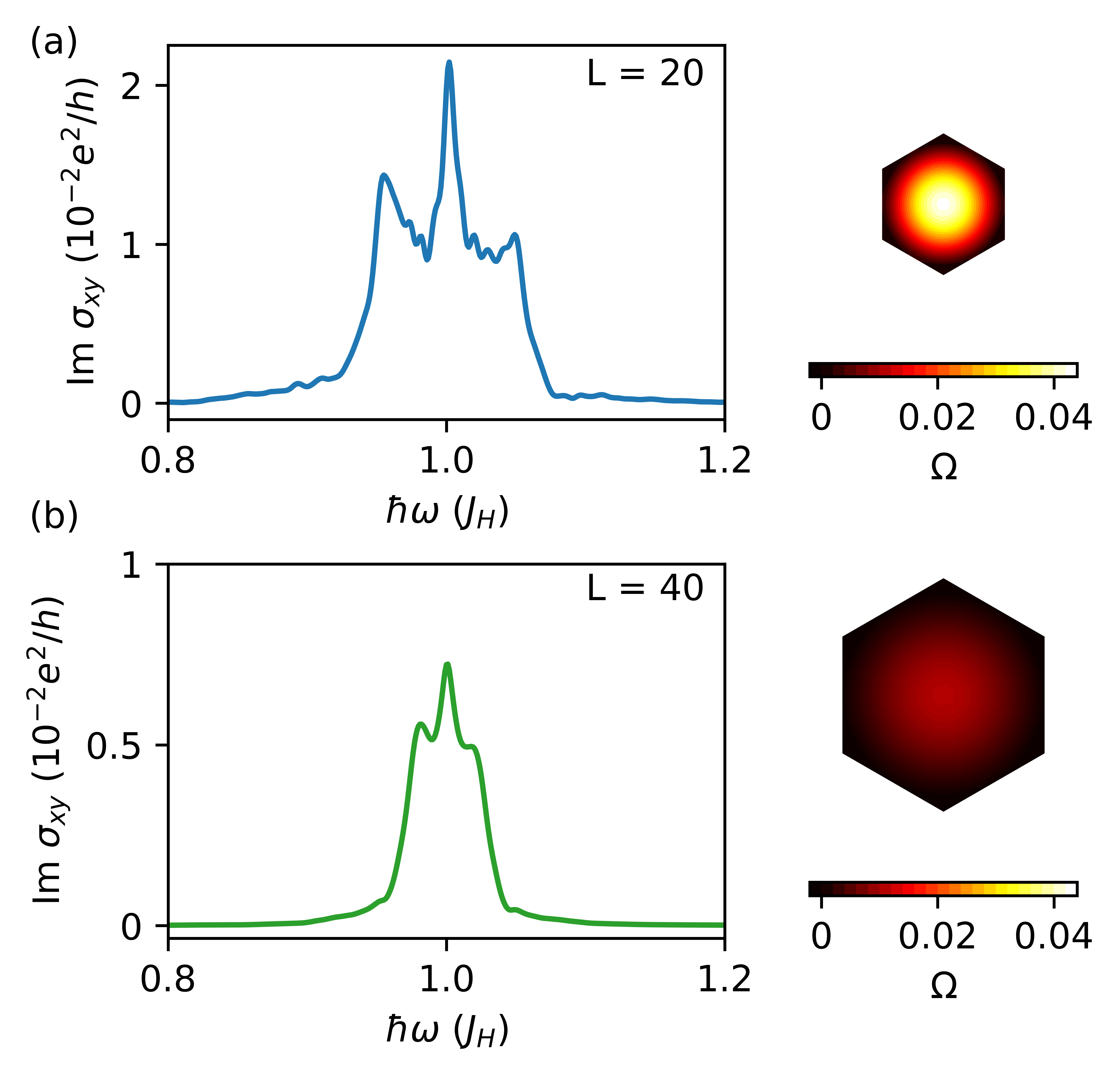}
\caption{Resonant feature in Im\,$\sigma_{xy}$ for a skyrmion in a hexagonal box with (a) width L = 20, (b) L = 40. 
The color plots on the right show the profiles of the solid angle of the spin textures at each elementary plaquette, depicting a large real-space 
Berry phase gradient when $L$ is small.
We observe side peaks which move towards and merge with the main peak at $J_H$ when $L$ increases. For the larger $L$, the result has a stronger resemblance to
the resonance in the SkX case. We attribute the side peaks to the inhomogeneity of the emergent magnetic field, as described in the text.}
\label{fig:flake}
\end{figure}

To demonstrate that the resonance feature near $J_H$ is present regardless of the crystal structure of the skyrmion spin texture, we consider the model $H_0$ in 
an open-boundary hexagonal box of width $L$, as defined in Sec.\ref{sec:model}. The spin texture contains a single skyrmion of radius $R=L/2$ centred in the middle of the box.

Figure \ref{fig:flake} shows the corresponding Im\,$\sigma_{xy}$ obtained from the real-space Kubo formula, demonstrating the presence of the resonance at $J_H$ even with a single-skyrmion spin texture. The results are shown for a small and a large confinement area with $L = 20$ and $40$ respectively. 
The color plots to the right display the profile of the solid angle subtended by the three spins at each elementary triangular plaquette of the spin texture, a measure of the local scalar spin chirality, depicting a more gradual variation of the real-space Berry flux density (i.e., the emergent magnetic field) in the larger box. 
We observe from these results that there is a central peak which agrees with the resonant feature seen in the SkX.
In addition, we observe side peaks away from the main resonant peak at $J_H$; these side peaks appear to merge into the resonant feature at $J_H$ 
upon increasing the confinement area and the skyrmion size, thus better resembling the resonant peak of the SkX in Fig.\ref{fig:spec_large}(a). 
We attribute these side peaks to the inhomogeneity of the emergent magnetic field of the skyrmion.
We have found that when we cut out hexagonal unit cells centered at different points from the SkX spin texture in Sec.\ref{sec:model}, the side peaks become even more pronounced when the box is centered at the collinear regions where three skyrmions meet in Fig.\ref{fig:neel_sk}(a). 
In the collinear region, the emergent magnetic field is zero, whereas it becomes nonzero in the surrounding region which lies within the skyrmion core. This pattern of the emergent magnetic field resembles what has been studied in ``magnetic anti-dot'' structures, where the field is absent inside the dot region and nonzero outside. Such anti-dots support current-carrying localized states with a discrete energy-level spectrum \cite{Sim1998, Matulis1999, Cserti2008}. The number of localized states trapped at the anti-dot and their degree of localization are enhanced when the field strength contrast between the inside and the outside is large, which occurs in our case when the real-space Berry phase gradient is big, namely small $L$. In our spinful electron model coupled to a spin texture, these energy levels appear in both the parallel and anti-parallel sectors. It is thus plausible that the observed side peaks originate from inter-sector transitions between these energy levels, which can 
occur at a frequency slightly different from $J_H$.

The persistence of the Hall resonance here implies that it is observable even when the texture is no longer a SkX, e.g. a disordered array of skyrmions. In fact, the resonance can be seen even in a minimal three-site system hosting a noncoplanar spin configuration, which will be studied next.

\section{Hall resonance in a spin trimer}
\label{sec:trimer}

\begin{figure}[t]
\centering
\includegraphics[width=0.48\textwidth]{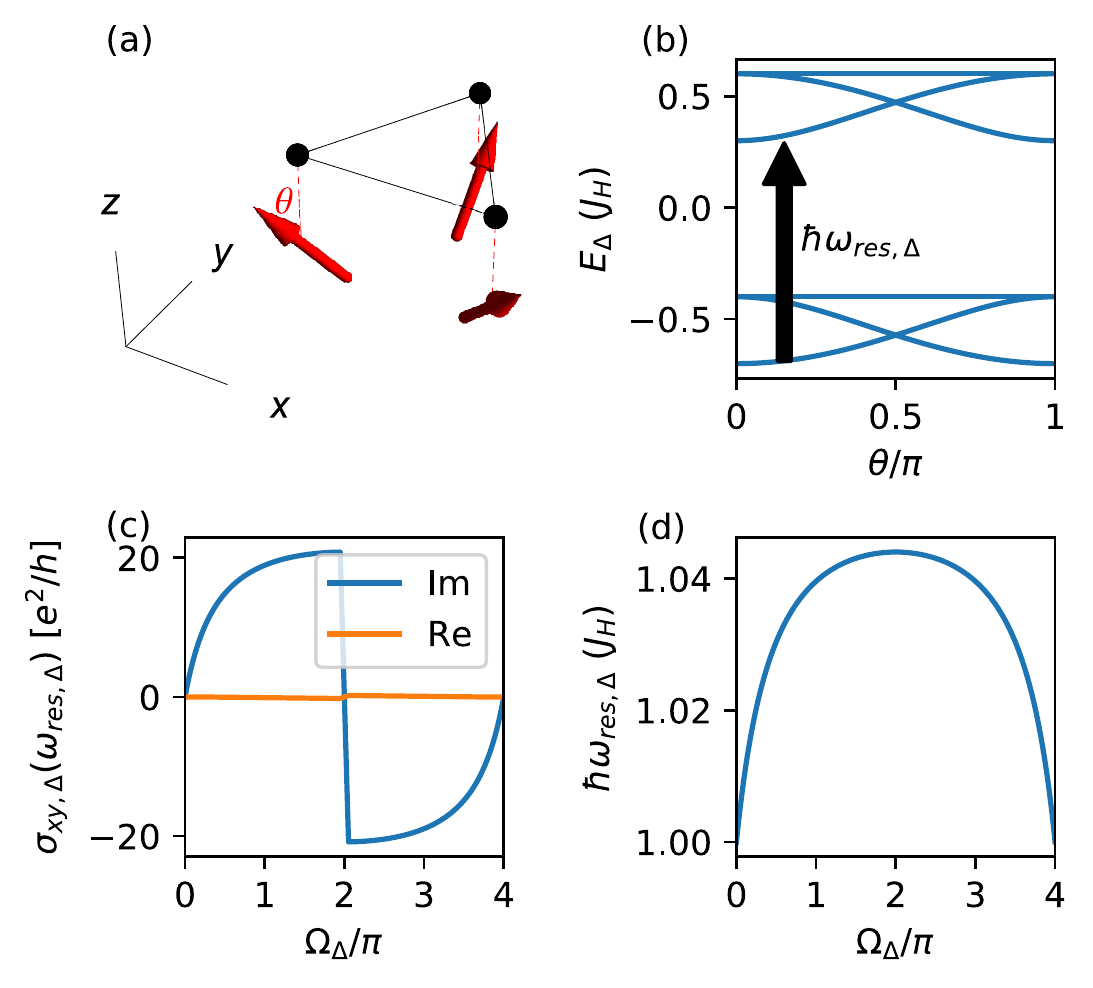}
\caption{(a) Schematic illustration of the trimer model. Red arrows denote the spin vectors on a triangular plaquette $\Delta$ which form a noncoplanar spin configuration when the polar angle $\theta$ is nonzero (mod $\pi$). (b) Energy levels of trimer model as a function of $\theta$ showing the presence of pairs of level whose energies differ by roughly $J_H$. This leads to a resonance in $\sigma_{xy,\Delta}$ at $\hbar\omega_{\text{res}, \Delta}\approx J_H$. (c) $\sigma_{xy,\Delta}$ at resonance as a function of the solid angle $\Omega_{\Delta}$ subtended by the spins. (d) Small change in resonance frequency $\omega_{\text{res}, \Delta}$ as a function of $\Omega_{\Delta}$.}
\label{fig:trimer}
\end{figure}

To illustrate the persistence of the Hall resonance in a minimal three-site system, we consider the following trimer model, analogous to Eq.\ref{eq:model}, defined on a triangular plaquette $\Delta$ with three spins $\{\bS_i\}$ for $i = 1, 2, 3 \in \Delta$ forming a non-zero solid angle $\Omega_{\Delta}$.
 \bea
H_{\Delta} &=& - t\sum_{\langle ij\rangle \in \Delta} (c^{\dagger}_{i\sigma}c_{j\sigma}+h.c.) - J_H\sum_{i\in \Delta} \bs_i \cdot \bS_i.
\eea
For simplicity, we suppose that the three spins are related to one another by a $2\pi/3$ rotation around the z-axis as the following.
\bea
\bS_1 &=& (\sin \theta \cos \varphi, \sin \theta \sin \varphi, \cos \theta),\nonumber\\
\bS_2 &=& (\sin \theta \cos (\varphi+2\pi/3), \sin \theta \sin (\varphi+2\pi/3), \cos \theta), \nonumber\\
\bS_3 &=& (\sin \theta \cos (\varphi+4\pi/3), \sin \theta \sin (\varphi+4\pi/3), \cos \theta),\nonumber
\eea
where $\theta \in [0, \pi]$ and $\varphi = -\pi/6$. This corresponds to the spin configuration in Fig.\ref{fig:trimer}(a). The model is invariant under the $2\pi/3$ rotation, which admits an analytical expression for $\sigma_{xy, \Delta}$ at resonance.

In the basis of $\mathbf{C} = (c_{1\uparrow}, c_{1\downarrow}, c_{2\uparrow}, c_{2\downarrow}, c_{3\uparrow}, c_{3\downarrow})^T$, the trimer model is represented by
\bea
\mathcal{H} &=& \mathcal{H}_0 + \mathcal{H}_1 + \mathcal{H}_2 + \mathcal{H}_3,\\
\mathcal{H}_0 &=& -t \begin{pmatrix}
0 & 1 & 1\\
1 & 0 & 1\\
1 & 1 & 0
\end{pmatrix} \otimes \mathbb{1}_{\text{spin}},\\
\mathcal{H}_1 &=& -\frac{J_H}{2} \begin{pmatrix}
1 & 0 & 0\\
0 & 0 & 0\\
0 & 0 & 0
\end{pmatrix} \otimes \bS_1 \cdot \bsigma,\\
\mathcal{H}_2 &=& -\frac{J_H}{2} \begin{pmatrix}
0 & 0 & 0\\
0 & 1 & 0\\
0 & 0 & 0
\end{pmatrix} \otimes \bS_2 \cdot \bsigma,\\
\mathcal{H}_3 &=& -\frac{J_H}{2} \begin{pmatrix}
0 & 0 & 0\\
0 & 0 & 0\\
0 & 0 & 1
\end{pmatrix} \otimes \bS_3 \cdot \bsigma.
\eea
$\mathcal{H}$ commutes with the three-fold rotational operator $\mathcal{Z} = \mathcal{W} \otimes \mathcal{U}$, where $\mathcal{W}$ and $\mathcal{U}$ are expressed below.
\bea
\mathcal{W} &=& \begin{pmatrix}
0 & 0 & 1\\
1 & 0 & 0\\
0 & 1 & 0
\end{pmatrix},\\
\mathcal{U} &=& \begin{pmatrix}
e^{\I \pi / 3} & 0\\
0 & e^{-\I \pi / 3}
\end{pmatrix}.
\eea
$\mathcal{W}$ has three eigenvalues $\{1, \nu, \nu^2\}$, where $\nu = e^{\I 2 \pi /3}$, and the corresponding eigenvectors are given by
\bea
\ket{\xi_1} &=& \frac{1}{\sqrt{3}}\begin{pmatrix}
1 & 1 & 1
\end{pmatrix}^T,\\
\ket{\xi_{\nu}} &=& \frac{1}{\sqrt{3}} \begin{pmatrix}
1 & \nu^2 & \nu
\end{pmatrix}^T,\\
\ket{\xi_{\nu^2}} &=& \frac{1}{\sqrt{3}} \begin{pmatrix}
1 & \nu & \nu^2
\end{pmatrix}^T,
\eea
respectively.
$\mathcal{U}$ has two eigenvalues $\lambda = e^{\I\pi/3}$ and $\bar{\lambda} = e^{-\I\pi/3}$ corresponding to the eigenvectors
\bea
\ket{\chi_{\lambda}} &=& \begin{pmatrix}
1 & 0
\end{pmatrix}^T, \\
\ket{\chi_{\bar{\lambda}} } &=& \begin{pmatrix}
0 & 1
\end{pmatrix}^T.
\eea
One can check that $\mathcal{Z}$ has three eigenvalues $\{-1, \lambda, \bar{\lambda}\}$, each of which is two-fold degenerate. Denote the eigensubspaces of the three eigenvalues by $\{\ket{\psi_1}, \ket{\psi_2}\}$, $\{\ket{\psi_3}, \ket{\psi_4}\}$ and $\{\ket{\psi_5}, \ket{\psi_6}\}$ respectively, where
\bea
\ket{\psi_1} &=& \ket{\xi_{\nu} \otimes \chi_{\lambda}},\\
\ket{\psi_2} &=& \ket{\xi_{\nu^2} \otimes \chi_{\bar{\lambda}} },\\
\ket{\psi_3} &=& \ket{\xi_{1} \otimes \chi_{\lambda}},\\
\ket{\psi_4} &=& \ket{\xi_{\nu} \otimes \chi_{\bar{\lambda}}  }, \\
\ket{\psi_5} &=& \ket{\xi_{\nu^2} \otimes \chi_{\lambda} }, \\
\ket{\psi_6} &=& \ket{\xi_{1} \otimes \chi_{\bar{\lambda}} }.
\eea

In the $\{\psi_a\}$ basis, $\mathcal{H}$ is block diagonalized into three two-by-two blocks which can be further diagonalized straightforwardly. The spectrum of $\mathcal{H}$ contains six energy levels as shown in Fig.\ref{fig:trimer}(b), featuring pairs of energy levels differing in energy by roughly $J_H$. At $1/6$ filling, only the lowest level is occupied. At resonance $\hbar\omega \approx J_H$, the transition between the first and the forth energy levels, $E_1$ and $E_4$, is the only dominant contribution to the resonant peak $\sigma_{xy, \Delta}^{\text{peak}}$. For $\theta \in \,[0, \pi/2)$, these eigenstates are given by
\bea
\ket{E_1} &=& \frac{\zeta \ket{\psi_3} + \eta \ket{\psi_4}}{\sqrt{2}}\\
\ket{E_4} &=& \frac{\varrho \ket{\psi_6} + \kappa \ket{\psi_5}}{\sqrt{2}},\\
E_1 &=& -\frac{t}{2} - \frac{1}{2}\sqrt{J_H^2 + 9t^2 + 6J_Ht\cos\theta},\\
E_4 &=& -\frac{t}{2} + \frac{1}{2}\sqrt{J_H^2 + 9t^2 - 6J_Ht\cos\theta},
\eea
where the coefficients $\zeta, \eta, \varrho$ and $\kappa$ are functions of $t, J_H, \theta, \varphi$.
\bea
\zeta &=& - \frac{\xi + J_H\cos\theta + 3t}{\sqrt{\xi\left(\xi + J_H\cos\theta + 3t\right)}},\\
\eta &=& - \frac{J_H\sin\theta e^{\I \varphi}}{\sqrt{\xi\left(\xi + J_H\cos\theta + 3t\right)}},\\
\varrho &=& \frac{\xi' + J_H\cos\theta - 3t}{\sqrt{\xi'\left(\xi' + J_H\cos\theta - 3t\right)}},\\
\kappa &=& -\frac{J_H \sin\theta e^{-\I \varphi}}{\sqrt{\xi'\left(\xi' + J_H\cos\theta - 3t\right)}},\\
\xi &=& \sqrt{J_H^2 + 9t^2 + 6 J_Ht\cos\theta},\\
\xi' &=& \sqrt{J_H^2 + 9t^2 - 6 J_Ht\cos\theta}.
\eea

Using these eigenfunctions to compute the resonant peak $\sigma_{xy, \Delta}^{\text{peak}}$ and keeping only the interlevel contribution between $\ket{E_1}$ and $\ket{E_4}$, we obtain
\bea
\label{eq:trimer_sigmaxy}
\sigma_{xy, \Delta}^{\text{peak}} &\approx& \frac{\I \sqrt{3} t^2}{4\hbar\omega_{\text{res}, \Delta}\gamma} \frac{e^2}{\hbar} |\zeta^*\kappa - \varrho\eta^*|^2,
\eea
where $\hbar\omega_{\text{res}, \Delta} = E_4 - E_1$ changes slightly with $\theta$ and is plotted as a function of $\Omega_{\Delta}$ in Fig.\ref{fig:trimer}(d). The term $|\zeta^*\kappa - \varrho\eta^*|^2$ depends on $\theta$ as $\sim \sin^2\theta$ peaking as $\theta$ approaches $\pi/2$ from below,
in agreement with the monotonic increasing of Im $\sigma_{xy, \Delta}^{\text{peak}}$ with $\Omega_{\Delta}$, as shown in Fig.\ref{fig:trimer}(c), for $\Omega_{\Delta} \in \,[0, 2\pi)$. While crossing from  $\theta \in [0, \pi/2)$ to $(\pi/2, \pi]$, the energy levels can cross one another as shown in Fig.\ref{fig:trimer}(b). Im $\sigma_{xy, \Delta}^{\text{peak}}$ switches sign abruptly and scales with $\theta$ like $ \sim - \sin^2 \theta$ instead, which can be confirmed by recalculating Eq.\ref{eq:trimer_sigmaxy} taking into account the level crossing. 
The sharp jump at $(\theta, \Omega_{\Delta}) = (\pi/2, 2\pi)$ is expected to be smoothed out by thermal broadening as the temperature rises.
We have also checked that the analytical expression agrees very well with the numerical results which incorporate other less dominant inter-level contributions to Hall conductivity, i.e. the transition involving $\ket{E_1}$ and $\ket{E_a}$ for $a\neq 1, 4$. Our key observation is that Im $\sigma_{xy, \Delta}^{\text{peak}}$ grows monotonically with the solid angle $\Omega_{\Delta}$, at least for small $\Omega_{\Delta}$.
An implication of this finding is that the Hall resonance can be utilized in a local optical probe which can distinguish regions with spin noncoplanarity from those without the noncoplanarity, thereby enabling a visualization of a spin texture hosting skyrmions and perhaps other noncoplanar magnetic objects.

\begin{figure}[t]
\centering
\includegraphics[width=0.4\textwidth]{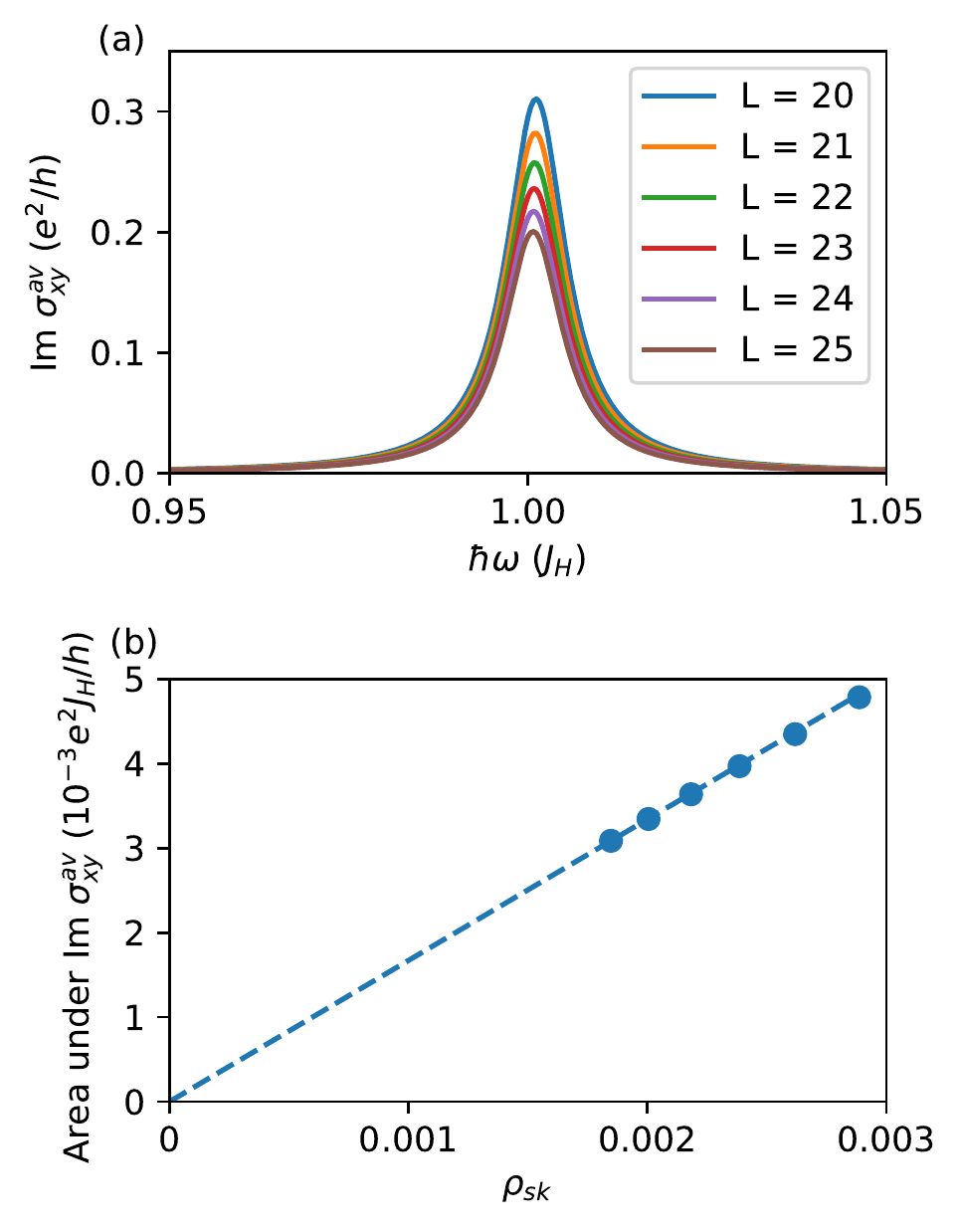}
\caption{Results from the local approximation: (a) Lorentzian-like resonance of Im $\sigma_{xy}^{\text{av}}$ near $\hbar\omega = J_H$ for various SkX's with different skyrmion densities $\rho_{sk} = 2/\sqrt{3}L^2$.
(b) A linear relation between $\rho_{sk}$ and the area under Im $\sigma_{xy}^{\text{av}}(\omega)$ in (a). $\rho_{sk} = 0.002$ corresponds to $L=24$.}
\label{fig:trimer_skx}
\end{figure}

Before concluding the section, we examine a possible connection for a given spin texture between the net Hall response and the local Hall response $\sigma_{xy, \Delta}$ associated with a local $\Omega_{\Delta}$.
Under a high-frequency applied electric field, the electronic response is expected to be local. That is, in an inhomogeneous system, the overall response function can be approximated by the local response function associated with the local property, that exhibits the inhomogeneity, averaged over the system, e.g. magnetization in inhomogeneous magnetic domain configuration \cite{Bartram2020} or the real-space Berry phase associated with $\Omega_{\Delta}$ in the case of skyrmion spin texture. A supporting argument for this is to regard the electrons as semi-classical objects which traverse only a small distance after several cycles of the high-frequency applied field, thereby sensing only the local properties \cite{Bartram2020}
\footnote{This statement depends on a few variables including the actual frequency of the applied field, the electron group velocity, and the degree of inhomogeneity in the system. A sufficient condition is that the distance travelled by the electron in a few cycles is small compared with the length scale characterizing the variation of the inhomogeneity, e.g. periodicity $L$ in the case of a SkX.}. 
For a given skyrmion spin texture, the Hall conductivity obtained from averaging the local Hall conductivity is given by
\bea
\label{eq:local_approx}
\sigma_{xy}^{\text{av}} &=& \frac{1}{N_{\Delta}} \sum_{\Delta} \sigma_{xy, \Delta},
\eea
where the local Hall response $\sigma_{xy, \Delta}$ can be computed using the trimer model, $N_{\Delta}$ is the number of the plaquette, and the sum is carried over all the triangular plaquettes.

Figure \ref{fig:trimer_skx}(a) shows the frequency dependence of Im $\sigma_{xy}^{\text{av}}$ featuring a resonance near $J_H$ for a series of SkX studied in Sec.\ref{sec:scaling}.
We observe a very sharp resonance whose height is an order of magnitude higher than that in Fig.\ref{fig:kubo_skx}(a), whereas the resonance width is an order of magnitude smaller. This is the consequence of having an enormous amount of nearly degenerate atomic-like energy levels due to the absence of the inter-trimer hopping. The hopping is expected to lift the degeneracy, which permits other transition channels at frequencies different from $J_H$, thereby broadening the resonance width to resemble Fig.\ref{fig:kubo_skx}(a) better.
Figure \ref{fig:trimer_skx}(b) shows the scaling between the area under the resonance and $\rho_{sk}$, which exhibits a linear relation identical to Fig.\ref{fig:kubo_skx}(b), except that the values are three times larger. We attribute this overestimate to the fact that the electrons are highly confined to each plaquette, which provides a better chance for the inter-sector transitions to occur compared with when the electrons are more delocalized in a lattice environment. Therefore, maintaining a certain degree of delocalization by enlarging the number of sites beyond the trimer is a promising way to reduce the discrepancy between the local approximation and the actual results.

\section{Impact of spin-orbit coupling}
\label{sec:soc}

\begin{figure}[t]
\centering
\includegraphics[width=0.45\textwidth]{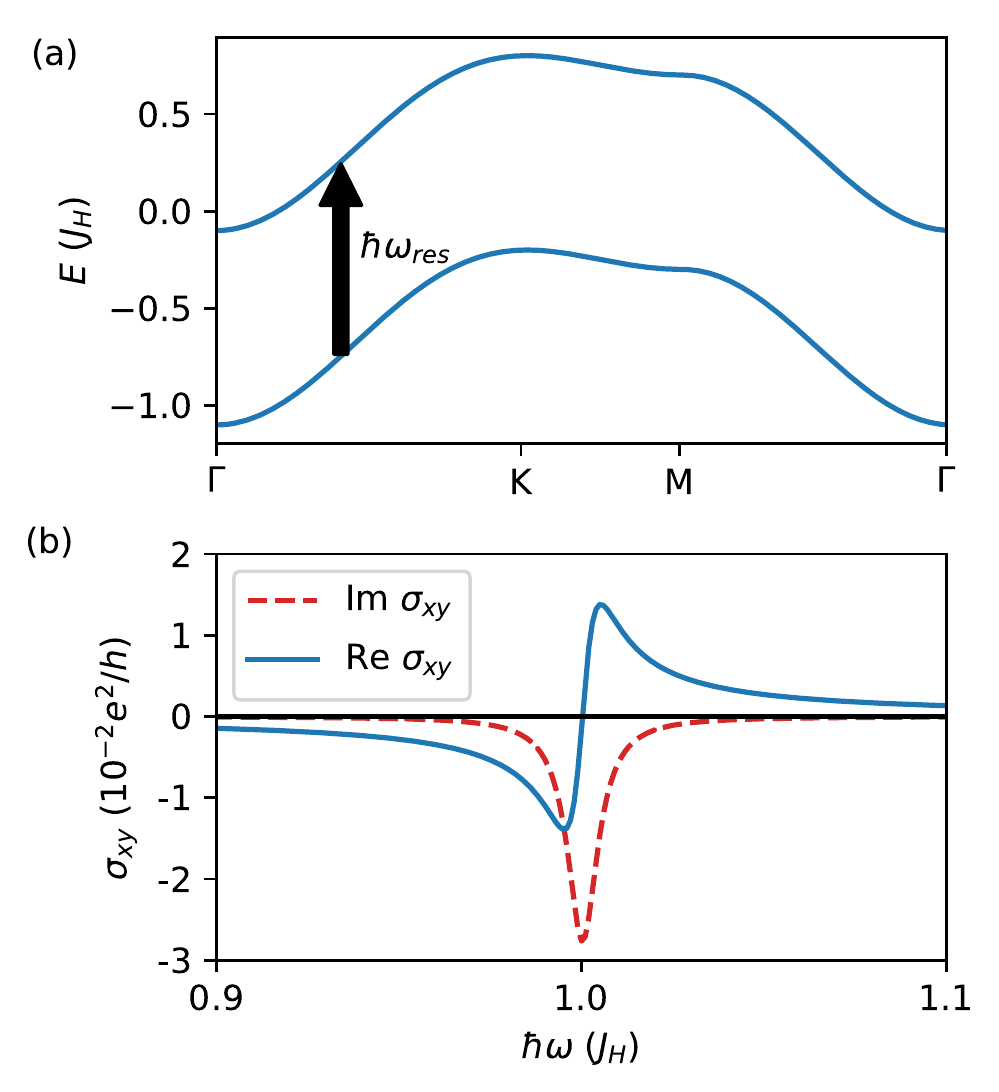}
\caption{(a) Band structure of the model with Rashba SOC, $H_0 + H_R$, for a ferromagnetic spin texture $\bS_i = -\hat{z}$, featuring similarly dispersing bands differing in energy by approximately $J_H$. (b) A resonance feature near $\hbar\omega = J_H$ arises from the Rashba SOC and the similarly dispersing bands.}
\label{fig:rashba}
\end{figure}

So far, the effect of spin-orbit coupling (SOC) has been ignored. SOC, like skyrmion, can also produce a resonance in Hall conductivity near $\hbar\omega=J_H$, which can already be seen even in a simple ferromagnetic spin texture. To demonstrate this, we introduce a Rashba hopping term $H_R$, defined below, into Eq.\ref{eq:model} and set $\bS_{i} = -\hat{z}$; this may be thought of as modelling the response of a system at saturation magnetization \cite{Banerjee2014}, while the SkX in 
Section \ref{sec:model} is an intermediate-field phase with $\langle \bS \rangle = - \epsilon \hat{z}$, for $0 < \epsilon < 1$.
\bea
H_R &=& \sum_{\langle ij \rangle} \I \chi_R c^{\dagger}_{i\alpha}\left(\hat{z}\times\hat{\br}_{ij}\cdot \bsigma_{\alpha\beta} \right)c_{j\beta} + h.c.,
\eea
where $\hat{\br}_{ij}$ is the unit vector of $\br_{ij} = \br_j - \br_i$, and $\chi_R$ is the strength of the Rashba SOC. Figure \ref{fig:rashba}(a) shows the band structure of $H_0 + H_R$ for $\chi_R = 0.05t$ \footnote{The sign of $\chi_R$ is chosen following Ref.\onlinecite{Banerjee2014}}, featuring similarly dispersing bands differing in energy by roughly $J_H$. As a result, a resonant feature is obtained near $J_H$, as shown in Fig.\ref{fig:rashba}(b) in the same frequency window as that for skyrmion spin textures. 
Such a resonant feature persists despite the variation of $\chi_R$ in the large-$J_H$ limit since the presence of the pair of similarly dispersing bands is rather robust. 
A notable distinction from the results of the previous sections lies in that the width of the resonance here is narrower since it involves interband transitions between only a single pair of similarly dispersing bands.
When the texture contains skyrmions, we also expect SOC to generate a resonance at $J_H$, and experiments will detect the effect of both.
The analogous roles of spin-orbit coupling and coupling to skyrmions can be understood as two ways to cause the same mixing effect between the orbital and the spin degree of freedom of the conduction electron. While the spin-orbit coupling achieves this in momentum space, the coupling to skyrmions does this in real-space \cite{Rashba1964, Rashba2020, Egorov2021}.

\begin{figure}[t]
\centering
\includegraphics[width=0.98\columnwidth]{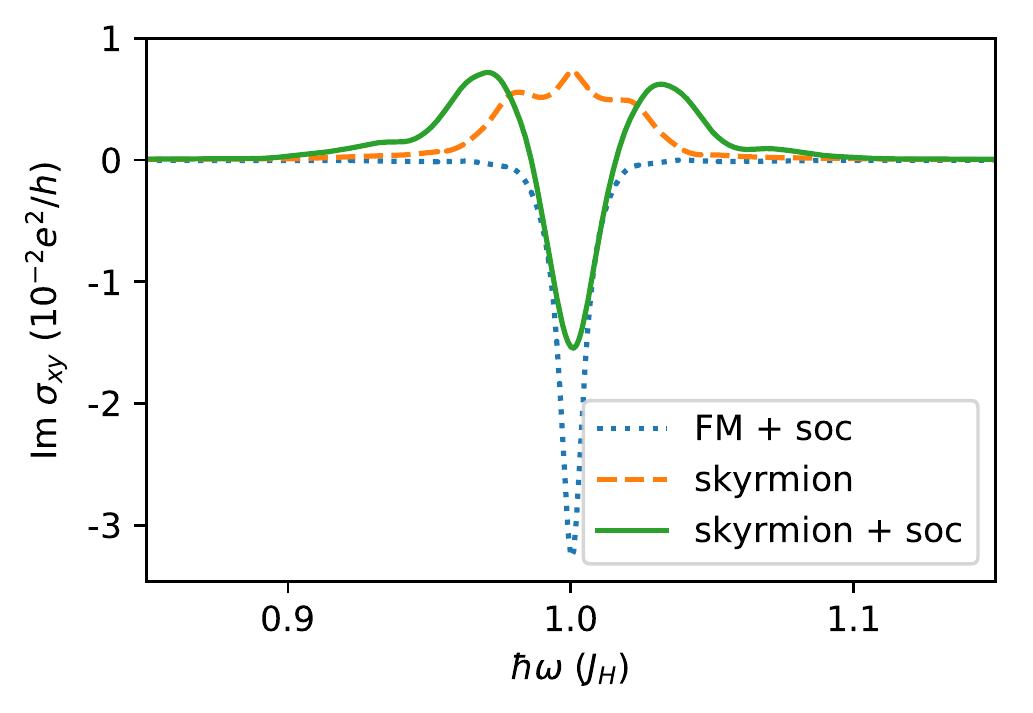}
\caption{Frequency dependence of Im $\sigma_{xy}$ near resonance obtained from a hexagonal-box calculation, as in Section \ref{sec:flake}, with $L = 40$ for three cases: (i) $\bS_i = - \hat{z}$ with Rashba hopping $\chi_R = 0.05 t$, (ii) skyrmion in a box with $\chi_R = 0$, and (iii) skyrmion in a box with $\chi_R = 0.05t$. The results indicate the difference in optical Hall conductivity between a ferromagnetic spin order and a skyrmion spin order.}
\label{fig:flake_soc}
\end{figure}

Recent works on d.c. Hall effect have made it increasingly apparent that SOC and non-collinearity of a spin texture intertwine to produce a resultant Hall effect which is not simply the addition of the individual contributions \cite{Lux2020, Batista2020, Blugel2021}. For instance, there can be extra Hall effect contributions arising only in the simultaneous presence of SOC and non-collinear magnetic order \cite{Lux2020, Blugel2021}. It is very likely that the intertwined effect also appears in the non-zero-frequency Hall effect. Indeed, a recent study has shown that an interplay between SOC and coplanar magnetic orders or magnetic
multipolar orders leads to distinct Hall conductivity spectra, which depend on the detailed arrangement of 
magnetic dipoles, despite the similar electronic band structures in the explored cases \cite{Motome2021}.

Therefore, a study where skyrmion spin texture and SOC are simultaneously treated is generally needed to make a \emph{quantitative} 
comparison with experiments. This is left for the future.
Even so, based on our results in the previous section, we expect the presence of skyrmions to be detectable using optical Hall measurements since it gives rise to a distinct optical Hall response between a spin texture containing skyrmions and other ordered phases, e.g. ferromagnetic order. Such a difference is expected to be pronounced when the skyrmion density is large. Figure \ref{fig:flake_soc} highlights this point. It illustrates the difference in the Hall conductivity spectrum between a ferromagnetic order and a skyrmion spin order in the presence of Rashba hopping.
The results are obtained from hexagonal-box calculations as in Section \ref{sec:flake}. The result from Fig.\ref{fig:flake}(b) is also included here for a comparison.

\section{Conclusion}
\label{sec:conclusion}

\begin{figure}[t]
\centering
\includegraphics[width=0.48\textwidth]{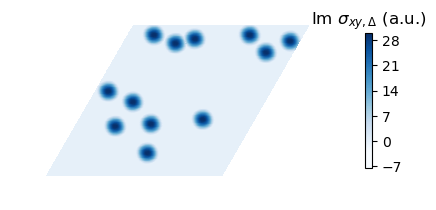}
\caption{Real-space profile of Im $\sigma_{xy, \Delta}$ for a random skyrmion array at resonance $\hbar\omega \approx J_H$ obtained from the local approximation. The dark droplets corresponding to a large nonzero-frequency topological Hall conductivity are the skyrmions, while the bright regions with Im\,$\sigma_{xy, \Delta}=0$
correspond to the ferromagnetic background. The color scale in this spatial map is in arbitrary units since the value of Im\,$\sigma_{xy, \Delta}$ is 
sensitive to the broadening $\gamma$.
Such a profile could potentially be mapped out by magneto-optical Kerr microscopy which is known to probe the local Hall conductivity, enabling the resonant optical Hall conductivity as a tool to visualize skyrmions.}
\label{fig:microscopy}
\end{figure}

We have shown that in a 2D model of conduction electrons coupled to skyrmion spin textures via a Hund's coupling, a high-frequency resonance in Hall conductivity arises at a characteristic frequency set by the Hund's coupling. For SkX spin textures, the resonance originates from transitions between many pairs of topological Chern bands which disperse similarly and differ in energy by a similar amount $\approx J_H$. 
Its presence does not depend on whether the spin texture is a crystal of skyrmions or a single skyrmion.
A linear relation between the skyrmion density and the area under the Hall resonance, Im $\sigma_{xy}(\omega)$, is found and is explained using a smooth texture approximation and a local approximation.
Probes such as the magneto-optical Kerr effect and Faraday effect, which are known to track non-zero-frequency Hall conductivity, 
may be suitable experimental techniques for detecting the optical topological Hall conductivity and its resonance. 
Near the resonant frequency, a real-space profile of a skyrmion spin texture may be mapped out using Kerr microscopy technique, as theoretically shown in Fig.\ref{fig:microscopy}. Our results can be tested in materials hosting a large density of skyrmions such as Gd$_2$PdSi$_3$\cite{Kurumaji2019}, Gd$_3$Ru$_4$Al$_{12}$ \cite{Hirschberger2019}, GdRu$_2$Si$_2$ \cite{Khanh2020}, and MnGe \cite{Tokura2012}. We also point out how SOC can produce a similar resonance and how it affects the experimental manifestation of the resonant topological Hall conductivity. Similar to the d.c. Hall effect \cite{Lux2020, Blugel2021}, a \emph{quantitative} comparison with experiments when skyrmions and significant SOC coexist generally requires a study where both skyrmions and SOC are simultaneously treated. Studying this would be the next logical step in exploring optical probes of skyrmions. It is also intriguing to explore whether the Hall resonance, involving pairs of topological 
Chern bands, can be understood using the state-pairwise geometrical construction formulated in Ref.\cite{Ashvin2021}.

\begin{acknowledgments}
We thank Michael Bartram for a related collaboration and discussions. This work was funded by NSERC of Canada. This research was  enabled  in  part  by  support  provided  by  WestGrid  (www.westgrid.ca) and Compute Canada  Calcul Canada (www.computecanada.ca).
\end{acknowledgments}

\appendix

\begin{widetext}

\section{Simplifying the Kubo formula for sum rule and resonant Hall conductivity}
\label{app:matrix_element}
\begin{figure}[t]
\centering
\includegraphics[width=0.6 \textwidth]{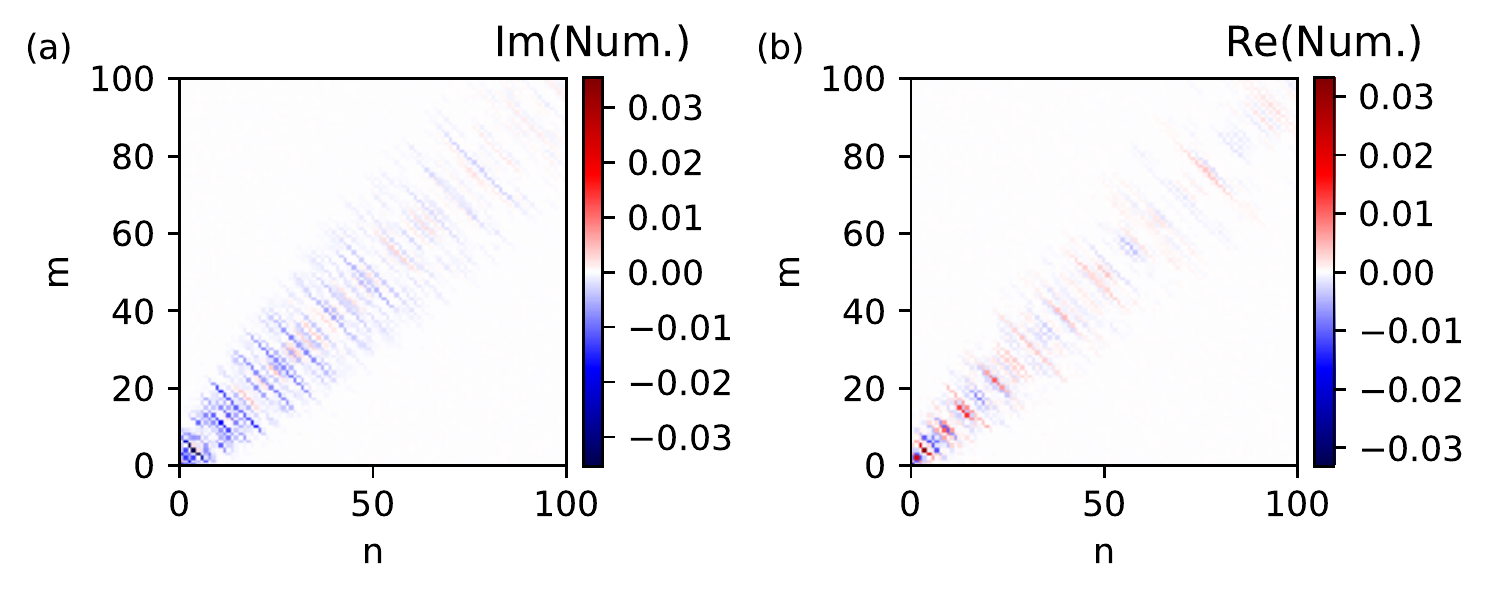}
\caption{Color plots illustrating that the numerator in the Kubo formula in Eq.\ref{eq:kubo1}, involving current operator matrix elements, 
has its magnitude peaking around $n = m$ and decaying quickly with increasing $|n - m|$. The result is obtained at a generic crystal momentum 
$\bk$ and for $L = 20$.}
\label{fig:matrix_element}
\end{figure}

In this appendix, we will show that the area under the resonance of Im $\sigma_{xy}$, $\mathcal{S}$, is proportional to $\rho_{sk}$.

\bea
\mathcal{S} &=& \int \td \omega \text{ Im }\sigma^{\rm res}_{xy}(\omega),\nonumber\\
&\approx& \frac{\pi}{\mathcal{A}J_H} \text{Im} \sum_{\bk}\sum_{E^P_{\bk n} < \mu} \sum_{m} 
\bra{P \bk n} j^x_{pa} \ket{A \bk m}\bra{A \bk m} j^y_{ap} \ket{P \bk n},\\
&\equiv& \frac{\pi}{\mathcal{A}J_H} \text{Im } \mathcal{N}.
\eea
The magnitude of the matrix element $\bra{P \bk n} j^x_{pa} \ket{A \bk m}\bra{A \bk m} j^y_{ap} \ket{P \bk n}$ is found to generally peak around $n \approx m$, and it decays abruptly as $|n - m|$ increases, as illustrated by Fig.\ref{fig:matrix_element}. This sharp feature is used to shed light on the Lorentzian shape of the Hall resonance in the main text. The Hall resonance curve $\sigma_{xy}^{\rm res}(\omega)$ in Eq.\ref{eq:res} and the area $\mathcal{S}$ above can be obtained by calculating 
$\mathcal{N}$, which can be simplified by the following observations.

Observation (1): we can use the Bloch theorem to convert the matrix elements of $j^x_{pa}$ and $j^y_{ap}$, which involve the whole lattice summation, into summations over a chosen magnetic unit cell, i.e.
\bea
\bra{P \bk n} j^x_{pa}\ket{A \bk m} &=& -\frac{\I e N_{sk}}{\hbar}\sum_{i \in u.c.}\sum_{\bdelta_a} T_{pa, i(i+\bdelta_a)} \bra{P \bk n} p^{\dagger}_ia_{i+\bdelta_a}+p_{i+\bdelta_a}^{\dagger}a_i\ket{A \bk m}\bdelta_{a, x},\\
\bra{A \bk m} j^y_{ap}\ket{P \bk n} & = &-\frac{\I e N_{sk}}{\hbar}\sum_{i\in u.c.}\sum_{\bdelta_a} T_{ap, i(i+\bdelta_a)} \bra{A \bk m} a^{\dagger}_ip_{i+\bdelta_a}+a_{i+\bdelta_a}^{\dagger}p_i\ket{P \bk n}\bdelta_{a, y},
\eea
where $N_{sk}$ denotes the number of skyrmion in the system which equals to the number of unit cell since we have one skyrmion per unit cell. $\bdelta_a$ are the nearest neighbor vectors.
\bea
\bdelta_1 &=& (1, 0),\\
\bdelta_2 &=& (1/2, \sqrt{3}/2),\\
\bdelta_3 &=& (-1/2, \sqrt{3}/2).
\eea

Observation (2): the operator $\sum_m \ket{A\bk m}\bra{A\bk m}$ can be shown to be diagonal in the sublattice index $s$.
\bea
\sum_m \ket{A\bk m}\bra{A\bk m} &=&\!\!\!\! \sum_{\bD, \bD', s} \!\!\! \frac{e^{\I \bk \cdot (\bD - \bD')}}{N_{sk}} \ket{A\bD s}\bra{A\bD' s}
\eea
where $s = 1, \cdots, L^2$ is the sublattice index, $\ket{A\bD s}$ is an anti-parallel electron state localized at a site whose position is given by $\bD + \bd_s$, where $\bD$ is the position of a reference point of a unit cell, and $\bd_s$ is the position of the site relative to the reference point. When this is used concurrently with observation (1), the phase factor becomes unity as the unit cell is fixed to be the same for the matrix elements of $j^x_{pa}$ and $j^y_{ap}$. 

For a given bond $\langle i (i+\bdelta_a)\rangle$ in the $j^x_{pa}$ matrix element, the $s$-diagonal feature selects only several bonds $\langle i (i + \bdelta_a)\rangle$'s in $j^y_{ap}$ matrix element. Suppose that $\langle i_1 i_2 \rangle$ and $\langle i_3 i_4 \rangle$ are the relevant bonds from $j^x_{pa}$ and $j^y_{ap}$ respectively, the contribution to $\mathcal{N}$ becomes $\sim
 \bra{P\bk n} (p^{\dagger}_{i_1}a_{i_2}+p^{\dagger}_{i_2}a_{i_1})(a^{\dagger}_{i_3}p_{i_4}+a^{\dagger}_{i_4}p_{i_3})\ket{P\bk n} = \bra{P\bk n} p^{\dagger}_{i_5}p_{i_6}\ket{P\bk n}$, where $i_5$ and $i_6$ are two sites taken from $\{i_1, i_2, i_3, i_4\}$, and the precise answer depends on the type of bonds, which will be illustrated later. To arrive at this property, we recall that $a$ and $p$ operator annihilate $\ket{P\bk n}$ and $\ket{A \bk m}$ respectively.
 
 From these, $\mathcal{N}$ becomes
\bea
\mathcal{N} &=& N_{sk} \left(\frac{\I e}{\hbar}\right)^2 \sum_{\bk}\sum_{E^P_{\bk n} < \mu} \sum'_{\langle i_1 i_2\rangle}\sum''_{\langle i_3 i_4 \rangle} \bra{P\bk n} Q_{i_1i_2i_3i_4}^{i_5i_6} \ket{P \bk n},\\
Q_{i_1i_2i_3i_4}^{i_5i_6} &=& T_{pa, i_1i_2} T_{ap, i_3i_4} \br_{i_1i_2, x}\br_{i_3i_4,y} p^{\dagger}_{i_5}p_{i_6},
\eea
where the sum over $\langle i_1 i_2\rangle$ is the equivalence of $\sum_{i_1 \in u.c.}\sum_{\bdelta_a}$ with $i_2 = i_1 + \bdelta_a$. The sum over $\langle i_3 i_4 \rangle$ is similarly defined, except that it is done over a more restricted subset as mentioned earlier, and hence the double prime.
There are three types of $\langle i_1 i_2 \rangle$ bond associated with the nearest neighbor vectors.
However, there are only two types of $\langle i_3 i_4 \rangle$ associated with $\bdelta_2$ and $\bdelta_3$ since $\bdelta_1$ has a vanishing y-component.

\begin{figure}[h]
\centering
\includegraphics[width = 0.6\textwidth]{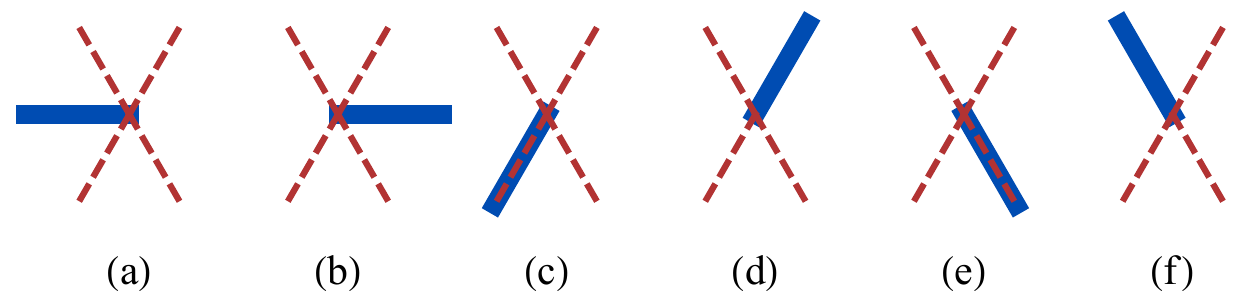}
\caption{Diagrams illustrating the three types of $\langle i_1 i_2\rangle$ bond in $j^x_{pa}$ corresponding to $\{(a), (b)\}$, $\{(c), (d)\}$ and $\{(e), (f)\}$. They are denoted by the solid blue lines. For a given solid bond, there are several $\langle i_3i_4 \rangle$ bonds in $j^y_{ap}$, which combine with $\langle i_1 i_2\rangle$ to give a nonzero contribution towards $\mathcal{N}$. They are denoted by red dashed lines.}
\label{fig:diagram}
\end{figure}

We illustrate the computation of $Q_{i_1i_2i_3i_4}^{i_5i_6}$ using the following concrete example.
Suppose that $\langle i_1i_2 \rangle = \bdelta_1$, i.e. $i_2 = i_1 + \bdelta_1$. There are only eight $\langle i_3i_4 \rangle$ bonds contributing to $\mathcal{S}$ as sketched in Fig.\ref{fig:diagram}(a-b). $\langle i_1 i_2 \rangle$ is denoted by the blue bonds, while $\langle i_3 i_4 \rangle$ is denoted by the red dashed bonds. The common feature is that either $i_1$ or $i_2$ is equal to $i_3$ or $i_4$.
For $i_3 = i_2$ and $i_4 = i_3 + \bdelta_3$, $i_5$ and $i_6$ correspond to the dangling end points, namely $(i_5, i_6) = (i_1, i_4)$ since $i_2$ is glued to $i_3$.

The hopping matrix element, $\bra{P\bk n} p^{\dagger}_{i_5}p_{i_6}\ket{P\bk n}$, is hard to compute analytically since it requires solving the Hofstadter model for the parallel sector. 
Therefore, for the sake of deriving the scaling relation, we assume that it is given by a value corresponding to a uniform ferromagnetic spin texture with $\bS_i = \hat{z}$, where $\ket{P\bk n} \rightarrow \ket{\bp \uparrow}$ and $p^{\dagger}_i \rightarrow c^{\dagger}_{i\uparrow}$. $\bp$ can be determined by $\bk$ when viewing that the SkX BZ is obtained by folding the BZ of the Bravais triangular lattice to which $\bp$ belong.
\bea
\bra{P \bk n} p^{\dagger}_{i_5}p_{i_6} \ket{P \bk n} \rightarrow \bra{\bp \uparrow} c^{\dagger}_{i_5 \uparrow}c_{i_6\uparrow} \ket{\bp \uparrow} = \frac{e^{\I \bp \cdot \bdelta_2}}{\mathcal{A}}.
\eea
This approximation works well when the SkX spin texture varies slowly in space, namely large $L$, such that the neighbouring spins are almost parallel. 
Since the result is site-independent, the sum over $\langle i_1 i_2 \rangle$ contains only $T_{pa, i_1i_2}$ and $T_{ap, i_3i_4}$, namely
\bea
\sum_{i_1 \in u.c.} T_{pa, i_1(i_1 + \bdelta_1)} T_{ap, (i_1+\bdelta_1)(i_1+\bdelta_1 +\bdelta_3)} \bdelta_{1, x}\bdelta_{3,y}.
\eea

The following observation is useful for computing the sum.
The hopping integral
\bea
T_{pa, i_1i_2} &=& -t \left(e^{-\I\phi_{i_2}}\cos\frac{\theta_{i_1}}{2}\sin\frac{\theta_{i_2}}{2} - e^{-\I\phi_{i_1}}\cos\frac{\theta_{i_2}}{2}\sin\frac{\theta_{i_1}}{2}\right),\nonumber\\
&=& -t \left[m_{i_1}^z(m_{i_2}^x - \I m_{i_2}^y) - m_{i_2}^z (m_{i_1}^x - \I m_{i_1}^y)\right], 
\eea
where we have re-expressed it in terms of a slowly varying auxiliary ``vector field" $\bfm_{i_1}$ defined on each site as
\bea
\bfm_{i_1} &=& \left(\sin\frac{\theta_{i_1}}{2}\cos\phi_{i_1}, \sin\frac{\theta_{i_1}}{2}\sin\phi_{i_1}, \cos\frac{\theta_{i_1}}{2}\right).
\eea
We have checked that, at large $L$, the SkX spin texture defined in Section \ref{sec:model} indeed leads to a smooth $\{\bfm_{i_1}\}$ which can be approximated by a continuum vector field $\bfm(\br)$, where
\bea
\bfm_{i_1} &\approx & \bfm - \bdelta_1 \cdot \boldsymbol{\nabla} \bfm,\\
\bfm_{i_2} &\approx & \bfm,\\
\bfm_{i_3} &\approx & \bfm,\\
\bfm_{i_4} &\approx & \bfm + \bdelta_3 \cdot \boldsymbol{\nabla} \bfm.
\eea
The Taylor expansion is expanded around the common site $i_2 = i_3$. After converted into an integral, the summation, therefore, becomes
\bea
\frac{2t^2}{\sqrt{3}} \int_{u.c.} \td \br \frac{\sqrt{3}}{2}\left(-\bdelta_1 \cdot \boldsymbol{\nabla} m^z (m^x - \I m^y) + m^z \bdelta_1 \cdot \boldsymbol{\nabla}  (m^x - \I m^y) \right)  \left(\bdelta_3 \cdot \boldsymbol{\nabla} m^z (m^x + \I m^y) - m^z \bdelta_3 \cdot \boldsymbol{\nabla}  (m^x + \I m^y)\right).\nonumber
\eea
This integral can be computed using the circular ansatz defined in the main text. After including the contributions from other bond combinations listed in Fig.\ref{fig:diagram}, we obtain the final result for $\mathcal{N}$ which is purely imaginary:
\bea
\mathcal{N} &=& \I \pi\sqrt{3}N_{sk} \left(\frac{te}{\hbar}\right)^2 \int_{E^{FM}_{\bp \uparrow} < \mu_{FM}} \frac{\td \bp}{(2\pi)^2} \left[\cos p_x + \cos \sqrt{3}p_y + 2 \cos \frac{p_y \sqrt{3}}{2}\left(\cos \frac{p_x}{2} + \cos \frac{3p_x}{2}\right)\right],
\eea
where we have converted $\sum_{\bp} \rightarrow \mathcal{A} \int \frac{\td \bp}{(2\pi)^2}$. The summation over the occupied bands is estimated by the ferromagnetic value corresponding to the dispersion $E_{\bp \uparrow}^{FM}$ of the parallel sector and the chemical potential $\mu_{FM}$ determined by the electron density.

Therefore, the area $\mathcal{S}$ is indeed proportional to the skyrmion density $N_{sk} / \mathcal{A}$.
\bea
{\cal S} &\approx & \frac{e^2}{\hbar^2} \frac{\pi t^2}{J_H} \frac{N_{sk}}{\mathcal{A}} {\cal F},\\
{\cal F} &=& \pi \sqrt{3}\int_{E^{FM}_{\bp \uparrow} < \mu_{FM}} \frac{\td \bp}{(2\pi)^2} \left[\cos p_x + \cos \sqrt{3}p_y + 2 \cos \frac{p_y \sqrt{3}}{2}\left(\cos \frac{p_x}{2} + \cos \frac{3p_x}{2}\right)\right].
\eea

\end{widetext}

\bibliography{refs}

\begin{thebibliography}{78}%
\makeatletter
\providecommand \@ifxundefined [1]{%
 \@ifx{#1\undefined}
}%
\providecommand \@ifnum [1]{%
 \ifnum #1\expandafter \@firstoftwo
 \else \expandafter \@secondoftwo
 \fi
}%
\providecommand \@ifx [1]{%
 \ifx #1\expandafter \@firstoftwo
 \else \expandafter \@secondoftwo
 \fi
}%
\providecommand \natexlab [1]{#1}%
\providecommand \enquote  [1]{``#1''}%
\providecommand \bibnamefont  [1]{#1}%
\providecommand \bibfnamefont [1]{#1}%
\providecommand \citenamefont [1]{#1}%
\providecommand \href@noop [0]{\@secondoftwo}%
\providecommand \href [0]{\begingroup \@sanitize@url \@href}%
\providecommand \@href[1]{\@@startlink{#1}\@@href}%
\providecommand \@@href[1]{\endgroup#1\@@endlink}%
\providecommand \@sanitize@url [0]{\catcode `\\12\catcode `\$12\catcode
  `\&12\catcode `\#12\catcode `\^12\catcode `\_12\catcode `\%12\relax}%
\providecommand \@@startlink[1]{}%
\providecommand \@@endlink[0]{}%
\providecommand \url  [0]{\begingroup\@sanitize@url \@url }%
\providecommand \@url [1]{\endgroup\@href {#1}{\urlprefix }}%
\providecommand \urlprefix  [0]{URL }%
\providecommand \Eprint [0]{\href }%
\providecommand \doibase [0]{http://dx.doi.org/}%
\providecommand \selectlanguage [0]{\@gobble}%
\providecommand \bibinfo  [0]{\@secondoftwo}%
\providecommand \bibfield  [0]{\@secondoftwo}%
\providecommand \translation [1]{[#1]}%
\providecommand \BibitemOpen [0]{}%
\providecommand \bibitemStop [0]{}%
\providecommand \bibitemNoStop [0]{.\EOS\space}%
\providecommand \EOS [0]{\spacefactor3000\relax}%
\providecommand \BibitemShut  [1]{\csname bibitem#1\endcsname}%
\let\auto@bib@innerbib\@empty
%</preamble>
\bibitem [{\citenamefont {Skyrme}(1962)}]{skyrme}%
  \BibitemOpen
  \bibfield  {author} {\bibinfo {author} {\bibfnamefont {T.}~\bibnamefont
  {Skyrme}},\ }\href {\doibase https://doi.org/10.1016/0029-5582(62)90775-7}
  {\bibfield  {journal} {\bibinfo  {journal} {Nuclear Physics}\ }\textbf
  {\bibinfo {volume} {31}},\ \bibinfo {pages} {556} (\bibinfo {year}
  {1962})}\BibitemShut {NoStop}%
\bibitem [{\citenamefont {Nagaosa}\ and\ \citenamefont
  {Tokura}(2013)}]{Nagaosa2013}%
  \BibitemOpen
  \bibfield  {author} {\bibinfo {author} {\bibfnamefont {N.}~\bibnamefont
  {Nagaosa}}\ and\ \bibinfo {author} {\bibfnamefont {Y.}~\bibnamefont
  {Tokura}},\ }\href {\doibase 10.1038/nnano.2013.243} {\bibfield  {journal}
  {\bibinfo  {journal} {Nature Nanotechnology}\ }\textbf {\bibinfo {volume}
  {8}},\ \bibinfo {pages} {899} (\bibinfo {year} {2013})}\BibitemShut {NoStop}%
\bibitem [{\citenamefont {Fert}\ \emph {et~al.}(2013)\citenamefont {Fert},
  \citenamefont {Cros},\ and\ \citenamefont {Sampaio}}]{Fert2013}%
  \BibitemOpen
  \bibfield  {author} {\bibinfo {author} {\bibfnamefont {A.}~\bibnamefont
  {Fert}}, \bibinfo {author} {\bibfnamefont {V.}~\bibnamefont {Cros}}, \ and\
  \bibinfo {author} {\bibfnamefont {J.}~\bibnamefont {Sampaio}},\ }\href
  {\doibase 10.1038/nnano.2013.29} {\bibfield  {journal} {\bibinfo  {journal}
  {Nature Nanotechnology}\ }\textbf {\bibinfo {volume} {8}},\ \bibinfo {pages}
  {152} (\bibinfo {year} {2013})}\BibitemShut {NoStop}%
\bibitem [{\citenamefont {Jiang}\ \emph {et~al.}(2017)\citenamefont {Jiang},
  \citenamefont {Chen}, \citenamefont {Liu}, \citenamefont {Zang},
  \citenamefont {{te Velthuis}},\ and\ \citenamefont {Hoffmann}}]{Jiang2017}%
  \BibitemOpen
  \bibfield  {author} {\bibinfo {author} {\bibfnamefont {W.}~\bibnamefont
  {Jiang}}, \bibinfo {author} {\bibfnamefont {G.}~\bibnamefont {Chen}},
  \bibinfo {author} {\bibfnamefont {K.}~\bibnamefont {Liu}}, \bibinfo {author}
  {\bibfnamefont {J.}~\bibnamefont {Zang}}, \bibinfo {author} {\bibfnamefont
  {S.~G.}\ \bibnamefont {{te Velthuis}}}, \ and\ \bibinfo {author}
  {\bibfnamefont {A.}~\bibnamefont {Hoffmann}},\ }\href {\doibase
  https://doi.org/10.1016/j.physrep.2017.08.001} {\bibfield  {journal}
  {\bibinfo  {journal} {Physics Reports}\ }\textbf {\bibinfo {volume} {704}},\
  \bibinfo {pages} {1} (\bibinfo {year} {2017})},\ \bibinfo {note} {skyrmions
  in Magnetic Multilayers}\BibitemShut {NoStop}%
\bibitem [{\citenamefont {Back}\ \emph {et~al.}(2020)\citenamefont {Back},
  \citenamefont {Cros}, \citenamefont {Ebert}, \citenamefont {Everschor-Sitte},
  \citenamefont {Fert}, \citenamefont {Garst}, \citenamefont {Ma},
  \citenamefont {Mankovsky}, \citenamefont {Monchesky}, \citenamefont
  {Mostovoy}, \citenamefont {Nagaosa}, \citenamefont {Parkin}, \citenamefont
  {Pfleiderer}, \citenamefont {Reyren}, \citenamefont {Rosch}, \citenamefont
  {Taguchi}, \citenamefont {Tokura}, \citenamefont {von Bergmann},\ and\
  \citenamefont {Zang}}]{Back2020}%
  \BibitemOpen
  \bibfield  {author} {\bibinfo {author} {\bibfnamefont {C.}~\bibnamefont
  {Back}}, \bibinfo {author} {\bibfnamefont {V.}~\bibnamefont {Cros}}, \bibinfo
  {author} {\bibfnamefont {H.}~\bibnamefont {Ebert}}, \bibinfo {author}
  {\bibfnamefont {K.}~\bibnamefont {Everschor-Sitte}}, \bibinfo {author}
  {\bibfnamefont {A.}~\bibnamefont {Fert}}, \bibinfo {author} {\bibfnamefont
  {M.}~\bibnamefont {Garst}}, \bibinfo {author} {\bibfnamefont
  {T.}~\bibnamefont {Ma}}, \bibinfo {author} {\bibfnamefont {S.}~\bibnamefont
  {Mankovsky}}, \bibinfo {author} {\bibfnamefont {T.~L.}\ \bibnamefont
  {Monchesky}}, \bibinfo {author} {\bibfnamefont {M.}~\bibnamefont {Mostovoy}},
  \bibinfo {author} {\bibfnamefont {N.}~\bibnamefont {Nagaosa}}, \bibinfo
  {author} {\bibfnamefont {S.~S.~P.}\ \bibnamefont {Parkin}}, \bibinfo {author}
  {\bibfnamefont {C.}~\bibnamefont {Pfleiderer}}, \bibinfo {author}
  {\bibfnamefont {N.}~\bibnamefont {Reyren}}, \bibinfo {author} {\bibfnamefont
  {A.}~\bibnamefont {Rosch}}, \bibinfo {author} {\bibfnamefont
  {Y.}~\bibnamefont {Taguchi}}, \bibinfo {author} {\bibfnamefont
  {Y.}~\bibnamefont {Tokura}}, \bibinfo {author} {\bibfnamefont
  {K.}~\bibnamefont {von Bergmann}}, \ and\ \bibinfo {author} {\bibfnamefont
  {J.}~\bibnamefont {Zang}},\ }\href {\doibase 10.1088/1361-6463/ab8418}
  {\bibfield  {journal} {\bibinfo  {journal} {Journal of Physics D: Applied
  Physics}\ }\textbf {\bibinfo {volume} {53}},\ \bibinfo {pages} {363001}
  (\bibinfo {year} {2020})}\BibitemShut {NoStop}%
\bibitem [{\citenamefont {Tokura}\ and\ \citenamefont
  {Kanazawa}(2021)}]{Tokura2021}%
  \BibitemOpen
  \bibfield  {author} {\bibinfo {author} {\bibfnamefont {Y.}~\bibnamefont
  {Tokura}}\ and\ \bibinfo {author} {\bibfnamefont {N.}~\bibnamefont
  {Kanazawa}},\ }\href {\doibase 10.1021/acs.chemrev.0c00297} {\bibfield
  {journal} {\bibinfo  {journal} {Chemical Reviews}\ }\textbf {\bibinfo
  {volume} {121}},\ \bibinfo {pages} {2857} (\bibinfo {year}
  {2021})}\BibitemShut {NoStop}%
\bibitem [{\citenamefont {Barrett}\ \emph {et~al.}(1995)\citenamefont
  {Barrett}, \citenamefont {Dabbagh}, \citenamefont {Pfeiffer}, \citenamefont
  {West},\ and\ \citenamefont {Tycko}}]{Barrett1995}%
  \BibitemOpen
  \bibfield  {author} {\bibinfo {author} {\bibfnamefont {S.~E.}\ \bibnamefont
  {Barrett}}, \bibinfo {author} {\bibfnamefont {G.}~\bibnamefont {Dabbagh}},
  \bibinfo {author} {\bibfnamefont {L.~N.}\ \bibnamefont {Pfeiffer}}, \bibinfo
  {author} {\bibfnamefont {K.~W.}\ \bibnamefont {West}}, \ and\ \bibinfo
  {author} {\bibfnamefont {R.}~\bibnamefont {Tycko}},\ }\href {\doibase
  10.1103/PhysRevLett.74.5112} {\bibfield  {journal} {\bibinfo  {journal}
  {Phys. Rev. Lett.}\ }\textbf {\bibinfo {volume} {74}},\ \bibinfo {pages}
  {5112} (\bibinfo {year} {1995})}\BibitemShut {NoStop}%
\bibitem [{\citenamefont {Shkolnikov}\ \emph {et~al.}(2005)\citenamefont
  {Shkolnikov}, \citenamefont {Misra}, \citenamefont {Bishop}, \citenamefont
  {De~Poortere},\ and\ \citenamefont {Shayegan}}]{Shkolnikov2005}%
  \BibitemOpen
  \bibfield  {author} {\bibinfo {author} {\bibfnamefont {Y.~P.}\ \bibnamefont
  {Shkolnikov}}, \bibinfo {author} {\bibfnamefont {S.}~\bibnamefont {Misra}},
  \bibinfo {author} {\bibfnamefont {N.~C.}\ \bibnamefont {Bishop}}, \bibinfo
  {author} {\bibfnamefont {E.~P.}\ \bibnamefont {De~Poortere}}, \ and\ \bibinfo
  {author} {\bibfnamefont {M.}~\bibnamefont {Shayegan}},\ }\href {\doibase
  10.1103/PhysRevLett.95.066809} {\bibfield  {journal} {\bibinfo  {journal}
  {Phys. Rev. Lett.}\ }\textbf {\bibinfo {volume} {95}},\ \bibinfo {pages}
  {066809} (\bibinfo {year} {2005})}\BibitemShut {NoStop}%
\bibitem [{\citenamefont {Neubauer}\ \emph {et~al.}(2009)\citenamefont
  {Neubauer}, \citenamefont {Pfleiderer}, \citenamefont {Binz}, \citenamefont
  {Rosch}, \citenamefont {Ritz}, \citenamefont {Niklowitz},\ and\ \citenamefont
  {B\"oni}}]{Neubauer2009}%
  \BibitemOpen
  \bibfield  {author} {\bibinfo {author} {\bibfnamefont {A.}~\bibnamefont
  {Neubauer}}, \bibinfo {author} {\bibfnamefont {C.}~\bibnamefont
  {Pfleiderer}}, \bibinfo {author} {\bibfnamefont {B.}~\bibnamefont {Binz}},
  \bibinfo {author} {\bibfnamefont {A.}~\bibnamefont {Rosch}}, \bibinfo
  {author} {\bibfnamefont {R.}~\bibnamefont {Ritz}}, \bibinfo {author}
  {\bibfnamefont {P.~G.}\ \bibnamefont {Niklowitz}}, \ and\ \bibinfo {author}
  {\bibfnamefont {P.}~\bibnamefont {B\"oni}},\ }\href {\doibase
  10.1103/PhysRevLett.102.186602} {\bibfield  {journal} {\bibinfo  {journal}
  {Phys. Rev. Lett.}\ }\textbf {\bibinfo {volume} {102}},\ \bibinfo {pages}
  {186602} (\bibinfo {year} {2009})}\BibitemShut {NoStop}%
\bibitem [{\citenamefont {Lee}\ \emph {et~al.}(2009)\citenamefont {Lee},
  \citenamefont {Kang}, \citenamefont {Onose}, \citenamefont {Tokura},\ and\
  \citenamefont {Ong}}]{Lee2009}%
  \BibitemOpen
  \bibfield  {author} {\bibinfo {author} {\bibfnamefont {M.}~\bibnamefont
  {Lee}}, \bibinfo {author} {\bibfnamefont {W.}~\bibnamefont {Kang}}, \bibinfo
  {author} {\bibfnamefont {Y.}~\bibnamefont {Onose}}, \bibinfo {author}
  {\bibfnamefont {Y.}~\bibnamefont {Tokura}}, \ and\ \bibinfo {author}
  {\bibfnamefont {N.~P.}\ \bibnamefont {Ong}},\ }\href {\doibase
  10.1103/PhysRevLett.102.186601} {\bibfield  {journal} {\bibinfo  {journal}
  {Phys. Rev. Lett.}\ }\textbf {\bibinfo {volume} {102}},\ \bibinfo {pages}
  {186601} (\bibinfo {year} {2009})}\BibitemShut {NoStop}%
\bibitem [{\citenamefont {M{\"u}hlbauer}\ \emph {et~al.}(2009)\citenamefont
  {M{\"u}hlbauer}, \citenamefont {Binz}, \citenamefont {Jonietz}, \citenamefont
  {Pfleiderer}, \citenamefont {Rosch}, \citenamefont {Neubauer}, \citenamefont
  {Georgii},\ and\ \citenamefont {B{\"o}ni}}]{Muhlbauer2009}%
  \BibitemOpen
  \bibfield  {author} {\bibinfo {author} {\bibfnamefont {S.}~\bibnamefont
  {M{\"u}hlbauer}}, \bibinfo {author} {\bibfnamefont {B.}~\bibnamefont {Binz}},
  \bibinfo {author} {\bibfnamefont {F.}~\bibnamefont {Jonietz}}, \bibinfo
  {author} {\bibfnamefont {C.}~\bibnamefont {Pfleiderer}}, \bibinfo {author}
  {\bibfnamefont {A.}~\bibnamefont {Rosch}}, \bibinfo {author} {\bibfnamefont
  {A.}~\bibnamefont {Neubauer}}, \bibinfo {author} {\bibfnamefont
  {R.}~\bibnamefont {Georgii}}, \ and\ \bibinfo {author} {\bibfnamefont
  {P.}~\bibnamefont {B{\"o}ni}},\ }\href {\doibase 10.1126/science.1166767}
  {\bibfield  {journal} {\bibinfo  {journal} {Science}\ }\textbf {\bibinfo
  {volume} {323}},\ \bibinfo {pages} {915} (\bibinfo {year}
  {2009})}\BibitemShut {NoStop}%
\bibitem [{\citenamefont {Kanazawa}\ \emph {et~al.}(2012)\citenamefont
  {Kanazawa}, \citenamefont {Kim}, \citenamefont {Inosov}, \citenamefont
  {White}, \citenamefont {Egetenmeyer}, \citenamefont {Gavilano}, \citenamefont
  {Ishiwata}, \citenamefont {Onose}, \citenamefont {Arima}, \citenamefont
  {Keimer},\ and\ \citenamefont {Tokura}}]{Tokura2012}%
  \BibitemOpen
  \bibfield  {author} {\bibinfo {author} {\bibfnamefont {N.}~\bibnamefont
  {Kanazawa}}, \bibinfo {author} {\bibfnamefont {J.-H.}\ \bibnamefont {Kim}},
  \bibinfo {author} {\bibfnamefont {D.~S.}\ \bibnamefont {Inosov}}, \bibinfo
  {author} {\bibfnamefont {J.~S.}\ \bibnamefont {White}}, \bibinfo {author}
  {\bibfnamefont {N.}~\bibnamefont {Egetenmeyer}}, \bibinfo {author}
  {\bibfnamefont {J.~L.}\ \bibnamefont {Gavilano}}, \bibinfo {author}
  {\bibfnamefont {S.}~\bibnamefont {Ishiwata}}, \bibinfo {author}
  {\bibfnamefont {Y.}~\bibnamefont {Onose}}, \bibinfo {author} {\bibfnamefont
  {T.}~\bibnamefont {Arima}}, \bibinfo {author} {\bibfnamefont
  {B.}~\bibnamefont {Keimer}}, \ and\ \bibinfo {author} {\bibfnamefont
  {Y.}~\bibnamefont {Tokura}},\ }\href {\doibase 10.1103/PhysRevB.86.134425}
  {\bibfield  {journal} {\bibinfo  {journal} {Phys. Rev. B}\ }\textbf {\bibinfo
  {volume} {86}},\ \bibinfo {pages} {134425} (\bibinfo {year}
  {2012})}\BibitemShut {NoStop}%
\bibitem [{\citenamefont {Yu}\ \emph {et~al.}(2010)\citenamefont {Yu},
  \citenamefont {Onose}, \citenamefont {Kanazawa}, \citenamefont {Park},
  \citenamefont {Han}, \citenamefont {Matsui}, \citenamefont {Nagaosa},\ and\
  \citenamefont {Tokura}}]{Yu2010}%
  \BibitemOpen
  \bibfield  {author} {\bibinfo {author} {\bibfnamefont {X.~Z.}\ \bibnamefont
  {Yu}}, \bibinfo {author} {\bibfnamefont {Y.}~\bibnamefont {Onose}}, \bibinfo
  {author} {\bibfnamefont {N.}~\bibnamefont {Kanazawa}}, \bibinfo {author}
  {\bibfnamefont {J.~H.}\ \bibnamefont {Park}}, \bibinfo {author}
  {\bibfnamefont {J.~H.}\ \bibnamefont {Han}}, \bibinfo {author} {\bibfnamefont
  {Y.}~\bibnamefont {Matsui}}, \bibinfo {author} {\bibfnamefont
  {N.}~\bibnamefont {Nagaosa}}, \ and\ \bibinfo {author} {\bibfnamefont
  {Y.}~\bibnamefont {Tokura}},\ }\href {\doibase 10.1038/nature09124}
  {\bibfield  {journal} {\bibinfo  {journal} {Nature}\ }\textbf {\bibinfo
  {volume} {465}},\ \bibinfo {pages} {901} (\bibinfo {year}
  {2010})}\BibitemShut {NoStop}%
\bibitem [{\citenamefont {Seki}\ \emph {et~al.}(2012)\citenamefont {Seki},
  \citenamefont {Yu}, \citenamefont {Ishiwata},\ and\ \citenamefont
  {Tokura}}]{Seki2012}%
  \BibitemOpen
  \bibfield  {author} {\bibinfo {author} {\bibfnamefont {S.}~\bibnamefont
  {Seki}}, \bibinfo {author} {\bibfnamefont {X.~Z.}\ \bibnamefont {Yu}},
  \bibinfo {author} {\bibfnamefont {S.}~\bibnamefont {Ishiwata}}, \ and\
  \bibinfo {author} {\bibfnamefont {Y.}~\bibnamefont {Tokura}},\ }\href
  {\doibase 10.1126/science.1214143} {\bibfield  {journal} {\bibinfo  {journal}
  {Science}\ }\textbf {\bibinfo {volume} {336}},\ \bibinfo {pages} {198}
  (\bibinfo {year} {2012})}\BibitemShut {NoStop}%
\bibitem [{\citenamefont {Moreau-Luchaire}\ \emph {et~al.}(2016)\citenamefont
  {Moreau-Luchaire}, \citenamefont {Moutafis}, \citenamefont {Reyren},
  \citenamefont {Sampaio}, \citenamefont {Vaz}, \citenamefont {Van~Horne},
  \citenamefont {Bouzehouane}, \citenamefont {Garcia}, \citenamefont
  {Deranlot}, \citenamefont {Warnicke}, \citenamefont {Wohlh{\"u}ter},
  \citenamefont {George}, \citenamefont {Weigand}, \citenamefont {Raabe},
  \citenamefont {Cros},\ and\ \citenamefont {Fert}}]{Moreau2016}%
  \BibitemOpen
  \bibfield  {author} {\bibinfo {author} {\bibfnamefont {C.}~\bibnamefont
  {Moreau-Luchaire}}, \bibinfo {author} {\bibfnamefont {C.}~\bibnamefont
  {Moutafis}}, \bibinfo {author} {\bibfnamefont {N.}~\bibnamefont {Reyren}},
  \bibinfo {author} {\bibfnamefont {J.}~\bibnamefont {Sampaio}}, \bibinfo
  {author} {\bibfnamefont {C.~A.~F.}\ \bibnamefont {Vaz}}, \bibinfo {author}
  {\bibfnamefont {N.}~\bibnamefont {Van~Horne}}, \bibinfo {author}
  {\bibfnamefont {K.}~\bibnamefont {Bouzehouane}}, \bibinfo {author}
  {\bibfnamefont {K.}~\bibnamefont {Garcia}}, \bibinfo {author} {\bibfnamefont
  {C.}~\bibnamefont {Deranlot}}, \bibinfo {author} {\bibfnamefont
  {P.}~\bibnamefont {Warnicke}}, \bibinfo {author} {\bibfnamefont
  {P.}~\bibnamefont {Wohlh{\"u}ter}}, \bibinfo {author} {\bibfnamefont {J.-M.}\
  \bibnamefont {George}}, \bibinfo {author} {\bibfnamefont {M.}~\bibnamefont
  {Weigand}}, \bibinfo {author} {\bibfnamefont {J.}~\bibnamefont {Raabe}},
  \bibinfo {author} {\bibfnamefont {V.}~\bibnamefont {Cros}}, \ and\ \bibinfo
  {author} {\bibfnamefont {A.}~\bibnamefont {Fert}},\ }\href {\doibase
  10.1038/nnano.2015.313} {\bibfield  {journal} {\bibinfo  {journal} {Nature
  Nanotechnology}\ }\textbf {\bibinfo {volume} {11}},\ \bibinfo {pages} {444}
  (\bibinfo {year} {2016})}\BibitemShut {NoStop}%
\bibitem [{\citenamefont {Wang}\ \emph {et~al.}(2018)\citenamefont {Wang},
  \citenamefont {Feng}, \citenamefont {Kim}, \citenamefont {Kim}, \citenamefont
  {Lee}, \citenamefont {Pollard}, \citenamefont {Shin}, \citenamefont {Zhou},
  \citenamefont {Peng}, \citenamefont {Lee}, \citenamefont {Meng},
  \citenamefont {Yang}, \citenamefont {Han}, \citenamefont {Kim}, \citenamefont
  {Lu},\ and\ \citenamefont {Noh}}]{Wang2018}%
  \BibitemOpen
  \bibfield  {author} {\bibinfo {author} {\bibfnamefont {L.}~\bibnamefont
  {Wang}}, \bibinfo {author} {\bibfnamefont {Q.}~\bibnamefont {Feng}}, \bibinfo
  {author} {\bibfnamefont {Y.}~\bibnamefont {Kim}}, \bibinfo {author}
  {\bibfnamefont {R.}~\bibnamefont {Kim}}, \bibinfo {author} {\bibfnamefont
  {K.~H.}\ \bibnamefont {Lee}}, \bibinfo {author} {\bibfnamefont {S.~D.}\
  \bibnamefont {Pollard}}, \bibinfo {author} {\bibfnamefont {Y.~J.}\
  \bibnamefont {Shin}}, \bibinfo {author} {\bibfnamefont {H.}~\bibnamefont
  {Zhou}}, \bibinfo {author} {\bibfnamefont {W.}~\bibnamefont {Peng}}, \bibinfo
  {author} {\bibfnamefont {D.}~\bibnamefont {Lee}}, \bibinfo {author}
  {\bibfnamefont {W.}~\bibnamefont {Meng}}, \bibinfo {author} {\bibfnamefont
  {H.}~\bibnamefont {Yang}}, \bibinfo {author} {\bibfnamefont {J.~H.}\
  \bibnamefont {Han}}, \bibinfo {author} {\bibfnamefont {M.}~\bibnamefont
  {Kim}}, \bibinfo {author} {\bibfnamefont {Q.}~\bibnamefont {Lu}}, \ and\
  \bibinfo {author} {\bibfnamefont {T.~W.}\ \bibnamefont {Noh}},\ }\href
  {\doibase 10.1038/s41563-018-0204-4} {\bibfield  {journal} {\bibinfo
  {journal} {Nature Materials}\ }\textbf {\bibinfo {volume} {17}},\ \bibinfo
  {pages} {1087} (\bibinfo {year} {2018})}\BibitemShut {NoStop}%
\bibitem [{\citenamefont {Gao}\ \emph {et~al.}(2020)\citenamefont {Gao},
  \citenamefont {Rosales}, \citenamefont {G{\'o}mez~Albarrac{\'i}n},
  \citenamefont {Tsurkan}, \citenamefont {Kaur}, \citenamefont {Fennell},
  \citenamefont {Steffens}, \citenamefont {Boehm}, \citenamefont
  {{\v{C}}erm{\'a}k}, \citenamefont {Schneidewind}, \citenamefont {Ressouche},
  \citenamefont {Cabra}, \citenamefont {R{\"u}egg},\ and\ \citenamefont
  {Zaharko}}]{Gao2020}%
  \BibitemOpen
  \bibfield  {author} {\bibinfo {author} {\bibfnamefont {S.}~\bibnamefont
  {Gao}}, \bibinfo {author} {\bibfnamefont {H.~D.}\ \bibnamefont {Rosales}},
  \bibinfo {author} {\bibfnamefont {F.~A.}\ \bibnamefont
  {G{\'o}mez~Albarrac{\'i}n}}, \bibinfo {author} {\bibfnamefont
  {V.}~\bibnamefont {Tsurkan}}, \bibinfo {author} {\bibfnamefont
  {G.}~\bibnamefont {Kaur}}, \bibinfo {author} {\bibfnamefont {T.}~\bibnamefont
  {Fennell}}, \bibinfo {author} {\bibfnamefont {P.}~\bibnamefont {Steffens}},
  \bibinfo {author} {\bibfnamefont {M.}~\bibnamefont {Boehm}}, \bibinfo
  {author} {\bibfnamefont {P.}~\bibnamefont {{\v{C}}erm{\'a}k}}, \bibinfo
  {author} {\bibfnamefont {A.}~\bibnamefont {Schneidewind}}, \bibinfo {author}
  {\bibfnamefont {E.}~\bibnamefont {Ressouche}}, \bibinfo {author}
  {\bibfnamefont {D.~C.}\ \bibnamefont {Cabra}}, \bibinfo {author}
  {\bibfnamefont {C.}~\bibnamefont {R{\"u}egg}}, \ and\ \bibinfo {author}
  {\bibfnamefont {O.}~\bibnamefont {Zaharko}},\ }\href {\doibase
  10.1038/s41586-020-2716-8} {\bibfield  {journal} {\bibinfo  {journal}
  {Nature}\ }\textbf {\bibinfo {volume} {586}},\ \bibinfo {pages} {37}
  (\bibinfo {year} {2020})}\BibitemShut {NoStop}%
\bibitem [{\citenamefont {Okubo}\ \emph {et~al.}(2012)\citenamefont {Okubo},
  \citenamefont {Chung},\ and\ \citenamefont {Kawamura}}]{Okubo2012}%
  \BibitemOpen
  \bibfield  {author} {\bibinfo {author} {\bibfnamefont {T.}~\bibnamefont
  {Okubo}}, \bibinfo {author} {\bibfnamefont {S.}~\bibnamefont {Chung}}, \ and\
  \bibinfo {author} {\bibfnamefont {H.}~\bibnamefont {Kawamura}},\ }\href
  {\doibase 10.1103/PhysRevLett.108.017206} {\bibfield  {journal} {\bibinfo
  {journal} {Phys. Rev. Lett.}\ }\textbf {\bibinfo {volume} {108}},\ \bibinfo
  {pages} {017206} (\bibinfo {year} {2012})}\BibitemShut {NoStop}%
\bibitem [{\citenamefont {Kurumaji}\ \emph {et~al.}(2019)\citenamefont
  {Kurumaji}, \citenamefont {Nakajima}, \citenamefont {Hirschberger},
  \citenamefont {Kikkawa}, \citenamefont {Yamasaki}, \citenamefont {Sagayama},
  \citenamefont {Nakao}, \citenamefont {Taguchi}, \citenamefont {Arima},\ and\
  \citenamefont {Tokura}}]{Kurumaji2019}%
  \BibitemOpen
  \bibfield  {author} {\bibinfo {author} {\bibfnamefont {T.}~\bibnamefont
  {Kurumaji}}, \bibinfo {author} {\bibfnamefont {T.}~\bibnamefont {Nakajima}},
  \bibinfo {author} {\bibfnamefont {M.}~\bibnamefont {Hirschberger}}, \bibinfo
  {author} {\bibfnamefont {A.}~\bibnamefont {Kikkawa}}, \bibinfo {author}
  {\bibfnamefont {Y.}~\bibnamefont {Yamasaki}}, \bibinfo {author}
  {\bibfnamefont {H.}~\bibnamefont {Sagayama}}, \bibinfo {author}
  {\bibfnamefont {H.}~\bibnamefont {Nakao}}, \bibinfo {author} {\bibfnamefont
  {Y.}~\bibnamefont {Taguchi}}, \bibinfo {author} {\bibfnamefont {T.-h.}\
  \bibnamefont {Arima}}, \ and\ \bibinfo {author} {\bibfnamefont
  {Y.}~\bibnamefont {Tokura}},\ }\href {\doibase 10.1126/science.aau0968}
  {\bibfield  {journal} {\bibinfo  {journal} {Science}\ }\textbf {\bibinfo
  {volume} {365}},\ \bibinfo {pages} {914} (\bibinfo {year}
  {2019})}\BibitemShut {NoStop}%
\bibitem [{\citenamefont {Hirschberger}\ \emph {et~al.}(2019)\citenamefont
  {Hirschberger}, \citenamefont {Nakajima}, \citenamefont {Gao}, \citenamefont
  {Peng}, \citenamefont {Kikkawa}, \citenamefont {Kurumaji}, \citenamefont
  {Kriener}, \citenamefont {Yamasaki}, \citenamefont {Sagayama}, \citenamefont
  {Nakao}, \citenamefont {Ohishi}, \citenamefont {Kakurai}, \citenamefont
  {Taguchi}, \citenamefont {Yu}, \citenamefont {Arima},\ and\ \citenamefont
  {Tokura}}]{Hirschberger2019}%
  \BibitemOpen
  \bibfield  {author} {\bibinfo {author} {\bibfnamefont {M.}~\bibnamefont
  {Hirschberger}}, \bibinfo {author} {\bibfnamefont {T.}~\bibnamefont
  {Nakajima}}, \bibinfo {author} {\bibfnamefont {S.}~\bibnamefont {Gao}},
  \bibinfo {author} {\bibfnamefont {L.}~\bibnamefont {Peng}}, \bibinfo {author}
  {\bibfnamefont {A.}~\bibnamefont {Kikkawa}}, \bibinfo {author} {\bibfnamefont
  {T.}~\bibnamefont {Kurumaji}}, \bibinfo {author} {\bibfnamefont
  {M.}~\bibnamefont {Kriener}}, \bibinfo {author} {\bibfnamefont
  {Y.}~\bibnamefont {Yamasaki}}, \bibinfo {author} {\bibfnamefont
  {H.}~\bibnamefont {Sagayama}}, \bibinfo {author} {\bibfnamefont
  {H.}~\bibnamefont {Nakao}}, \bibinfo {author} {\bibfnamefont
  {K.}~\bibnamefont {Ohishi}}, \bibinfo {author} {\bibfnamefont
  {K.}~\bibnamefont {Kakurai}}, \bibinfo {author} {\bibfnamefont
  {Y.}~\bibnamefont {Taguchi}}, \bibinfo {author} {\bibfnamefont
  {X.}~\bibnamefont {Yu}}, \bibinfo {author} {\bibfnamefont {T.-h.}\
  \bibnamefont {Arima}}, \ and\ \bibinfo {author} {\bibfnamefont
  {Y.}~\bibnamefont {Tokura}},\ }\href {\doibase 10.1038/s41467-019-13675-4}
  {\bibfield  {journal} {\bibinfo  {journal} {Nature Communications}\ }\textbf
  {\bibinfo {volume} {10}},\ \bibinfo {pages} {5831} (\bibinfo {year}
  {2019})}\BibitemShut {NoStop}%
\bibitem [{\citenamefont {Khanh}\ \emph {et~al.}(2020)\citenamefont {Khanh},
  \citenamefont {Nakajima}, \citenamefont {Yu}, \citenamefont {Gao},
  \citenamefont {Shibata}, \citenamefont {Hirschberger}, \citenamefont
  {Yamasaki}, \citenamefont {Sagayama}, \citenamefont {Nakao}, \citenamefont
  {Peng}, \citenamefont {Nakajima}, \citenamefont {Takagi}, \citenamefont
  {Arima}, \citenamefont {Tokura},\ and\ \citenamefont {Seki}}]{Khanh2020}%
  \BibitemOpen
  \bibfield  {author} {\bibinfo {author} {\bibfnamefont {N.~D.}\ \bibnamefont
  {Khanh}}, \bibinfo {author} {\bibfnamefont {T.}~\bibnamefont {Nakajima}},
  \bibinfo {author} {\bibfnamefont {X.}~\bibnamefont {Yu}}, \bibinfo {author}
  {\bibfnamefont {S.}~\bibnamefont {Gao}}, \bibinfo {author} {\bibfnamefont
  {K.}~\bibnamefont {Shibata}}, \bibinfo {author} {\bibfnamefont
  {M.}~\bibnamefont {Hirschberger}}, \bibinfo {author} {\bibfnamefont
  {Y.}~\bibnamefont {Yamasaki}}, \bibinfo {author} {\bibfnamefont
  {H.}~\bibnamefont {Sagayama}}, \bibinfo {author} {\bibfnamefont
  {H.}~\bibnamefont {Nakao}}, \bibinfo {author} {\bibfnamefont
  {L.}~\bibnamefont {Peng}}, \bibinfo {author} {\bibfnamefont {K.}~\bibnamefont
  {Nakajima}}, \bibinfo {author} {\bibfnamefont {R.}~\bibnamefont {Takagi}},
  \bibinfo {author} {\bibfnamefont {T.-h.}\ \bibnamefont {Arima}}, \bibinfo
  {author} {\bibfnamefont {Y.}~\bibnamefont {Tokura}}, \ and\ \bibinfo {author}
  {\bibfnamefont {S.}~\bibnamefont {Seki}},\ }\href {\doibase
  10.1038/s41565-020-0684-7} {\bibfield  {journal} {\bibinfo  {journal} {Nature
  Nanotechnology}\ }\textbf {\bibinfo {volume} {15}},\ \bibinfo {pages} {444}
  (\bibinfo {year} {2020})}\BibitemShut {NoStop}%
\bibitem [{\citenamefont {Wugalter}\ \emph {et~al.}(2020)\citenamefont
  {Wugalter}, \citenamefont {Komijani},\ and\ \citenamefont
  {Coleman}}]{Coleman2020}%
  \BibitemOpen
  \bibfield  {author} {\bibinfo {author} {\bibfnamefont {A.}~\bibnamefont
  {Wugalter}}, \bibinfo {author} {\bibfnamefont {Y.}~\bibnamefont {Komijani}},
  \ and\ \bibinfo {author} {\bibfnamefont {P.}~\bibnamefont {Coleman}},\ }\href
  {\doibase 10.1103/PhysRevB.101.075133} {\bibfield  {journal} {\bibinfo
  {journal} {Phys. Rev. B}\ }\textbf {\bibinfo {volume} {101}},\ \bibinfo
  {pages} {075133} (\bibinfo {year} {2020})}\BibitemShut {NoStop}%
\bibitem [{\citenamefont {Kornja{\v{c}}a}\ \emph {et~al.}(2021)\citenamefont
  {Kornja{\v{c}}a}, \citenamefont {Quito},\ and\ \citenamefont
  {Flint}}]{Flint2021}%
  \BibitemOpen
  \bibfield  {author} {\bibinfo {author} {\bibfnamefont {M.}~\bibnamefont
  {Kornja{\v{c}}a}}, \bibinfo {author} {\bibfnamefont {V.~L.}\ \bibnamefont
  {Quito}}, \ and\ \bibinfo {author} {\bibfnamefont {R.}~\bibnamefont
  {Flint}},\ }\href@noop {} {\bibfield  {journal} {\bibinfo  {journal} {arXiv
  preprint arXiv:2104.11173}\ } (\bibinfo {year} {2021})}\BibitemShut {NoStop}%
\bibitem [{\citenamefont {Khalaf}\ \emph {et~al.}(2021)\citenamefont {Khalaf},
  \citenamefont {Chatterjee}, \citenamefont {Bultinck}, \citenamefont
  {Zaletel},\ and\ \citenamefont {Vishwanath}}]{Vishwanath2021}%
  \BibitemOpen
  \bibfield  {author} {\bibinfo {author} {\bibfnamefont {E.}~\bibnamefont
  {Khalaf}}, \bibinfo {author} {\bibfnamefont {S.}~\bibnamefont {Chatterjee}},
  \bibinfo {author} {\bibfnamefont {N.}~\bibnamefont {Bultinck}}, \bibinfo
  {author} {\bibfnamefont {M.~P.}\ \bibnamefont {Zaletel}}, \ and\ \bibinfo
  {author} {\bibfnamefont {A.}~\bibnamefont {Vishwanath}},\ }\href {\doibase
  10.1126/sciadv.abf5299} {\bibfield  {journal} {\bibinfo  {journal} {Science
  Advances}\ }\textbf {\bibinfo {volume} {7}} (\bibinfo {year} {2021}),\
  10.1126/sciadv.abf5299}\BibitemShut {NoStop}%
\bibitem [{\citenamefont {Nomura}\ and\ \citenamefont
  {Nagaosa}(2010)}]{nomura2010}%
  \BibitemOpen
  \bibfield  {author} {\bibinfo {author} {\bibfnamefont {K.}~\bibnamefont
  {Nomura}}\ and\ \bibinfo {author} {\bibfnamefont {N.}~\bibnamefont
  {Nagaosa}},\ }\href {\doibase 10.1103/PhysRevB.82.161401} {\bibfield
  {journal} {\bibinfo  {journal} {Phys. Rev. B}\ }\textbf {\bibinfo {volume}
  {82}},\ \bibinfo {pages} {161401} (\bibinfo {year} {2010})}\BibitemShut
  {NoStop}%
\bibitem [{\citenamefont {Hurst}\ \emph {et~al.}(2015)\citenamefont {Hurst},
  \citenamefont {Efimkin}, \citenamefont {Zang},\ and\ \citenamefont
  {Galitski}}]{Hurst2015}%
  \BibitemOpen
  \bibfield  {author} {\bibinfo {author} {\bibfnamefont {H.~M.}\ \bibnamefont
  {Hurst}}, \bibinfo {author} {\bibfnamefont {D.~K.}\ \bibnamefont {Efimkin}},
  \bibinfo {author} {\bibfnamefont {J.}~\bibnamefont {Zang}}, \ and\ \bibinfo
  {author} {\bibfnamefont {V.}~\bibnamefont {Galitski}},\ }\href {\doibase
  10.1103/PhysRevB.91.060401} {\bibfield  {journal} {\bibinfo  {journal} {Phys.
  Rev. B}\ }\textbf {\bibinfo {volume} {91}},\ \bibinfo {pages} {060401}
  (\bibinfo {year} {2015})}\BibitemShut {NoStop}%
\bibitem [{\citenamefont {Lado}\ and\ \citenamefont
  {Fern\'andez-Rossier}(2015)}]{Lado2015}%
  \BibitemOpen
  \bibfield  {author} {\bibinfo {author} {\bibfnamefont {J.~L.}\ \bibnamefont
  {Lado}}\ and\ \bibinfo {author} {\bibfnamefont {J.}~\bibnamefont
  {Fern\'andez-Rossier}},\ }\href {\doibase 10.1103/PhysRevB.92.115433}
  {\bibfield  {journal} {\bibinfo  {journal} {Phys. Rev. B}\ }\textbf {\bibinfo
  {volume} {92}},\ \bibinfo {pages} {115433} (\bibinfo {year}
  {2015})}\BibitemShut {NoStop}%
\bibitem [{\citenamefont {Araki}\ and\ \citenamefont
  {Nomura}(2017)}]{Nomura2017}%
  \BibitemOpen
  \bibfield  {author} {\bibinfo {author} {\bibfnamefont {Y.}~\bibnamefont
  {Araki}}\ and\ \bibinfo {author} {\bibfnamefont {K.}~\bibnamefont {Nomura}},\
  }\href {\doibase 10.1103/PhysRevB.96.165303} {\bibfield  {journal} {\bibinfo
  {journal} {Phys. Rev. B}\ }\textbf {\bibinfo {volume} {96}},\ \bibinfo
  {pages} {165303} (\bibinfo {year} {2017})}\BibitemShut {NoStop}%
\bibitem [{\citenamefont {Nogueira}\ \emph {et~al.}(2018)\citenamefont
  {Nogueira}, \citenamefont {Eremin}, \citenamefont {Katmis}, \citenamefont
  {Moodera}, \citenamefont {van~den Brink},\ and\ \citenamefont
  {Kravchuk}}]{Nogueira2018}%
  \BibitemOpen
  \bibfield  {author} {\bibinfo {author} {\bibfnamefont {F.~S.}\ \bibnamefont
  {Nogueira}}, \bibinfo {author} {\bibfnamefont {I.}~\bibnamefont {Eremin}},
  \bibinfo {author} {\bibfnamefont {F.}~\bibnamefont {Katmis}}, \bibinfo
  {author} {\bibfnamefont {J.~S.}\ \bibnamefont {Moodera}}, \bibinfo {author}
  {\bibfnamefont {J.}~\bibnamefont {van~den Brink}}, \ and\ \bibinfo {author}
  {\bibfnamefont {V.~P.}\ \bibnamefont {Kravchuk}},\ }\href {\doibase
  10.1103/PhysRevB.98.060401} {\bibfield  {journal} {\bibinfo  {journal} {Phys.
  Rev. B}\ }\textbf {\bibinfo {volume} {98}},\ \bibinfo {pages} {060401}
  (\bibinfo {year} {2018})}\BibitemShut {NoStop}%
\bibitem [{\citenamefont {Tiwari}\ \emph {et~al.}(2019)\citenamefont {Tiwari},
  \citenamefont {Lavoie}, \citenamefont {Pereg-Barnea},\ and\ \citenamefont
  {Coish}}]{Tiwari2019}%
  \BibitemOpen
  \bibfield  {author} {\bibinfo {author} {\bibfnamefont {K.~L.}\ \bibnamefont
  {Tiwari}}, \bibinfo {author} {\bibfnamefont {J.}~\bibnamefont {Lavoie}},
  \bibinfo {author} {\bibfnamefont {T.}~\bibnamefont {Pereg-Barnea}}, \ and\
  \bibinfo {author} {\bibfnamefont {W.~A.}\ \bibnamefont {Coish}},\ }\href
  {\doibase 10.1103/PhysRevB.100.125414} {\bibfield  {journal} {\bibinfo
  {journal} {Phys. Rev. B}\ }\textbf {\bibinfo {volume} {100}},\ \bibinfo
  {pages} {125414} (\bibinfo {year} {2019})}\BibitemShut {NoStop}%
\bibitem [{\citenamefont {Wang}\ \emph
  {et~al.}(2020{\natexlab{a}})\citenamefont {Wang}, \citenamefont {Xu},\ and\
  \citenamefont {Lai}}]{wang2020scattering}%
  \BibitemOpen
  \bibfield  {author} {\bibinfo {author} {\bibfnamefont {C.-Z.}\ \bibnamefont
  {Wang}}, \bibinfo {author} {\bibfnamefont {H.-Y.}\ \bibnamefont {Xu}}, \ and\
  \bibinfo {author} {\bibfnamefont {Y.-C.}\ \bibnamefont {Lai}},\ }\href@noop
  {} {\enquote {\bibinfo {title} {Scattering of dirac electrons from a
  skyrmion: emergence of robust skew scattering},}\ } (\bibinfo {year}
  {2020}{\natexlab{a}}),\ \Eprint {http://arxiv.org/abs/2002.02944}
  {arXiv:2002.02944 [cond-mat.mes-hall]} \BibitemShut {NoStop}%
\bibitem [{\citenamefont {{Paul}}\ and\ \citenamefont {{Fu}}(2021)}]{Paul2021}%
  \BibitemOpen
  \bibfield  {author} {\bibinfo {author} {\bibfnamefont {N.}~\bibnamefont
  {{Paul}}}\ and\ \bibinfo {author} {\bibfnamefont {L.}~\bibnamefont {{Fu}}},\
  }\href@noop {} {\bibfield  {journal} {\bibinfo  {journal} {arXiv e-prints}\
  ,\ \bibinfo {eid} {arXiv:2103.02617}} (\bibinfo {year} {2021})},\ \Eprint
  {http://arxiv.org/abs/2103.02617} {arXiv:2103.02617 [cond-mat.mes-hall]}
  \BibitemShut {NoStop}%
\bibitem [{\citenamefont {Divic}\ \emph {et~al.}(2021)\citenamefont {Divic},
  \citenamefont {Ling}, \citenamefont {Pereg-Barnea},\ and\ \citenamefont
  {Paramekanti}}]{divic2021magnetic}%
  \BibitemOpen
  \bibfield  {author} {\bibinfo {author} {\bibfnamefont {S.}~\bibnamefont
  {Divic}}, \bibinfo {author} {\bibfnamefont {H.}~\bibnamefont {Ling}},
  \bibinfo {author} {\bibfnamefont {T.}~\bibnamefont {Pereg-Barnea}}, \ and\
  \bibinfo {author} {\bibfnamefont {A.}~\bibnamefont {Paramekanti}},\
  }\href@noop {} {\enquote {\bibinfo {title} {Magnetic skyrmion crystal at a
  topological insulator surface},}\ } (\bibinfo {year} {2021}),\ \Eprint
  {http://arxiv.org/abs/2103.15841} {arXiv:2103.15841 [cond-mat.mes-hall]}
  \BibitemShut {NoStop}%
\bibitem [{\citenamefont {Li}\ \emph {et~al.}(2021)\citenamefont {Li},
  \citenamefont {Yao},\ and\ \citenamefont {Chen}}]{li2021twisted}%
  \BibitemOpen
  \bibfield  {author} {\bibinfo {author} {\bibfnamefont {C.-K.}\ \bibnamefont
  {Li}}, \bibinfo {author} {\bibfnamefont {X.-P.}\ \bibnamefont {Yao}}, \ and\
  \bibinfo {author} {\bibfnamefont {G.}~\bibnamefont {Chen}},\ }\href@noop {}
  {\enquote {\bibinfo {title} {Twisted magnetic topological insulators},}\ }
  (\bibinfo {year} {2021}),\ \Eprint {http://arxiv.org/abs/2104.13235}
  {arXiv:2104.13235 [cond-mat.str-el]} \BibitemShut {NoStop}%
\bibitem [{\citenamefont {Soumyanarayanan}\ \emph {et~al.}(2017)\citenamefont
  {Soumyanarayanan}, \citenamefont {Raju}, \citenamefont {Gonzalez~Oyarce},
  \citenamefont {Tan}, \citenamefont {Im}, \citenamefont {Petrovi{\'{c}}},
  \citenamefont {Ho}, \citenamefont {Khoo}, \citenamefont {Tran}, \citenamefont
  {Gan}, \citenamefont {Ernult},\ and\ \citenamefont
  {Panagopoulos}}]{Soumyanarayanan2017}%
  \BibitemOpen
  \bibfield  {author} {\bibinfo {author} {\bibfnamefont {A.}~\bibnamefont
  {Soumyanarayanan}}, \bibinfo {author} {\bibfnamefont {M.}~\bibnamefont
  {Raju}}, \bibinfo {author} {\bibfnamefont {A.~L.}\ \bibnamefont
  {Gonzalez~Oyarce}}, \bibinfo {author} {\bibfnamefont {A.~K.~C.}\ \bibnamefont
  {Tan}}, \bibinfo {author} {\bibfnamefont {M.-Y.}\ \bibnamefont {Im}},
  \bibinfo {author} {\bibfnamefont {A.~P.}\ \bibnamefont {Petrovi{\'{c}}}},
  \bibinfo {author} {\bibfnamefont {P.}~\bibnamefont {Ho}}, \bibinfo {author}
  {\bibfnamefont {K.~H.}\ \bibnamefont {Khoo}}, \bibinfo {author}
  {\bibfnamefont {M.}~\bibnamefont {Tran}}, \bibinfo {author} {\bibfnamefont
  {C.~K.}\ \bibnamefont {Gan}}, \bibinfo {author} {\bibfnamefont
  {F.}~\bibnamefont {Ernult}}, \ and\ \bibinfo {author} {\bibfnamefont
  {C.}~\bibnamefont {Panagopoulos}},\ }\href {\doibase 10.1038/nmat4934}
  {\bibfield  {journal} {\bibinfo  {journal} {Nature Materials}\ }\textbf
  {\bibinfo {volume} {16}},\ \bibinfo {pages} {898} (\bibinfo {year}
  {2017})}\BibitemShut {NoStop}%
\bibitem [{\citenamefont {Maccariello}\ \emph {et~al.}(2018)\citenamefont
  {Maccariello}, \citenamefont {Legrand}, \citenamefont {Reyren}, \citenamefont
  {Garcia}, \citenamefont {Bouzehouane}, \citenamefont {Collin}, \citenamefont
  {Cros},\ and\ \citenamefont {Fert}}]{Maccariello2018}%
  \BibitemOpen
  \bibfield  {author} {\bibinfo {author} {\bibfnamefont {D.}~\bibnamefont
  {Maccariello}}, \bibinfo {author} {\bibfnamefont {W.}~\bibnamefont
  {Legrand}}, \bibinfo {author} {\bibfnamefont {N.}~\bibnamefont {Reyren}},
  \bibinfo {author} {\bibfnamefont {K.}~\bibnamefont {Garcia}}, \bibinfo
  {author} {\bibfnamefont {K.}~\bibnamefont {Bouzehouane}}, \bibinfo {author}
  {\bibfnamefont {S.}~\bibnamefont {Collin}}, \bibinfo {author} {\bibfnamefont
  {V.}~\bibnamefont {Cros}}, \ and\ \bibinfo {author} {\bibfnamefont
  {A.}~\bibnamefont {Fert}},\ }\href {\doibase 10.1038/s41565-017-0044-4}
  {\bibfield  {journal} {\bibinfo  {journal} {Nature Nanotechnology}\ }\textbf
  {\bibinfo {volume} {13}},\ \bibinfo {pages} {233} (\bibinfo {year}
  {2018})}\BibitemShut {NoStop}%
\bibitem [{\citenamefont {Yu}\ \emph {et~al.}(2011)\citenamefont {Yu},
  \citenamefont {Kanazawa}, \citenamefont {Onose}, \citenamefont {Kimoto},
  \citenamefont {Zhang}, \citenamefont {Ishiwata}, \citenamefont {Matsui},\
  and\ \citenamefont {Tokura}}]{Yu2011}%
  \BibitemOpen
  \bibfield  {author} {\bibinfo {author} {\bibfnamefont {X.~Z.}\ \bibnamefont
  {Yu}}, \bibinfo {author} {\bibfnamefont {N.}~\bibnamefont {Kanazawa}},
  \bibinfo {author} {\bibfnamefont {Y.}~\bibnamefont {Onose}}, \bibinfo
  {author} {\bibfnamefont {K.}~\bibnamefont {Kimoto}}, \bibinfo {author}
  {\bibfnamefont {W.~Z.}\ \bibnamefont {Zhang}}, \bibinfo {author}
  {\bibfnamefont {S.}~\bibnamefont {Ishiwata}}, \bibinfo {author}
  {\bibfnamefont {Y.}~\bibnamefont {Matsui}}, \ and\ \bibinfo {author}
  {\bibfnamefont {Y.}~\bibnamefont {Tokura}},\ }\href {\doibase
  10.1038/nmat2916} {\bibfield  {journal} {\bibinfo  {journal} {Nature
  Materials}\ }\textbf {\bibinfo {volume} {10}},\ \bibinfo {pages} {106}
  (\bibinfo {year} {2011})}\BibitemShut {NoStop}%
\bibitem [{\citenamefont {Ye}\ \emph {et~al.}(1999)\citenamefont {Ye},
  \citenamefont {Kim}, \citenamefont {Millis}, \citenamefont {Shraiman},
  \citenamefont {Majumdar},\ and\ \citenamefont {Te\ifmmode \check{s}\else
  \v{s}\fi{}anovi\ifmmode~\acute{c}\else \'{c}\fi{}}}]{He1999}%
  \BibitemOpen
  \bibfield  {author} {\bibinfo {author} {\bibfnamefont {J.}~\bibnamefont
  {Ye}}, \bibinfo {author} {\bibfnamefont {Y.~B.}\ \bibnamefont {Kim}},
  \bibinfo {author} {\bibfnamefont {A.~J.}\ \bibnamefont {Millis}}, \bibinfo
  {author} {\bibfnamefont {B.~I.}\ \bibnamefont {Shraiman}}, \bibinfo {author}
  {\bibfnamefont {P.}~\bibnamefont {Majumdar}}, \ and\ \bibinfo {author}
  {\bibfnamefont {Z.}~\bibnamefont {Te\ifmmode \check{s}\else
  \v{s}\fi{}anovi\ifmmode~\acute{c}\else \'{c}\fi{}}},\ }\href {\doibase
  10.1103/PhysRevLett.83.3737} {\bibfield  {journal} {\bibinfo  {journal}
  {Phys. Rev. Lett.}\ }\textbf {\bibinfo {volume} {83}},\ \bibinfo {pages}
  {3737} (\bibinfo {year} {1999})}\BibitemShut {NoStop}%
\bibitem [{\citenamefont {Schulz}\ \emph {et~al.}(2012)\citenamefont {Schulz},
  \citenamefont {Ritz}, \citenamefont {Bauer}, \citenamefont {Halder},
  \citenamefont {Wagner}, \citenamefont {Franz}, \citenamefont {Pfleiderer},
  \citenamefont {Everschor}, \citenamefont {Garst},\ and\ \citenamefont
  {Rosch}}]{Schulz2012}%
  \BibitemOpen
  \bibfield  {author} {\bibinfo {author} {\bibfnamefont {T.}~\bibnamefont
  {Schulz}}, \bibinfo {author} {\bibfnamefont {R.}~\bibnamefont {Ritz}},
  \bibinfo {author} {\bibfnamefont {A.}~\bibnamefont {Bauer}}, \bibinfo
  {author} {\bibfnamefont {M.}~\bibnamefont {Halder}}, \bibinfo {author}
  {\bibfnamefont {M.}~\bibnamefont {Wagner}}, \bibinfo {author} {\bibfnamefont
  {C.}~\bibnamefont {Franz}}, \bibinfo {author} {\bibfnamefont
  {C.}~\bibnamefont {Pfleiderer}}, \bibinfo {author} {\bibfnamefont
  {K.}~\bibnamefont {Everschor}}, \bibinfo {author} {\bibfnamefont
  {M.}~\bibnamefont {Garst}}, \ and\ \bibinfo {author} {\bibfnamefont
  {A.}~\bibnamefont {Rosch}},\ }\href {\doibase 10.1038/nphys2231} {\bibfield
  {journal} {\bibinfo  {journal} {Nature Physics}\ }\textbf {\bibinfo {volume}
  {8}},\ \bibinfo {pages} {301} (\bibinfo {year} {2012})}\BibitemShut {NoStop}%
\bibitem [{\citenamefont {Shiomi}\ \emph {et~al.}(2013)\citenamefont {Shiomi},
  \citenamefont {Kanazawa}, \citenamefont {Shibata}, \citenamefont {Onose},\
  and\ \citenamefont {Tokura}}]{Tokura2013}%
  \BibitemOpen
  \bibfield  {author} {\bibinfo {author} {\bibfnamefont {Y.}~\bibnamefont
  {Shiomi}}, \bibinfo {author} {\bibfnamefont {N.}~\bibnamefont {Kanazawa}},
  \bibinfo {author} {\bibfnamefont {K.}~\bibnamefont {Shibata}}, \bibinfo
  {author} {\bibfnamefont {Y.}~\bibnamefont {Onose}}, \ and\ \bibinfo {author}
  {\bibfnamefont {Y.}~\bibnamefont {Tokura}},\ }\href {\doibase
  10.1103/PhysRevB.88.064409} {\bibfield  {journal} {\bibinfo  {journal} {Phys.
  Rev. B}\ }\textbf {\bibinfo {volume} {88}},\ \bibinfo {pages} {064409}
  (\bibinfo {year} {2013})}\BibitemShut {NoStop}%
\bibitem [{\citenamefont {Hirschberger}\ \emph {et~al.}(2020)\citenamefont
  {Hirschberger}, \citenamefont {Spitz}, \citenamefont {Nomoto}, \citenamefont
  {Kurumaji}, \citenamefont {Gao}, \citenamefont {Masell}, \citenamefont
  {Nakajima}, \citenamefont {Kikkawa}, \citenamefont {Yamasaki}, \citenamefont
  {Sagayama}, \citenamefont {Nakao}, \citenamefont {Taguchi}, \citenamefont
  {Arita}, \citenamefont {Arima},\ and\ \citenamefont {Tokura}}]{Tokura2020}%
  \BibitemOpen
  \bibfield  {author} {\bibinfo {author} {\bibfnamefont {M.}~\bibnamefont
  {Hirschberger}}, \bibinfo {author} {\bibfnamefont {L.}~\bibnamefont {Spitz}},
  \bibinfo {author} {\bibfnamefont {T.}~\bibnamefont {Nomoto}}, \bibinfo
  {author} {\bibfnamefont {T.}~\bibnamefont {Kurumaji}}, \bibinfo {author}
  {\bibfnamefont {S.}~\bibnamefont {Gao}}, \bibinfo {author} {\bibfnamefont
  {J.}~\bibnamefont {Masell}}, \bibinfo {author} {\bibfnamefont
  {T.}~\bibnamefont {Nakajima}}, \bibinfo {author} {\bibfnamefont
  {A.}~\bibnamefont {Kikkawa}}, \bibinfo {author} {\bibfnamefont
  {Y.}~\bibnamefont {Yamasaki}}, \bibinfo {author} {\bibfnamefont
  {H.}~\bibnamefont {Sagayama}}, \bibinfo {author} {\bibfnamefont
  {H.}~\bibnamefont {Nakao}}, \bibinfo {author} {\bibfnamefont
  {Y.}~\bibnamefont {Taguchi}}, \bibinfo {author} {\bibfnamefont
  {R.}~\bibnamefont {Arita}}, \bibinfo {author} {\bibfnamefont {T.-h.}\
  \bibnamefont {Arima}}, \ and\ \bibinfo {author} {\bibfnamefont
  {Y.}~\bibnamefont {Tokura}},\ }\href {\doibase
  10.1103/PhysRevLett.125.076602} {\bibfield  {journal} {\bibinfo  {journal}
  {Phys. Rev. Lett.}\ }\textbf {\bibinfo {volume} {125}},\ \bibinfo {pages}
  {076602} (\bibinfo {year} {2020})}\BibitemShut {NoStop}%
\bibitem [{\citenamefont {Kim}\ \emph {et~al.}(2019)\citenamefont {Kim},
  \citenamefont {Nakata}, \citenamefont {Loss},\ and\ \citenamefont
  {Tserkovnyak}}]{Kim2019}%
  \BibitemOpen
  \bibfield  {author} {\bibinfo {author} {\bibfnamefont {S.~K.}\ \bibnamefont
  {Kim}}, \bibinfo {author} {\bibfnamefont {K.}~\bibnamefont {Nakata}},
  \bibinfo {author} {\bibfnamefont {D.}~\bibnamefont {Loss}}, \ and\ \bibinfo
  {author} {\bibfnamefont {Y.}~\bibnamefont {Tserkovnyak}},\ }\href {\doibase
  10.1103/PhysRevLett.122.057204} {\bibfield  {journal} {\bibinfo  {journal}
  {Phys. Rev. Lett.}\ }\textbf {\bibinfo {volume} {122}},\ \bibinfo {pages}
  {057204} (\bibinfo {year} {2019})}\BibitemShut {NoStop}%
\bibitem [{\citenamefont {Kan}\ \emph {et~al.}(2018)\citenamefont {Kan},
  \citenamefont {Moriyama}, \citenamefont {Kobayashi},\ and\ \citenamefont
  {Shimakawa}}]{Kan2018}%
  \BibitemOpen
  \bibfield  {author} {\bibinfo {author} {\bibfnamefont {D.}~\bibnamefont
  {Kan}}, \bibinfo {author} {\bibfnamefont {T.}~\bibnamefont {Moriyama}},
  \bibinfo {author} {\bibfnamefont {K.}~\bibnamefont {Kobayashi}}, \ and\
  \bibinfo {author} {\bibfnamefont {Y.}~\bibnamefont {Shimakawa}},\ }\href
  {\doibase 10.1103/PhysRevB.98.180408} {\bibfield  {journal} {\bibinfo
  {journal} {Phys. Rev. B}\ }\textbf {\bibinfo {volume} {98}},\ \bibinfo
  {pages} {180408} (\bibinfo {year} {2018})}\BibitemShut {NoStop}%
\bibitem [{\citenamefont {Gerber}(2018)}]{Gerber2018}%
  \BibitemOpen
  \bibfield  {author} {\bibinfo {author} {\bibfnamefont {A.}~\bibnamefont
  {Gerber}},\ }\href {\doibase 10.1103/PhysRevB.98.214440} {\bibfield
  {journal} {\bibinfo  {journal} {Phys. Rev. B}\ }\textbf {\bibinfo {volume}
  {98}},\ \bibinfo {pages} {214440} (\bibinfo {year} {2018})}\BibitemShut
  {NoStop}%
\bibitem [{\citenamefont {Wang}\ \emph
  {et~al.}(2020{\natexlab{b}})\citenamefont {Wang}, \citenamefont {Feng},
  \citenamefont {Lee}, \citenamefont {Ko}, \citenamefont {Lu},\ and\
  \citenamefont {Noh}}]{Wang2020}%
  \BibitemOpen
  \bibfield  {author} {\bibinfo {author} {\bibfnamefont {L.}~\bibnamefont
  {Wang}}, \bibinfo {author} {\bibfnamefont {Q.}~\bibnamefont {Feng}}, \bibinfo
  {author} {\bibfnamefont {H.~G.}\ \bibnamefont {Lee}}, \bibinfo {author}
  {\bibfnamefont {E.~K.}\ \bibnamefont {Ko}}, \bibinfo {author} {\bibfnamefont
  {Q.}~\bibnamefont {Lu}}, \ and\ \bibinfo {author} {\bibfnamefont {T.~W.}\
  \bibnamefont {Noh}},\ }\href {\doibase 10.1021/acs.nanolett.9b05206}
  {\bibfield  {journal} {\bibinfo  {journal} {Nano Letters}\ }\textbf {\bibinfo
  {volume} {20}},\ \bibinfo {pages} {2468} (\bibinfo {year}
  {2020}{\natexlab{b}})}\BibitemShut {NoStop}%
\bibitem [{\citenamefont {Kim}\ \emph {et~al.}(2020)\citenamefont {Kim},
  \citenamefont {Son}, \citenamefont {Suyolcu}, \citenamefont {Miao},
  \citenamefont {Schreiber}, \citenamefont {Nair}, \citenamefont {Putzky},
  \citenamefont {Minola}, \citenamefont {Christiani}, \citenamefont {van Aken},
  \citenamefont {Shen}, \citenamefont {Schlom}, \citenamefont {Logvenov},\ and\
  \citenamefont {Keimer}}]{Keimer2020}%
  \BibitemOpen
  \bibfield  {author} {\bibinfo {author} {\bibfnamefont {G.}~\bibnamefont
  {Kim}}, \bibinfo {author} {\bibfnamefont {K.}~\bibnamefont {Son}}, \bibinfo
  {author} {\bibfnamefont {Y.~E.}\ \bibnamefont {Suyolcu}}, \bibinfo {author}
  {\bibfnamefont {L.}~\bibnamefont {Miao}}, \bibinfo {author} {\bibfnamefont
  {N.~J.}\ \bibnamefont {Schreiber}}, \bibinfo {author} {\bibfnamefont {H.~P.}\
  \bibnamefont {Nair}}, \bibinfo {author} {\bibfnamefont {D.}~\bibnamefont
  {Putzky}}, \bibinfo {author} {\bibfnamefont {M.}~\bibnamefont {Minola}},
  \bibinfo {author} {\bibfnamefont {G.}~\bibnamefont {Christiani}}, \bibinfo
  {author} {\bibfnamefont {P.~A.}\ \bibnamefont {van Aken}}, \bibinfo {author}
  {\bibfnamefont {K.~M.}\ \bibnamefont {Shen}}, \bibinfo {author}
  {\bibfnamefont {D.~G.}\ \bibnamefont {Schlom}}, \bibinfo {author}
  {\bibfnamefont {G.}~\bibnamefont {Logvenov}}, \ and\ \bibinfo {author}
  {\bibfnamefont {B.}~\bibnamefont {Keimer}},\ }\href {\doibase
  10.1103/PhysRevMaterials.4.104410} {\bibfield  {journal} {\bibinfo  {journal}
  {Phys. Rev. Materials}\ }\textbf {\bibinfo {volume} {4}},\ \bibinfo {pages}
  {104410} (\bibinfo {year} {2020})}\BibitemShut {NoStop}%
\bibitem [{\citenamefont {Wu}\ \emph {et~al.}(2020)\citenamefont {Wu},
  \citenamefont {Wen}, \citenamefont {Fu}, \citenamefont {Wilson},
  \citenamefont {Liu}, \citenamefont {Zhang}, \citenamefont {Vasiukov},
  \citenamefont {Kareev}, \citenamefont {Pixley},\ and\ \citenamefont
  {Chakhalian}}]{Wu2020}%
  \BibitemOpen
  \bibfield  {author} {\bibinfo {author} {\bibfnamefont {L.}~\bibnamefont
  {Wu}}, \bibinfo {author} {\bibfnamefont {F.}~\bibnamefont {Wen}}, \bibinfo
  {author} {\bibfnamefont {Y.}~\bibnamefont {Fu}}, \bibinfo {author}
  {\bibfnamefont {J.~H.}\ \bibnamefont {Wilson}}, \bibinfo {author}
  {\bibfnamefont {X.}~\bibnamefont {Liu}}, \bibinfo {author} {\bibfnamefont
  {Y.}~\bibnamefont {Zhang}}, \bibinfo {author} {\bibfnamefont {D.~M.}\
  \bibnamefont {Vasiukov}}, \bibinfo {author} {\bibfnamefont {M.~S.}\
  \bibnamefont {Kareev}}, \bibinfo {author} {\bibfnamefont {J.~H.}\
  \bibnamefont {Pixley}}, \ and\ \bibinfo {author} {\bibfnamefont
  {J.}~\bibnamefont {Chakhalian}},\ }\href {\doibase
  10.1103/PhysRevB.102.220406} {\bibfield  {journal} {\bibinfo  {journal}
  {Phys. Rev. B}\ }\textbf {\bibinfo {volume} {102}},\ \bibinfo {pages}
  {220406} (\bibinfo {year} {2020})}\BibitemShut {NoStop}%
\bibitem [{\citenamefont {Bartram}\ \emph {et~al.}(2020)\citenamefont
  {Bartram}, \citenamefont {Sorn}, \citenamefont {Li}, \citenamefont {Hwangbo},
  \citenamefont {Shen}, \citenamefont {Frontini}, \citenamefont {He},
  \citenamefont {Yu}, \citenamefont {Paramekanti},\ and\ \citenamefont
  {Yang}}]{Bartram2020}%
  \BibitemOpen
  \bibfield  {author} {\bibinfo {author} {\bibfnamefont {F.~M.}\ \bibnamefont
  {Bartram}}, \bibinfo {author} {\bibfnamefont {S.}~\bibnamefont {Sorn}},
  \bibinfo {author} {\bibfnamefont {Z.}~\bibnamefont {Li}}, \bibinfo {author}
  {\bibfnamefont {K.}~\bibnamefont {Hwangbo}}, \bibinfo {author} {\bibfnamefont
  {S.}~\bibnamefont {Shen}}, \bibinfo {author} {\bibfnamefont {F.}~\bibnamefont
  {Frontini}}, \bibinfo {author} {\bibfnamefont {L.}~\bibnamefont {He}},
  \bibinfo {author} {\bibfnamefont {P.}~\bibnamefont {Yu}}, \bibinfo {author}
  {\bibfnamefont {A.}~\bibnamefont {Paramekanti}}, \ and\ \bibinfo {author}
  {\bibfnamefont {L.}~\bibnamefont {Yang}},\ }\href {\doibase
  10.1103/PhysRevB.102.140408} {\bibfield  {journal} {\bibinfo  {journal}
  {Phys. Rev. B}\ }\textbf {\bibinfo {volume} {102}},\ \bibinfo {pages}
  {140408} (\bibinfo {year} {2020})}\BibitemShut {NoStop}%
\bibitem [{\citenamefont {Sorn}\ and\ \citenamefont
  {Paramekanti}(2021)}]{Sorn2021}%
  \BibitemOpen
  \bibfield  {author} {\bibinfo {author} {\bibfnamefont {S.}~\bibnamefont
  {Sorn}}\ and\ \bibinfo {author} {\bibfnamefont {A.}~\bibnamefont
  {Paramekanti}},\ }\href {\doibase 10.1103/PhysRevB.103.104413} {\bibfield
  {journal} {\bibinfo  {journal} {Phys. Rev. B}\ }\textbf {\bibinfo {volume}
  {103}},\ \bibinfo {pages} {104413} (\bibinfo {year} {2021})}\BibitemShut
  {NoStop}%
\bibitem [{\citenamefont {Pekar}\ and\ \citenamefont
  {Rashba}(1964)}]{Rashba1964}%
  \BibitemOpen
  \bibfield  {author} {\bibinfo {author} {\bibfnamefont {S.}~\bibnamefont
  {Pekar}}\ and\ \bibinfo {author} {\bibfnamefont {E.}~\bibnamefont {Rashba}},\
  }\href@noop {} {\bibfield  {journal} {\bibinfo  {journal} {Zh. Eksperim. i
  Teor. Fiz.}\ }\textbf {\bibinfo {volume} {47}} (\bibinfo {year}
  {1964})}\BibitemShut {NoStop}%
\bibitem [{\citenamefont {Yuan}\ \emph {et~al.}(2020)\citenamefont {Yuan},
  \citenamefont {Wang}, \citenamefont {Luo}, \citenamefont {Rashba},\ and\
  \citenamefont {Zunger}}]{Rashba2020}%
  \BibitemOpen
  \bibfield  {author} {\bibinfo {author} {\bibfnamefont {L.-D.}\ \bibnamefont
  {Yuan}}, \bibinfo {author} {\bibfnamefont {Z.}~\bibnamefont {Wang}}, \bibinfo
  {author} {\bibfnamefont {J.-W.}\ \bibnamefont {Luo}}, \bibinfo {author}
  {\bibfnamefont {E.~I.}\ \bibnamefont {Rashba}}, \ and\ \bibinfo {author}
  {\bibfnamefont {A.}~\bibnamefont {Zunger}},\ }\href {\doibase
  10.1103/PhysRevB.102.014422} {\bibfield  {journal} {\bibinfo  {journal}
  {Phys. Rev. B}\ }\textbf {\bibinfo {volume} {102}},\ \bibinfo {pages}
  {014422} (\bibinfo {year} {2020})}\BibitemShut {NoStop}%
\bibitem [{\citenamefont {Egorov}\ and\ \citenamefont
  {Evarestov}(2021)}]{Egorov2021}%
  \BibitemOpen
  \bibfield  {author} {\bibinfo {author} {\bibfnamefont {S.~A.}\ \bibnamefont
  {Egorov}}\ and\ \bibinfo {author} {\bibfnamefont {R.~A.}\ \bibnamefont
  {Evarestov}},\ }\href {\doibase 10.1021/acs.jpclett.1c00282} {\bibfield
  {journal} {\bibinfo  {journal} {The Journal of Physical Chemistry Letters}\
  }\textbf {\bibinfo {volume} {12}},\ \bibinfo {pages} {2363} (\bibinfo {year}
  {2021})}\BibitemShut {NoStop}%
\bibitem [{\citenamefont {Wilson}\ \emph {et~al.}(2014)\citenamefont {Wilson},
  \citenamefont {Efimkin},\ and\ \citenamefont {Galitski}}]{Wilson2014}%
  \BibitemOpen
  \bibfield  {author} {\bibinfo {author} {\bibfnamefont {J.~H.}\ \bibnamefont
  {Wilson}}, \bibinfo {author} {\bibfnamefont {D.~K.}\ \bibnamefont {Efimkin}},
  \ and\ \bibinfo {author} {\bibfnamefont {V.~M.}\ \bibnamefont {Galitski}},\
  }\href {\doibase 10.1103/PhysRevB.90.205432} {\bibfield  {journal} {\bibinfo
  {journal} {Phys. Rev. B}\ }\textbf {\bibinfo {volume} {90}},\ \bibinfo
  {pages} {205432} (\bibinfo {year} {2014})}\BibitemShut {NoStop}%
\bibitem [{\citenamefont {Sato}\ \emph {et~al.}(2021)\citenamefont {Sato},
  \citenamefont {Umimoto}, \citenamefont {Sugita}, \citenamefont {Kato},\ and\
  \citenamefont {Motome}}]{Motome2021}%
  \BibitemOpen
  \bibfield  {author} {\bibinfo {author} {\bibfnamefont {T.}~\bibnamefont
  {Sato}}, \bibinfo {author} {\bibfnamefont {Y.}~\bibnamefont {Umimoto}},
  \bibinfo {author} {\bibfnamefont {Y.}~\bibnamefont {Sugita}}, \bibinfo
  {author} {\bibfnamefont {Y.}~\bibnamefont {Kato}}, \ and\ \bibinfo {author}
  {\bibfnamefont {Y.}~\bibnamefont {Motome}},\ }\href {\doibase
  10.1103/PhysRevB.103.054416} {\bibfield  {journal} {\bibinfo  {journal}
  {Phys. Rev. B}\ }\textbf {\bibinfo {volume} {103}},\ \bibinfo {pages}
  {054416} (\bibinfo {year} {2021})}\BibitemShut {NoStop}%
\bibitem [{\citenamefont {Anderson}\ and\ \citenamefont
  {Hasegawa}(1955)}]{Anderson1955}%
  \BibitemOpen
  \bibfield  {author} {\bibinfo {author} {\bibfnamefont {P.~W.}\ \bibnamefont
  {Anderson}}\ and\ \bibinfo {author} {\bibfnamefont {H.}~\bibnamefont
  {Hasegawa}},\ }\href {\doibase 10.1103/PhysRev.100.675} {\bibfield  {journal}
  {\bibinfo  {journal} {Phys. Rev.}\ }\textbf {\bibinfo {volume} {100}},\
  \bibinfo {pages} {675} (\bibinfo {year} {1955})}\BibitemShut {NoStop}%
\bibitem [{\citenamefont {Ohgushi}\ \emph {et~al.}(2000)\citenamefont
  {Ohgushi}, \citenamefont {Murakami},\ and\ \citenamefont
  {Nagaosa}}]{Nagaosa2000}%
  \BibitemOpen
  \bibfield  {author} {\bibinfo {author} {\bibfnamefont {K.}~\bibnamefont
  {Ohgushi}}, \bibinfo {author} {\bibfnamefont {S.}~\bibnamefont {Murakami}}, \
  and\ \bibinfo {author} {\bibfnamefont {N.}~\bibnamefont {Nagaosa}},\ }\href
  {\doibase 10.1103/PhysRevB.62.R6065} {\bibfield  {journal} {\bibinfo
  {journal} {Phys. Rev. B}\ }\textbf {\bibinfo {volume} {62}},\ \bibinfo
  {pages} {R6065} (\bibinfo {year} {2000})}\BibitemShut {NoStop}%
\bibitem [{\citenamefont {Hamamoto}\ \emph {et~al.}(2015)\citenamefont
  {Hamamoto}, \citenamefont {Ezawa},\ and\ \citenamefont
  {Nagaosa}}]{Nagaosa2015}%
  \BibitemOpen
  \bibfield  {author} {\bibinfo {author} {\bibfnamefont {K.}~\bibnamefont
  {Hamamoto}}, \bibinfo {author} {\bibfnamefont {M.}~\bibnamefont {Ezawa}}, \
  and\ \bibinfo {author} {\bibfnamefont {N.}~\bibnamefont {Nagaosa}},\ }\href
  {\doibase 10.1103/PhysRevB.92.115417} {\bibfield  {journal} {\bibinfo
  {journal} {Phys. Rev. B}\ }\textbf {\bibinfo {volume} {92}},\ \bibinfo
  {pages} {115417} (\bibinfo {year} {2015})}\BibitemShut {NoStop}%
\bibitem [{\citenamefont {Banerjee}\ \emph {et~al.}(2014)\citenamefont
  {Banerjee}, \citenamefont {Rowland}, \citenamefont {Erten},\ and\
  \citenamefont {Randeria}}]{Banerjee2014}%
  \BibitemOpen
  \bibfield  {author} {\bibinfo {author} {\bibfnamefont {S.}~\bibnamefont
  {Banerjee}}, \bibinfo {author} {\bibfnamefont {J.}~\bibnamefont {Rowland}},
  \bibinfo {author} {\bibfnamefont {O.}~\bibnamefont {Erten}}, \ and\ \bibinfo
  {author} {\bibfnamefont {M.}~\bibnamefont {Randeria}},\ }\href {\doibase
  10.1103/PhysRevX.4.031045} {\bibfield  {journal} {\bibinfo  {journal} {Phys.
  Rev. X}\ }\textbf {\bibinfo {volume} {4}},\ \bibinfo {pages} {031045}
  (\bibinfo {year} {2014})}\BibitemShut {NoStop}%
\bibitem [{\citenamefont {Hofstadter}(1976)}]{Hofstadter1976}%
  \BibitemOpen
  \bibfield  {author} {\bibinfo {author} {\bibfnamefont {D.~R.}\ \bibnamefont
  {Hofstadter}},\ }\href {\doibase 10.1103/PhysRevB.14.2239} {\bibfield
  {journal} {\bibinfo  {journal} {Phys. Rev. B}\ }\textbf {\bibinfo {volume}
  {14}},\ \bibinfo {pages} {2239} (\bibinfo {year} {1976})}\BibitemShut
  {NoStop}%
\bibitem [{\citenamefont {G\"obel}\ \emph {et~al.}(2017)\citenamefont
  {G\"obel}, \citenamefont {Mook}, \citenamefont {Henk},\ and\ \citenamefont
  {Mertig}}]{Gobel2017}%
  \BibitemOpen
  \bibfield  {author} {\bibinfo {author} {\bibfnamefont {B.}~\bibnamefont
  {G\"obel}}, \bibinfo {author} {\bibfnamefont {A.}~\bibnamefont {Mook}},
  \bibinfo {author} {\bibfnamefont {J.}~\bibnamefont {Henk}}, \ and\ \bibinfo
  {author} {\bibfnamefont {I.}~\bibnamefont {Mertig}},\ }\href {\doibase
  10.1103/PhysRevB.95.094413} {\bibfield  {journal} {\bibinfo  {journal} {Phys.
  Rev. B}\ }\textbf {\bibinfo {volume} {95}},\ \bibinfo {pages} {094413}
  (\bibinfo {year} {2017})}\BibitemShut {NoStop}%
\bibitem [{\citenamefont {G{\"o}bel}\ \emph {et~al.}(2018)\citenamefont
  {G{\"o}bel}, \citenamefont {Mook}, \citenamefont {Henk},\ and\ \citenamefont
  {Mertig}}]{Gobel2018}%
  \BibitemOpen
  \bibfield  {author} {\bibinfo {author} {\bibfnamefont {B.}~\bibnamefont
  {G{\"o}bel}}, \bibinfo {author} {\bibfnamefont {A.}~\bibnamefont {Mook}},
  \bibinfo {author} {\bibfnamefont {J.}~\bibnamefont {Henk}}, \ and\ \bibinfo
  {author} {\bibfnamefont {I.}~\bibnamefont {Mertig}},\ }\href {\doibase
  10.1140/epjb/e2018-90090-0} {\bibfield  {journal} {\bibinfo  {journal} {The
  European Physical Journal B}\ }\textbf {\bibinfo {volume} {91}},\ \bibinfo
  {pages} {179} (\bibinfo {year} {2018})}\BibitemShut {NoStop}%
\bibitem [{\citenamefont {Sorn}\ \emph {et~al.}(2019)\citenamefont {Sorn},
  \citenamefont {Divic},\ and\ \citenamefont {Paramekanti}}]{Sorn2019}%
  \BibitemOpen
  \bibfield  {author} {\bibinfo {author} {\bibfnamefont {S.}~\bibnamefont
  {Sorn}}, \bibinfo {author} {\bibfnamefont {S.}~\bibnamefont {Divic}}, \ and\
  \bibinfo {author} {\bibfnamefont {A.}~\bibnamefont {Paramekanti}},\ }\href
  {\doibase 10.1103/PhysRevB.100.174411} {\bibfield  {journal} {\bibinfo
  {journal} {Phys. Rev. B}\ }\textbf {\bibinfo {volume} {100}},\ \bibinfo
  {pages} {174411} (\bibinfo {year} {2019})}\BibitemShut {NoStop}%
\bibitem [{\citenamefont {Mahan}(2000)}]{Mahan2000}%
  \BibitemOpen
  \bibfield  {author} {\bibinfo {author} {\bibfnamefont {G.}~\bibnamefont
  {Mahan}},\ }\href@noop {} {\emph {\bibinfo {title} {Many-Particle
  Physics}}},\ Physics of Solids and Liquids\ (\bibinfo  {publisher} {Springer
  US},\ \bibinfo {year} {2000})\BibitemShut {NoStop}%
\bibitem [{\citenamefont {Coleman}(2015)}]{Coleman2015}%
  \BibitemOpen
  \bibfield  {author} {\bibinfo {author} {\bibfnamefont {P.}~\bibnamefont
  {Coleman}},\ }\href@noop {} {\emph {\bibinfo {title} {Introduction to
  many-body physics}}}\ (\bibinfo  {publisher} {Cambridge University Press},\
  \bibinfo {year} {2015})\BibitemShut {NoStop}%
\bibitem [{\citenamefont {Choi}\ \emph {et~al.}(2019)\citenamefont {Choi},
  \citenamefont {Tai},\ and\ \citenamefont {Zhu}}]{Zhu2019}%
  \BibitemOpen
  \bibfield  {author} {\bibinfo {author} {\bibfnamefont {H.}~\bibnamefont
  {Choi}}, \bibinfo {author} {\bibfnamefont {Y.-Y.}\ \bibnamefont {Tai}}, \
  and\ \bibinfo {author} {\bibfnamefont {J.-X.}\ \bibnamefont {Zhu}},\ }\href
  {\doibase 10.1103/PhysRevB.99.134437} {\bibfield  {journal} {\bibinfo
  {journal} {Phys. Rev. B}\ }\textbf {\bibinfo {volume} {99}},\ \bibinfo
  {pages} {134437} (\bibinfo {year} {2019})}\BibitemShut {NoStop}%
\bibitem [{\citenamefont {Tong}(2016)}]{Tong2016}%
  \BibitemOpen
  \bibfield  {author} {\bibinfo {author} {\bibfnamefont {D.}~\bibnamefont
  {Tong}},\ }\href@noop {} {\bibfield  {journal} {\bibinfo  {journal} {arXiv
  preprint arXiv:1606.06687}\ } (\bibinfo {year} {2016})}\BibitemShut {NoStop}%
\bibitem [{\citenamefont {Tatara}\ and\ \citenamefont
  {Kawamura}(2002)}]{Kawamura2002}%
  \BibitemOpen
  \bibfield  {author} {\bibinfo {author} {\bibfnamefont {G.}~\bibnamefont
  {Tatara}}\ and\ \bibinfo {author} {\bibfnamefont {H.}~\bibnamefont
  {Kawamura}},\ }\href {\doibase 10.1143/JPSJ.71.2613} {\bibfield  {journal}
  {\bibinfo  {journal} {Journal of the Physical Society of Japan}\ }\textbf
  {\bibinfo {volume} {71}},\ \bibinfo {pages} {2613} (\bibinfo {year}
  {2002})},\ \Eprint
  {http://arxiv.org/abs/https://doi.org/10.1143/JPSJ.71.2613}
  {https://doi.org/10.1143/JPSJ.71.2613} \BibitemShut {NoStop}%
\bibitem [{\citenamefont {Nakazawa}\ and\ \citenamefont
  {Kohno}(2014)}]{Kohno2014}%
  \BibitemOpen
  \bibfield  {author} {\bibinfo {author} {\bibfnamefont {K.}~\bibnamefont
  {Nakazawa}}\ and\ \bibinfo {author} {\bibfnamefont {H.}~\bibnamefont
  {Kohno}},\ }\href {\doibase 10.7566/JPSJ.83.073707} {\bibfield  {journal}
  {\bibinfo  {journal} {Journal of the Physical Society of Japan}\ }\textbf
  {\bibinfo {volume} {83}},\ \bibinfo {pages} {073707} (\bibinfo {year}
  {2014})},\ \Eprint
  {http://arxiv.org/abs/https://doi.org/10.7566/JPSJ.83.073707}
  {https://doi.org/10.7566/JPSJ.83.073707} \BibitemShut {NoStop}%
\bibitem [{\citenamefont {Denisov}\ \emph {et~al.}(2016)\citenamefont
  {Denisov}, \citenamefont {Rozhansky}, \citenamefont {Averkiev},\ and\
  \citenamefont {Lahderanta}}]{Denisov2016}%
  \BibitemOpen
  \bibfield  {author} {\bibinfo {author} {\bibfnamefont {K.~S.}\ \bibnamefont
  {Denisov}}, \bibinfo {author} {\bibfnamefont {I.~V.}\ \bibnamefont
  {Rozhansky}}, \bibinfo {author} {\bibfnamefont {N.~S.}\ \bibnamefont
  {Averkiev}}, \ and\ \bibinfo {author} {\bibfnamefont {E.}~\bibnamefont
  {Lahderanta}},\ }\href {\doibase 10.1103/PhysRevLett.117.027202} {\bibfield
  {journal} {\bibinfo  {journal} {Phys. Rev. Lett.}\ }\textbf {\bibinfo
  {volume} {117}},\ \bibinfo {pages} {027202} (\bibinfo {year}
  {2016})}\BibitemShut {NoStop}%
\bibitem [{\citenamefont {Sim}\ \emph {et~al.}(1998)\citenamefont {Sim},
  \citenamefont {Ahn}, \citenamefont {Chang}, \citenamefont {Ihm},
  \citenamefont {Kim},\ and\ \citenamefont {Lee}}]{Sim1998}%
  \BibitemOpen
  \bibfield  {author} {\bibinfo {author} {\bibfnamefont {H.-S.}\ \bibnamefont
  {Sim}}, \bibinfo {author} {\bibfnamefont {K.-H.}\ \bibnamefont {Ahn}},
  \bibinfo {author} {\bibfnamefont {K.~J.}\ \bibnamefont {Chang}}, \bibinfo
  {author} {\bibfnamefont {G.}~\bibnamefont {Ihm}}, \bibinfo {author}
  {\bibfnamefont {N.}~\bibnamefont {Kim}}, \ and\ \bibinfo {author}
  {\bibfnamefont {S.~J.}\ \bibnamefont {Lee}},\ }\href {\doibase
  10.1103/PhysRevLett.80.1501} {\bibfield  {journal} {\bibinfo  {journal}
  {Phys. Rev. Lett.}\ }\textbf {\bibinfo {volume} {80}},\ \bibinfo {pages}
  {1501} (\bibinfo {year} {1998})}\BibitemShut {NoStop}%
\bibitem [{\citenamefont {Reijniers}\ \emph {et~al.}(1999)\citenamefont
  {Reijniers}, \citenamefont {Peeters},\ and\ \citenamefont
  {Matulis}}]{Matulis1999}%
  \BibitemOpen
  \bibfield  {author} {\bibinfo {author} {\bibfnamefont {J.}~\bibnamefont
  {Reijniers}}, \bibinfo {author} {\bibfnamefont {F.~M.}\ \bibnamefont
  {Peeters}}, \ and\ \bibinfo {author} {\bibfnamefont {A.}~\bibnamefont
  {Matulis}},\ }\href {\doibase 10.1103/PhysRevB.59.2817} {\bibfield  {journal}
  {\bibinfo  {journal} {Phys. Rev. B}\ }\textbf {\bibinfo {volume} {59}},\
  \bibinfo {pages} {2817} (\bibinfo {year} {1999})}\BibitemShut {NoStop}%
\bibitem [{\citenamefont {Korm\'anyos}\ \emph {et~al.}(2008)\citenamefont
  {Korm\'anyos}, \citenamefont {Rakyta}, \citenamefont {Oroszl\'any},\ and\
  \citenamefont {Cserti}}]{Cserti2008}%
  \BibitemOpen
  \bibfield  {author} {\bibinfo {author} {\bibfnamefont {A.}~\bibnamefont
  {Korm\'anyos}}, \bibinfo {author} {\bibfnamefont {P.}~\bibnamefont {Rakyta}},
  \bibinfo {author} {\bibfnamefont {L.}~\bibnamefont {Oroszl\'any}}, \ and\
  \bibinfo {author} {\bibfnamefont {J.}~\bibnamefont {Cserti}},\ }\href
  {\doibase 10.1103/PhysRevB.78.045430} {\bibfield  {journal} {\bibinfo
  {journal} {Phys. Rev. B}\ }\textbf {\bibinfo {volume} {78}},\ \bibinfo
  {pages} {045430} (\bibinfo {year} {2008})}\BibitemShut {NoStop}%
\bibitem [{Note1()}]{Note1}%
  \BibitemOpen
  \bibinfo {note} {This statement depends on a few variables including the
  actual frequency of the applied field, the electron group velocity, and the
  degree of inhomogeneity in the system. A sufficient condition is that the
  distance travelled by the electron in a few cycles is small compared with the
  length scale characterizing the variation of the inhomogeneity, e.g.
  periodicity $L$ in the case of a SkX.}\BibitemShut {Stop}%
\bibitem [{Note2()}]{Note2}%
  \BibitemOpen
  \bibinfo {note} {The sign of $\chi _R$ is chosen following Ref.\protect
  \rev@citealp {Banerjee2014}}\BibitemShut {NoStop}%
\bibitem [{\citenamefont {Lux}\ \emph {et~al.}(2020)\citenamefont {Lux},
  \citenamefont {Freimuth}, \citenamefont {Bl\"ugel},\ and\ \citenamefont
  {Mokrousov}}]{Lux2020}%
  \BibitemOpen
  \bibfield  {author} {\bibinfo {author} {\bibfnamefont {F.~R.}\ \bibnamefont
  {Lux}}, \bibinfo {author} {\bibfnamefont {F.}~\bibnamefont {Freimuth}},
  \bibinfo {author} {\bibfnamefont {S.}~\bibnamefont {Bl\"ugel}}, \ and\
  \bibinfo {author} {\bibfnamefont {Y.}~\bibnamefont {Mokrousov}},\ }\href
  {\doibase 10.1103/PhysRevLett.124.096602} {\bibfield  {journal} {\bibinfo
  {journal} {Phys. Rev. Lett.}\ }\textbf {\bibinfo {volume} {124}},\ \bibinfo
  {pages} {096602} (\bibinfo {year} {2020})}\BibitemShut {NoStop}%
\bibitem [{\citenamefont {Zhang}\ \emph {et~al.}(2020)\citenamefont {Zhang},
  \citenamefont {Ishizuka}, \citenamefont {Zhang}, \citenamefont {Hal\'asz},\
  and\ \citenamefont {Batista}}]{Batista2020}%
  \BibitemOpen
  \bibfield  {author} {\bibinfo {author} {\bibfnamefont {S.-S.}\ \bibnamefont
  {Zhang}}, \bibinfo {author} {\bibfnamefont {H.}~\bibnamefont {Ishizuka}},
  \bibinfo {author} {\bibfnamefont {H.}~\bibnamefont {Zhang}}, \bibinfo
  {author} {\bibfnamefont {G.~B.}\ \bibnamefont {Hal\'asz}}, \ and\ \bibinfo
  {author} {\bibfnamefont {C.~D.}\ \bibnamefont {Batista}},\ }\href {\doibase
  10.1103/PhysRevB.101.024420} {\bibfield  {journal} {\bibinfo  {journal}
  {Phys. Rev. B}\ }\textbf {\bibinfo {volume} {101}},\ \bibinfo {pages}
  {024420} (\bibinfo {year} {2020})}\BibitemShut {NoStop}%
\bibitem [{\citenamefont {Bouaziz}\ \emph {et~al.}(2021)\citenamefont
  {Bouaziz}, \citenamefont {Ishida}, \citenamefont {Lounis},\ and\
  \citenamefont {Bl\"ugel}}]{Blugel2021}%
  \BibitemOpen
  \bibfield  {author} {\bibinfo {author} {\bibfnamefont {J.}~\bibnamefont
  {Bouaziz}}, \bibinfo {author} {\bibfnamefont {H.}~\bibnamefont {Ishida}},
  \bibinfo {author} {\bibfnamefont {S.}~\bibnamefont {Lounis}}, \ and\ \bibinfo
  {author} {\bibfnamefont {S.}~\bibnamefont {Bl\"ugel}},\ }\href {\doibase
  10.1103/PhysRevLett.126.147203} {\bibfield  {journal} {\bibinfo  {journal}
  {Phys. Rev. Lett.}\ }\textbf {\bibinfo {volume} {126}},\ \bibinfo {pages}
  {147203} (\bibinfo {year} {2021})}\BibitemShut {NoStop}%
\bibitem [{\citenamefont {Ahn}\ \emph {et~al.}(2021)\citenamefont {Ahn},
  \citenamefont {Guo}, \citenamefont {Nagaosa},\ and\ \citenamefont
  {Vishwanath}}]{Ashvin2021}%
  \BibitemOpen
  \bibfield  {author} {\bibinfo {author} {\bibfnamefont {J.}~\bibnamefont
  {Ahn}}, \bibinfo {author} {\bibfnamefont {G.-Y.}\ \bibnamefont {Guo}},
  \bibinfo {author} {\bibfnamefont {N.}~\bibnamefont {Nagaosa}}, \ and\
  \bibinfo {author} {\bibfnamefont {A.}~\bibnamefont {Vishwanath}},\
  }\href@noop {} {\bibfield  {journal} {\bibinfo  {journal} {arXiv preprint
  arXiv:2103.01241}\ } (\bibinfo {year} {2021})}\BibitemShut {NoStop}%
\end{thebibliography}%

\end{document}